\documentclass[useAMS,usenatbib]{mn2e}
\usepackage{amssymb,graphicx,amsmath}
\usepackage{color, soul}	
\usepackage{todonotes}	
\setstcolor{red}
\usepackage[pass,letterpaper]{geometry}	
\usepackage{pdflscape}	

\def\araa{{\it ARA\&A}~} 
\def\aap{{\it A\&A}~} 
\def\aapr{{\it A\&AR}~} 
\def\aj{{\it AJ}~} 
\def\apj{{\it ApJ}~} 
\def\apjl{{\it ApJ}~} 
\def\apjs{{\it ApJS}~} 
\def\mnras{{\it MNRAS}~} 
\def\physrep{{\it Phys. Rep.}~}

\title[Spectral Models for LLAGNs in LINERs]{Spectral Models for Low-luminosity Active Galactic Nuclei in LINERs: The Role of Advection-dominated Accretion and Jets} 
\author[Nemmen et al.]{Rodrigo S. Nemmen,$^{1}$\thanks{E-mail: rodrigo.nemmen@nasa.gov} Thaisa Storchi-Bergmann,$^2$ Michael Eracleous,$^3$ \\
$^{1}$NASA Goddard Space Flight Center, Greenbelt, MD 20771, USA \\
$^{2}$Instituto de F\'isica, Universidade Federal do Rio Grande do Sul, Campus do Vale, Porto Alegre, RS, Brazil \\
$^3$Department of Astronomy and Astrophysics, Pennsylvania State University, 525 Davey Lab, University Park, PA 16802
}

\begin{document}

\date{Accepted 2013 June 18. Received 2013 June 17; in original form 2013 March 31}

\pagerange{\pageref{firstpage}--\pageref{lastpage}} \pubyear{2013}

\maketitle

\label{firstpage}

\begin{abstract}
We perform an exploratory study of the physical properties of accretion
flows and jets in low-luminosity active galactic nuclei (LLAGNs) by
modeling the spectral energy distributions (SEDs) of 12 LLAGNs in
low-ionization nuclear emission-line regions (LINERs). These SEDs we
constructed from high-resolution radio, X-ray and optical/UV
observations of the immediate vicinity of the black hole. We adopt a
coupled accretion-jet model comprising an inner advection-dominated
accretion flow (ADAF) and an outer standard thin disk. We present
best-fit models in which either the ADAF or the jet dominate the X-ray
emission.  Six sources in our sample display an optical-UV excess with
respect to ADAF and jet models; this excess can be explained as
emission from the truncated disk with transition radii $30-225~R_S$ in
four of them. In almost all sources the optical emission can also be
attributed to unresolved, old stellar clusters with masses $\sim
10^7-10^8~M_\odot$.  We find evidence for a correlation between the
accretion rate and jet power and an anti-correlation between the
radio-loudness and the accretion rate.  We confirm previous findings
that the radio emission is severely underpredicted by ADAF models and
explained by the relativistic jet. We find evidence for a nonlinear
relation between the X-ray and bolometric luminosities and a slight IR
excess in the average model SED compared to that of quasars.  We
suggest that the hardness of the X-ray spectrum can be used to identify
the X-ray emission mechanism and discuss directions for progress in
understanding the origin of the X-rays.
\end{abstract}


\section{Introduction}

In the present-day universe, supermassive black holes (SMBHs) are underfed compared to the ones at high redshifts and are ``sleeping''. Most of SMBH activity at low $z$ is dominated by the weak end of the AGN luminosity function in the form of low-luminosity AGNs (LLAGNs; \citealt{Ho95,Ho97v,Nagar05,Ho08}). 
LLAGNs are extremely sub-Eddington systems which are many orders of magnitude less luminous than quasars, with average bolometric luminosities $L_{\rm bol} \lesssim 10^{42} \ {\rm erg \ s}^{-1}$ and an average Eddington ratio of $L_{\rm bol}/L_{\rm Edd} \sim 10^{-5}$ \citep{Ho09} where $L_{\rm Edd}$ is the Eddington luminosity.  Though the LLAGN phase dominates the time evolution of SMBHs, it contributes little to their mass growth \citep{Hopkins06llagn,Sijacki07,Merloni08,Xu10}.
The bulk of the LLAGN population ($\approx 2/3$; \citealt{Ho08,Ho09}) reside in low-ionization nuclear emission-line regions (LINERs; \citealt{Heckman80,Ho97iii}). 

The observational properties of LLAGNs are quite different from those of more luminous AGNs. Regarding the broadband spectral energy distributions (SEDs), LLAGNs seem not to have the thermal continuum prominence in the ultraviolet (UV) -- the ``big blue bump'' -- which is one of the signatures of the presence of an optically thick, geometrically thin accretion disk \citep{Ho99, Nemmen06, Wu07, Ho08, Eracleous10}. Regarding the emission-lines, LLAGNs typically have weak and narrow Fe K$\alpha$ emission \citep{Terashima02} and a handful of LINERs display broad double-peaked H$\alpha$ lines (e.g., \citealt{Storchi-Bergmann03}); these properties of the emission-line spectrum are consistent with the absence of a thin accretion disk, or a thin accretion disk whose inner radius is truncated at $\gtrsim 100 GM/c^2$ \citep{Chen89, Chen89b}. Last but not least, with the typical gas supply available via ordinary mass loss from evolved stars and gravitational capture of gas from the hot interstellar medium (i.e. Bondi accretion) in nearby galaxies, LLAGNs would be expected to produce much higher luminosities than observed on the assumption of standard thin disks with a $10\%$ radiative efficiency \citep{Soria06b, Ho09}.
Taken together, this set of observational properties favors the scenario in which the accretion flow in LLAGNs is advection-dominated or radiatively inefficient.

Advection-dominated accretion flows (ADAFs\footnote{In this paper, we consider ADAFs and radiatively inefficient accretion flows (RIAFs) to be the same kind of accretion flow solution. For a clarification regarding the terminology, see \citealt{Yu11}.}; 
for reviews see \citealt{Narayan98rev,Yuan07,Narayan08}) are very hot, geometrically thick, optically thin flows which are typified by low radiative efficiencies ($L \ll 0.1 \dot{M} c^2$) and occur at low accretion rates ($\dot{M} \lesssim 0.01 \dot{M}_{\rm Edd}$). SMBHs are thought to spend most of their lives in the ADAF state \citep{Hopkins06old,Xu10}, the best studied individual case being Sgr A* (e.g., \citealt{Yuan07}). 

In many LLAGNs, another component in the accretion flow besides the ADAF is required in order to account for a number of observations, including a prominence in the mid or near-IR and steep fall-off of the spectrum in the optical-UV region -- a ``red bump'' \citep{Lasota96,Quataert99,Nemmen06,Ho08,Yu11,Wu13} -- as well as the presence of double-peaked Balmer emission lines (e.g., \citealt{Storchi-Bergmann03,Eracleous09}): the emission from a thin accretion disk whose inner radius is truncated at the outer radius of the ADAF. The accretion flow may begin as a standard thin disk but somehow at a certain transition radius it gradually switches from a cold to a hot ADAF mode. The details of how this transition might happen are still not well understood \citep{Manmoto00,Yuan04,Narayan08}, but it seems to be analogous to the transition between the different spectral states in black hole binary systems \citep{Remillard06,Done07}.

\citet{Maoz07} challenged the scenario of the central engines of LLAGNs consisting of ADAFs and truncated thin disks. Maoz argues that LLAGNs in LINERs have UV/X-ray luminosity ratios similar on average to those of brighter Seyfert 1s and based on that observation he posits that thin disks extending all the way down to the radius of the innermost stable circular orbits (ISCO) persist even for LLAGNs, despite their smaller accretion rates. \citet{Yu11} showed that the SEDs compiled by \citet{Maoz07} are naturally fitted by ADAF models with the addition of a truncated thin disk and also discussed on theoretical grounds why the ADAF model has a superior explanation capability than the pure thin disk model. The ADAF model is the only model that can naturally account for the set of observational properties of LLAGNs within a self-contained theoretical framework. Hence, it is the physical scenario adopted in this work.

ADAFs are relevant to the understanding of AGN feedback since they are quite efficient at producing powerful outflows and jets, as suggested by theoretical studies including analytical theory \citep{Narayan94,Nemmen07,Begelman12} and numerical simulations \citep{McKinney04,Hawley06,Sasha11,McKinney12,Yuan12b}. This is in line with the different observational studies that demonstrate that LLAGNs are generally radio-loud \citep{Nagar00,Nagar01,Ho02, Ho08,Younes12} and accompanied by powerful jets (e.g., \citealt{Heinz07, Merloni08}). 

It is clear then that an understanding of the physical nature of LLAGNs in LINERs will shed light on the nature of black hole accretion, outflows and consequently black hole growth and AGN feedback in present-day galaxies. One of the best ways of exploring the astrophysics of black hole accretion and outflows is by using multiwavelength observations of black hole systems and comparing the spectra predicted by specific models with the data. The goal of this work is therefore to probe the physics of accretion and ejection in the LLAGN population, by carrying out an exploratory modeling of their nuclear, broadband, radio to X-rays SEDs which provide constraints on physical models for the emission of the accretion flow and the jet.
Furthermore, with a large enough sample of systems, we can derive from the fits to the data the parameters that characterize the central engines and build a census of the ``astrophysical diet'' of low-state AGNs.

The structure of this paper is as follows. In Section \ref{sec:sample} we describe the sample of 21 LLAGNs in LINERs that we used, the data and the criteria that we use to select the 12 best-sampled SEDs. Section \ref{sec:models} describes the physical model that we adopted in order to interpret and fit the SEDs and derive the central engine parameters. In Section \ref{sec:seds} we describe the model fits to the SEDs.  In Section \ref{sec:pars} we describe the parameters resulting from our model fits and possible correlations between accretion and jet production. We present the average SED resulting from our fits in \textsection \ref{sec:avgsed}, including the predicted emission in wavebands which can be observed with future facilities such as ALMA. 
We discuss the role of the ADAF and jet in explaining the X-ray emission, the constraints on the transition radius and the limitations of our models in \textsection \ref{sec:disc}, outlining along the way the directions for progress on these issues. We conclude by presenting a summary of our results in \textsection \ref{sec:end}. 
In the Appendix, we discuss the uncertainties in the model parameters that we derive and present illustrative model fits to the 9 sparsely-sampled SEDs which were left out of the main analysis.

\section{Sample} \label{sec:sample}

\citet{Eracleous10} (hereafter EHF10) compiled a sample of 35 SEDs of LLAGNs found in LINERs which include high spatial resolution optical and UV observations with the \textit{Hubble Space Telescope} (HST), as well as X-ray observations with \emph{Chandra} and high-resolution radio observations with the Very Large Array -- VLA -- or VLBA/VLBI.  

LINERs constitute quite a heterogeneous class of objects. For instance, their nature and particularly the nature of the source of power for their line emission has been repeatedly questioned (e.g. \citealt{Eracleous10budget,Sarzi10,Yan12,Singh13}). For example, \cite{Yan12} concluded that the line emission observed in large-aperture ($>100$ pc) spectroscopy in most of the LINER-like galaxies in their sample is not primarily powered by an accreting black hole and thus urge caution in associating the corresponding line emission with the AGN bolometric luminosity. Therefore, in order to select from the sample of EHF10 the sources for which the nuclear emission is most probably associated with black hole accretion -- thus providing the best SEDs to be compared against accretion flow and jet models -- we carefully applied the following selection criteria. 

\noindent
{\it Availability of black hole mass estimates --}
We limited ourselves only to galaxies with black hole mass estimates since the mass is one of the fundamental input parameters to the spectral models.

\noindent
{\it Presence of a compact nuclear radio core and availability of at least one high-resolution radio detection with VLA or VLBA/VLBI --}
Radio measurements constrain the relative importance of the synchrotron emission from the jet and ADAF.

\noindent
{\it Presence of nuclear X-ray point source and availability of its X-ray spectra --}
Given that the X-ray spectrum is a crucial constraint for the models, we only selected objects for which there is detected nuclear X-ray emission as opposed to upper limits.

\noindent
{\it Availability of at least one high spatial resolution nuclear observation in the $1 \, \mu{\rm m}-1000 \ {\rm \AA}$ waveband --}
Measurements in the optical/near-IR potentially constrain the emission of the truncated thin accretion disk, whereas observations in optical/UV can constrain the inverse Compton emission produced in the inner regions of the ADAF. Therefore, we selected only sources which have at least one observation at high spatial resolution ($<1''$) in the $1 \, \mu{\rm m}-1000 \ {\rm \AA}$ waveband.

\noindent
{\it Absence of prominent absorption features from hot stars in the nuclear UV spectrum (when available) --}
We discarded from our analysis three sources (NGC 404, NGC 5055 and NGC 6500) from the EHF10 sample which display prominent absorption lines in their UV spectra, suggestive that stellar emission dominates the UV light (cf. \citealt{Maoz98uv}; EHF10 for more details).

These selection criteria leave us with 12 LINERs which are listed in Table \ref{tab:sample}. This table also lists their corresponding Hubble and LINER Types, black hole masses, bolometric luminosities and Eddington ratios. 

\begin{table*}
\centering
\caption{Sample of galaxies and their basic properties$^{\rm a}$.}
\begin{tabular}{@{}llrccccc@{}}
\hline
Galaxy & Hubble & Distance$^{\rm b}$ & $\log$             & LINER & $L_{\rm X}$ & $L_{\rm bol}$ & $L_{\rm bol}/L_{\rm Edd}$ \\
       & Type   & (Mpc)    & $(M_{\rm BH}/M_\odot)$ & Type  & (erg s$^{-1}$)$^{\rm c}$ & (erg s$^{-1}$)$^{\rm d}$ &   \\
\hline
NGC 1097                 &  SB(rl)b    & 14.5 (1) & 8.1 & L1      & $4.3 \times 10^{40}$ & $8.5 \times 10^{41}$ & $5 \times 10^{-5}$  \\
NGC 3031 (M81)           &  SA(s)ab    &  3.6 (3) & 7.8 & S1.5/L1 & $1.9 \times 10^{40}$ & $2.1 \times 10^{41}$ & $3 \times 10^{-5}$  \\
NGC 3998                 &  SA(r)0     & 13.1 (2) & 8.9 & L1      & $2.6 \times 10^{41}$ & $1.4 \times 10^{43}$ & $1 \times 10^{-4}$  \\
NGC 4143                 &  SAB(s)0    & 14.8 (2) & 8.3 & L1      & $1.1 \times 10^{40}$ & $3.2 \times 10^{41}$ & $1 \times 10^{-5}$  \\
NGC 4261                 &  E2-3       & 31.6 (2) & 8.7 & L2      & $1.0 \times 10^{41}$ & $6.8 \times 10^{41}$ & $1 \times 10^{-5}$  \\
NGC 4278                 &  E1-2       & 14.9 (2) & 8.6 & L1      & $9.1 \times 10^{39}$ & $2.7 \times 10^{41}$ & $5 \times 10^{-6}$  \\
NGC 4374 (M84, 3C 272.1) &  E1         & 17.1 (2) & 8.9 & L2      & $3.5 \times 10^{39}$ & $5.0 \times 10^{41}$ & $5 \times 10^{-6}$  \\
NGC 4486 (M87, 3C 274)   &  E0-1       & 14.9 (2) & 9.8 & L2      & $1.6 \times 10^{40}$ & $9.8 \times 10^{41}$ & $7 \times 10^{-6}$  \\
NGC 4552 (M89)           &  E          & 14.3 (2) & 8.2 & T2      & $2.6 \times 10^{39}$ & $7.8 \times 10^{40}$ & $4 \times 10^{-6}$  \\
NGC 4579 (M58)           &  SAB(rs)b   & 21.0 (4) & 7.8 & L2      & $1.8 \times 10^{41}$ & $1.0 \times 10^{42}$ & $1 \times 10^{-4}$  \\
NGC 4594 (M104)          &  SA(s)a     &  9.1 (2) & 8.5 & L2      & $1.6 \times 10^{40}$ & $4.8 \times 10^{41}$ & $1 \times 10^{-5}$  \\
NGC 4736 (M94)           &  (R)SA(r)ab &  4.8 (2) & 7.1 & L2      & $5.9 \times 10^{38}$ & $1.8 \times 10^{40}$ & $1 \times 10^{-5}$  \\
\hline
\end{tabular}
\begin{flushleft}
\footnotesize 
Notes: \\
(a) The information in this table was obtained from EHF10, see text.\\
(b) The number in parenthesis gives the source and
  method of the distance measurement, as follows: (1) From the catalog
  of \citet{Tully88}, determined from a model for peculiar velocities and
  assuming $H_0=75{\rm\; km\;s^{-1}\;Mpc^{-1}}$; (2) From \citet{Tonry01}, who used the surface brightness fluctuation method. Following \citet{Jensen03}, the distance modulus
  reported by \citet{Tonry01} was corrected by subtracting 0.16
  mag; (3) From \citet{Freedman94,Freedman01}, who used Cepheid
  variables; (4) From \citet{Gavazzi99} who used the Tully-Fisher
  method.\\
(c) $L_{\rm X}$ is the X-ray luminosity in the $2-10$ keV range.\\
(d) The bolometric luminosities of NGC 1097, NGC 3031,
  NGC 3998, NGC 4374, NGC 4486, NGC 4579 and NGC 4594 were estimated
  by integrating the SEDs. For all other galaxies $L_{\rm bol}$ was
  determined by scaling $L_{\rm X}$ as described in \textsection \ref{sec:sample}.\\
\end{flushleft}
\label{tab:sample}
\end{table*}

We treat the observations in the IR band taken with lower spatial resolutions ($>1''$) -- i.e. larger apertures -- as upper limits because they include considerable contamination from the emission of the host galaxy. 

The masses of the SMBHs were estimated from the stellar velocity dispersions using the $M_{\rm BH}-\sigma$ relationship \citep{Ferrarese00, Gebhardt00, Tremaine02}, with the exception of NGC 3031 (M81) and NGC 4486 (M87) whose black hole masses were estimated from the stellar and/or gas kinematics (\citealt{Bower00,Devereux03,Gebhardt09}; see Table \ref{tab:sample}). 
As EHF10 note, for the objects with multiple mass determinations using different methods, the estimated masses are consistent with each other within a factor of 2 or better. 

The optical-UV data of all objects were corrected for extinction (EHF10), unlike previous similar SED modeling efforts (e.g. \citealt{Wu07,Yuan09,Yu11}). 
In order to compute the bolometric luminosities from the SEDs, EHF10 used two methods. For the objects with the best sampled SEDs, they computed $L_{\rm bol}$ by integrating the SEDs directly, ignoring upper limit data points. These objects are NGC 3031, NGC 3998, NGC 4374, NGC 4486, NGC 4579 and NGC 4594. For these objects they assumed that pairs of neighboring points in the SEDs defined a power law, integrated each segment individually and summed the segments to obtain $L_{\rm bol}$. From this set of best sampled SEDs they estimated the average ``bolometric correction'' from the 2-10 keV luminosity to the bolometric luminosity ($L_{\rm bol}=50 L_X$), which they used to obtain $L_{\rm bol}$ for the remaining objects, for which the SEDs are not as well sampled. These bolometric luminosities are listed in Table \ref{tab:sample}. 

The sources in our sample were selected based on the availability of data. As discussed by EHF10, there may be some biases inherited from the surveys from which the data was obtained which rely on objects bright in the radio, UV and X-ray bands. For instance, transition objects are under-represented while type 1.9 LINERs are over-represented. It is not clear at present whether the relative number of LINER types in our sample could impact our modeling results. Therefore, the resulting sample cannot be considered complete since it only consists of objects for which the necessary data are available. The biases resulting from this selection are hard to quantify but we refer the reader to EHF10 for further discussion.

\section{Models for the accretion flow and jet} \label{sec:models}

\begin{figure}
\centering
\includegraphics[scale=0.25]{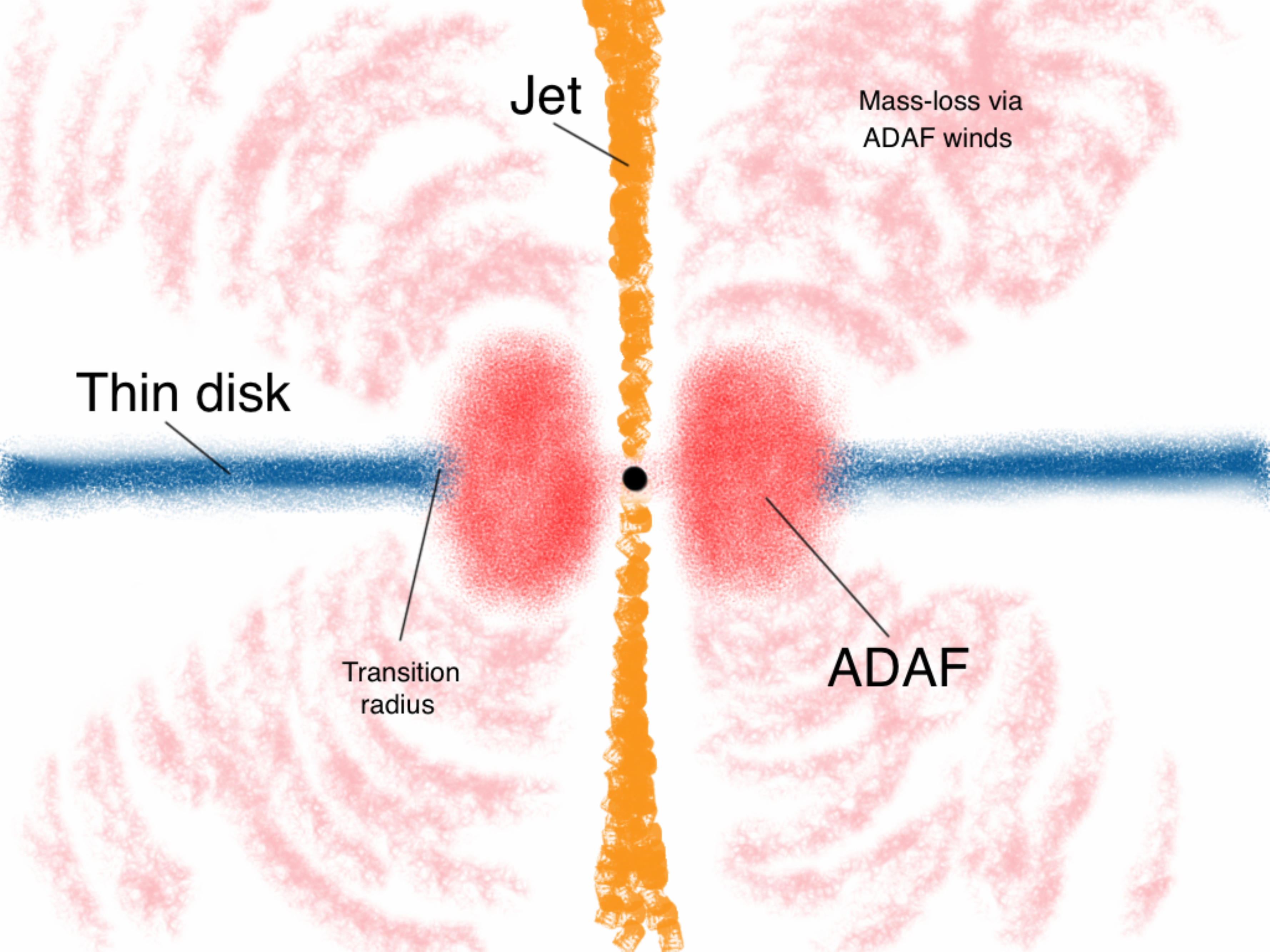}
\caption{Cartoon illustrating the model for the central engines of LLAGNs. It consists of three components: an inner ADAF, an outer truncated thin disk and a relativistic jet.}
\label{fig:cartoon}
\end{figure}

We fit the observed broadband SEDs of the LLAGNs in our sample using a model which consists of three components: an inner ADAF, an outer truncated thin accretion disk and a jet. The components of the model are illustrated in Figure \ref{fig:cartoon}. We describe here the main features of this model. 

\subsection{ADAF component}

The inner part of the accretion flow is in the form of an ADAF which is a hot, geometrically thick, optically thin two-temperature accretion flow, which has low radiative efficiency (e.g., \citealt{Narayan98rev, Kato98}). ADAFs are characterized by the presence of outflows or winds, which prevent a considerable fraction of the gas that is available at large radii from being accreted onto the black hole. This has been suggested by numerical simulations \citep{Stone01,Hawley01,Igumenshchev03,De-Villiers03,Proga03,McKinney04,Yuan12b} and analytical work \citep{Narayan94,Blandford99,Narayan00,Begelman12} (cf. \citealt{Narayan12grmhd} for an alternative view). 
In order to take this mass-loss into account, we follow \citet{Blandford99} and introduce the parameter $s$ by 
\begin{equation}	\label{eq:massloss}
\dot{M}=\dot{M}_{\rm o} \left( \frac{R}{R_{\rm o}} \right)^{s}, 
\end{equation}
to describe the radial variation of the accretion rate $\dot{M}_{\rm o}$ measured at the outer radius of the ADAF, $R_{\rm o}$. The results of the numerical simulations of the dynamics of ADAFs previously mentioned as well as \emph{Chandra} X-ray studies of NGC 3115 and Sgr A* (e.g., \citealt{Wong11,Wang13}) together with submillimeter polarization and Faraday rotation measurements of Sgr A* \citep{Marrone07} suggest that $0.3 \lesssim s \lesssim 1$ (the lower bound is estimated from fitting the SED of Sgr A*; cf. \citealt{Yuan03,Yuan06}). Following these results, in our models we conservatively adopt $s=0.3$ unless otherwise noted.

The other parameters that describe the ADAF solution are the black hole mass $M$; the viscosity parameter $\alpha$; the modified plasma $\beta$ parameter, defined as the ratio between the gas and total pressures, $\beta=P_g/P_{\rm tot}$; $\delta$, the fraction of energy dissipated via turbulence that directly heats electrons; and the adiabatic index $\gamma$. Following \cite{Nemmen06}, in our calculations we adopt $\alpha=0.3$, $\beta=0.9$ and $\gamma=1.5$. Traditional ADAF models adopted $\delta$ to be small ($\delta \lesssim 0.01$; e.g., \citealt{Narayan95}). On the other hand, it has been argued that the value of $\delta$ can be potentially increased due to different physical processes - such as magnetic reconnection - that affect the heating of protons and electrons in hot plasmas (e.g., \citealt{Quataert99turb,Sharma07}). Given the theoretical uncertainty related to the value of $\delta$, we allow it to vary over the range $0.01 \leq \delta \leq 0.5$. 

The cooling mechanisms incorporated in the calculations are synchrotron emission, bremsstrahlung and inverse Comptonization of the seed photons produced by the first two radiative processes. Given the values of the parameters of the ADAF, in order to compute its spectrum we first numerically solve for the global structure and dynamics of the flow, as outlined in \citet{Yuan00,Yuan03}. Obtaining the global solution of the differential equations for the structure of the accretion flow is a two-point boundary value problem. This problem is solved numerically using the shooting method, by varying the eigenvalue $j$ (the specific angular momentum of the flow at the horizon) until the sonic point condition at the sonic radius $R_s$ is satisfied, in addition to the outer boundary conditions \citep{Yuan03}. 

There are three outer boundary conditions that the ADAF solution must satisfy, specified in terms of the three variables of the problem: the ion temperature $T_{i}$, the electron temperature $T_{e}$ and the radial velocity $v$ (or equivalently the angular velocity $\Omega$). Following \citet{Yuan08}, when the outer boundary of the ADAF is at the radius $R_{\rm o}=10^4 R_S$ (where $R_S$ is the Schwarzschild radius) we adopt the outer boundary conditions $T_{\rm out,i}=0.2 T_{\rm vir}$, $T_{\rm out,e}=0.19 T_{\rm vir}$ and $\lambda_{\rm out}=0.2$, where the virial temperature is given by $T_{\rm vir}=3.6 \times 10^{12} (R_S/R) \ \rm K$, $\lambda \equiv v/c_s$ is the Mach number and $c_s$ is the adiabatic sound speed. When the outer boundary is at $R_{\rm o} \sim 10^2 R_S$ we adopt the boundary conditions $T_{\rm out,i}=0.6 T_{\rm vir}$, $T_{\rm out,e}=0.08 T_{\rm vir}$ and $\lambda_{\rm out}=0.5$. After the global solution is calculated, the spectrum of the accretion flow is obtained (see e.g., \citealt{Yuan03} for more details). We verified that if these boundary conditions are varied by a factor of a few, the resulting spectrum does not change much.

\subsection{Thin disk component}	\label{sec:disk}

Our model posits that outside the ADAF there is an outer thin accretion disk with an inner radius truncated at $R_{\rm tr}=R_{\rm o}$ and extending up to $10^5 R_S$ such that the outer radius of the ADAF corresponds to the transition radius to the thin disk.
The other parameters that describe the thin disk solution are the inclination angle $i$, the black hole mass and the accretion rate $\dot{M}_{\rm o}$ (the same as the accretion rate at the outer boundary of the ADAF). 

The thin disk emits locally as a black body and we take into account the reprocessing of the X-ray radiation from the ADAF. This reprocessing effect has only a little impact on the spectrum of the thin disk though, with the resulting SED being almost identical to that of a standard thin disk (e.g., \citealt{Frank02}). 

For sources without optical/UV constraints, we adopt $R_{\rm o} \sim 10^4 R_S$; in this case we simply ignore the contribution of the thin disk emission since for $R_{\rm tr} \sim 10^4 R_S$ the thin disk contributes very little to the emission compared to the ADAF. In sources for which we have available optical/UV data to constrain this component of the flow, we then explore models with $R_{\rm tr} < 10^4 R_S$. 

\subsection{Jet component}	\label{sec:jet}

The SEDs of LLAGNs are generally radio-loud (\citealt{Ho99,Ho01, Ho02, Terashima03}; EHF10); but see \citealt{Maoz07,Sikora07}). The radio prominence of these SEDs is usually explained by invoking the synchrotron emission of relativistic jets, since the the accretion flow does not produce enough radio emission to account for the observed radio luminosity (e.g., \citealt{Ulvestad01, Wu05, Nemmen06, Wu07, Yuan09}; but see \citealt{Liu13}). Some authors even argue that the entire SED of LLAGNs may be explained by the jet component (e.g., \citealt{Falcke04,Markoff08}). We therefore include in our modeling the contribution from the emission from a relativistic jet.

We adopt a jet model based on the internal shock scenario which is used to interpret gamma-ray burst afterglows (e.g., \citealt{Piran99, Spada01, Nemmen12}; see \citealt{Yuan05} for more details). This model has been adopted in previous works to understand the broadband SEDs of X-ray binaries and AGNs \citep{Spada01, Yuan05, Nemmen06, Wu07}. According to this jet model, some fraction of the material in the innermost regions of the accretion flow is transferred to the jet producing an outflow rate $\dot{M}_{\rm jet}$ and a standing shock wave in the region of the jet closest to the black hole is formed. This shock wave is created from the bending of the supersonic accretion flow near the black hole in the direction of the jet. We calculate the shock jump (Rankine-Hugoniot) conditions to find the electron and ion temperatures of the plasma ($T_e$ and $T_i$). We find that the jet spectrum is not very sensitive to changes in $T_e$ and $T_i$, since the emission is completely dominated by non-thermal electrons (see below). We therefore adopt $T_i=6.3 \times 10^{11} \ {\rm K}$ and $T_e=10^{9} \ {\rm K}$ in our calculations of the jet emission. 

The jet is modeled as having a conical geometry with half-opening angle $\phi$ and a bulk Lorentz factor $\Gamma_j$ which are independent of the distance from the black hole. The jet is along the axis of the ADAF and makes an angle $i$ with the line of sight. The internal shocks in the jet are presumably created by the collisions of shells of plasma with different velocities. These shocks accelerate a small fraction $\xi_e$ of the electrons into a power-law energy distribution with index $p$. The energy density of accelerated electrons and the amplified magnetic field are described by two free parameters, $\epsilon_e$ and $\epsilon_B$. 

Following \citet{Nemmen06,Wu07}, we adopt in our calculations of the jet emission the values $\phi=0.1$ radians, $\xi_e=10 \%$ and $\Gamma_j=2.3$, which corresponds to $v/c \approx 0.9$ (except in the case of M87, which has an independent estimate of $\Gamma_j$ available). For six sources we have independent constraints on the value of $i$ (cf. Table \ref{tab:models}). For the other six, we simply adopt $i=30^\circ$. Therefore, there are four free parameters in the jet model: $\dot{M}_{\rm jet}$, $p$, $\epsilon_e$ and $\epsilon_B$. We allow $p$ to vary in the range $2-3$ as expected from shock acceleration theory.
In our calculations we consider the synchrotron emission of the jet, with the optically thick part of the synchrotron spectrum contributing mainly in the radio and the optically thin part contributing mainly in the X-rays.

\subsection{Free parameters and fit procedure} \label{sec:fit}

Throughout this paper we will use the dimensionless mass accretion rates $\dot{m}=\dot{M}/\dot{M}_{\rm Edd}$, noting that the Eddington accretion rate is defined as
$\dot{M}_{\rm Edd} \equiv 22 M /(10^{9} M_\odot) \; M_\odot \, {\rm yr}^{-1}$ 
assuming a $10\%$ radiative efficiency. We will also express the black hole mass in terms of the mass of the sun, $m=M/M_{\odot}$, and the radius in terms of the Schwarzschild radius, $r=R/R_S$.

We have 8 free parameters in the SED fits with our coupled accretion-jet model. Four of these parameters describe the emission of the accretion flow: the accretion rate $\dot{M}_{\rm o}$, the fraction of viscously dissipated energy that directly heats the electrons $\delta$, the transition radius between the inner ADAF and the outer truncated thin disk $R_{\rm tr}$ and the ADAF mass-loss parameter $s$. The other four parameters characterize the jet emission: the mass-loss rate in the jet $\dot{M}_{\rm jet}$, the electron energy spectral index $p$, and the electron and magnetic energy parameters $\epsilon_e$ and $\epsilon_B$. 

Our fitting procedure can be summarized as follows. We begin with a model characterized by the initial values $\dot{m}_{\rm o}=0.001$, $r_{\rm tr}=10^4$, $\delta=0.1$, $s=0.3$, $\dot{m}_{\rm jet}=10^{-6}$, $p=2.2$, $\epsilon_e=0.1$ and $\epsilon_B=0.01$. We adopt an iterative procedure, in which we vary one parameter each time and keep the others fixed,  computing the total emitted spectrum resulting from the sum of the radiation from the jet, truncated disk and ADAF for each iteration. We repeat this process until we obtain an acceptable fit of the total SED to the radio, optical, UV and X-ray observations. 
We judge the goodness of fit of the models by eye instead of using an automatic model fitting technique (e.g. maximum likelihood, Bayesian analysis etc). The main difficulty with implementing a more rigorous spectral model fitting is that computing the dynamics and radiative transfer for the ADAF is computationally demanding and hence the evaluation time and parameter space sampling of the models is very time consuming (see also \citealt{Yu11}). Therefore, our goal is to find models which are approximately consistent with the spectral data as opposed to exhaustively exploring the full parameter space of the models.

We stress that the model components are not independent from each other. For instance, since the ADAF is connected to the thin disk at $R_o$, the accretion rate at the ADAF outer radius is the same as the one in the truncated disk as we note in \textsection \ref{sec:disk}. Furthermore, $\dot{M}_{\rm o}$ and $\dot{M}_{\rm jet}$ are expected to be correlated and we require in our fits that $\dot{M}_{\rm jet} / \dot{M}_{\rm o} < 1$ for consistency. However, given our ignorance about the mechanism of jet formation, we vary $\dot{M}_{\rm jet}$ and $\dot{M}_{\rm o}$ independently in our fits. 

Regarding the contribution of the jet to the emission, this component produces X-rays through optically thin synchrotron emission with its hardness being controlled by the the parameter $p$; for instance, the X-ray photon index is given by $\Gamma_X=1+p/2$ (e.g., \citealt{Yuan09}). Since shock acceleration theory favors $2 \leq p \leq 3$ (e.g., \citealt{Bednarz98,Achterberg01,Lemoine03}), the jet model has difficulty explaining sources with $\Gamma_X < 2$. For such sources, we favor the ADAF as the X-ray production site. On the other hand, for LLAGNs with $\Gamma_X \geq 2$, we also consider fits to the X-ray spectrum in which the jet dominates the emission.

We have also searched the literature for independent constraints from other works on the values of the model parameters, such as the mass accretion rate, inclination angle, transition radius and jet power. We list such independent constraints in Table \ref{tab:models}.

\subsection{Contribution to the SED by stellar emission}	\label{sec:stars}
 
In the optical and near-IR, even though the HST aperture is small ($\approx 0.1''$), it corresponds to $\approx \, 2-15$ pc for the galaxies in our sample. A $\approx \, 10$ pc aperture will include up to a few $10^8$ stars, as estimated by adopting a typical bulge stellar density of $\sim 10^5$ stars per cubic parsec \citep{Kormendy04}. The enclosed mass is estimated roughly as 
\begin{equation}	\label{eq:bulge}
M_{\rm b} \sim 2 \times 10^8 \left( \frac{\rho_{\rm b}}{6 \times 10^4 M_\odot \, {\rm pc}^{-3}} \right) \left( \frac{R}{10 {\rm pc}} \right)^3 \, M_\odot,
\end{equation}
where $\rho_{\rm b} \sim 6 \times 10^4 M_\odot \, {\rm pc}^{-3}$ is the typical Milky Way bulge density (e.g., \citealt{Genzel10}).
Since typical, evolved stellar populations in bulges emit mostly in the near-IR and optical, it is prudent to consider the possible contribution of unresolved, nuclear stellar population at these wavebands in our sample. Thus, for the 12 LLAGNs for which we have optical observations available, we also consider the possibility that the nuclear optical data is reproduced with simple stellar population spectral models. 

We adopt a 10 Gyr-old stellar population with solar metallicity and a Salpeter initial mass function as representative of the typical evolved stellar populations in galactic bulges \citep{Wyse97}. We perform a least-squares fit of the compact nuclear stellar population spectral model \citep{Bruzual03} to the HST optical measurements (and to the optical spectrum in the case of NGC 1097) by simply varying the mass of the stellar population. The resulting stellar population spectral models are displayed alongside the accretion flow and jet models in Figures \ref{fig:n1097}-\ref{fig:n4594}, for illustration. We did not try to perform a detailed simultaneous fit of the accretion flow/jet models and the stellar population to the observations. We simply want to illustrate the potential of unresolved stellar populations to account for the optical observations.

Given that the luminosity of the stellar population falls quite rapidly for $\lambda < 4000 {\rm \AA}$ and that LINERs tend to be UV-variable with the UV flux setting a lower limit to the AGN luminosity \citep{Maoz05,Maoz07}, we only consider the high-resolution HST observations with $\lambda < 4000 {\rm \AA}$ in the stellar population spectral fits. 

We note that the IR upper limits in the SEDs are consistent with the contribution of an old stellar population to the near-IR spectral region due to the large apertures of these observations. This is illustrated in Fig. \ref{fig:n1097} but is also the case for the other galaxies with similar upper limits in our sample.

\section{SED fitting results} \label{sec:seds}

We describe in this section the results obtained from fitting our coupled accretion-jet model to the SEDs. In the SED plots that follow below, the points in the optical-UV waveband correspond to the data without any extinction correction while the upper bars represent the same observations after a maximal extinction correction (see EHF10); hence, these bars illustrate the uncertainty in the extinction corrections. 

The arrows represent upper limits to the nuclear luminosity. These upper limits are either a result of non-detections, or because the corresponding observations were taken through a large aperture ($>1''$). In the latter case, the upper limits include a potentially significant contribution by the emission from dust and the stellar population of the galaxy bulge as discussed in the previous section. In order to illustrate this point, we consider the case of NGC 1097. The IR observations for this LINER were taken through an aperture of diameter  $\approx 5''$ \citep{Nemmen06} which corresponds to $\approx 364$ pc. The bulge stellar mass enclosed within this aperture can be roughly estimated using equation \ref{eq:bulge} as $\sim 7 \times 10^9 M_\odot$. Figure \ref{fig:n1097} shows the spectrum of an old stellar population with this amount of mass, demonstrating that the bulge stellar emission can account for at least part of the IR upper limits in our sample. Therefore, even though the IR upper limits are not very useful to constrain the accretion and jet models, we display the corresponding data points in Figures \ref{fig:seds01}-\ref{fig:n4736} for completeness.
 
Table \ref{tab:models} lists the model parameters. We list whenever available the Bondi accretion rate and the corresponding reference from which $\dot{m}_{\rm Bondi}$ was taken. $P_{\rm jet}^{\rm obs}$ corresponds to the jet power estimated either from observations or using the correlation between jet power and radio luminosity of \citet{Merloni07}. $P_{\rm jet}^{\rm mod}$ represents the jet power resulting from our jet model, calculated as $P_{\rm jet} = \dot{M}_{\rm jet} \Gamma_j^2 c^2$. Table \ref{tab:star} shows the stellar mass potentially enclosed within $0.1''$ as obtained from the stellar populations fits to the HST optical observations.

For completeness, we present model fits to the 9 SEDs from EHF10 which did not pass our data quality cut in the Appendix. However, due to the lower quality of their SEDs, we refrain from drawing conclusions based on those sources.

We demonstrate in this section that there are two possible types of models which can accommodate the observed SED of M81 (and the SEDs of other LINERs in our sample as we will show below): In the first type, the emission from the ADAF dominates the observed X-rays; in the second type of model, the jet emission dominates the X-rays. We will hereafter use the abbreviations AD (as in \emph{ADAF-dominated}) and JD (as in \emph{jet-dominated}) when referring to the former and latter types of models, respectively.

\subsection{NGC 1097}	\label{sec:1097}

\begin{figure}
\centering
\includegraphics[scale=0.5,trim=30 0 10 10,clip=true]{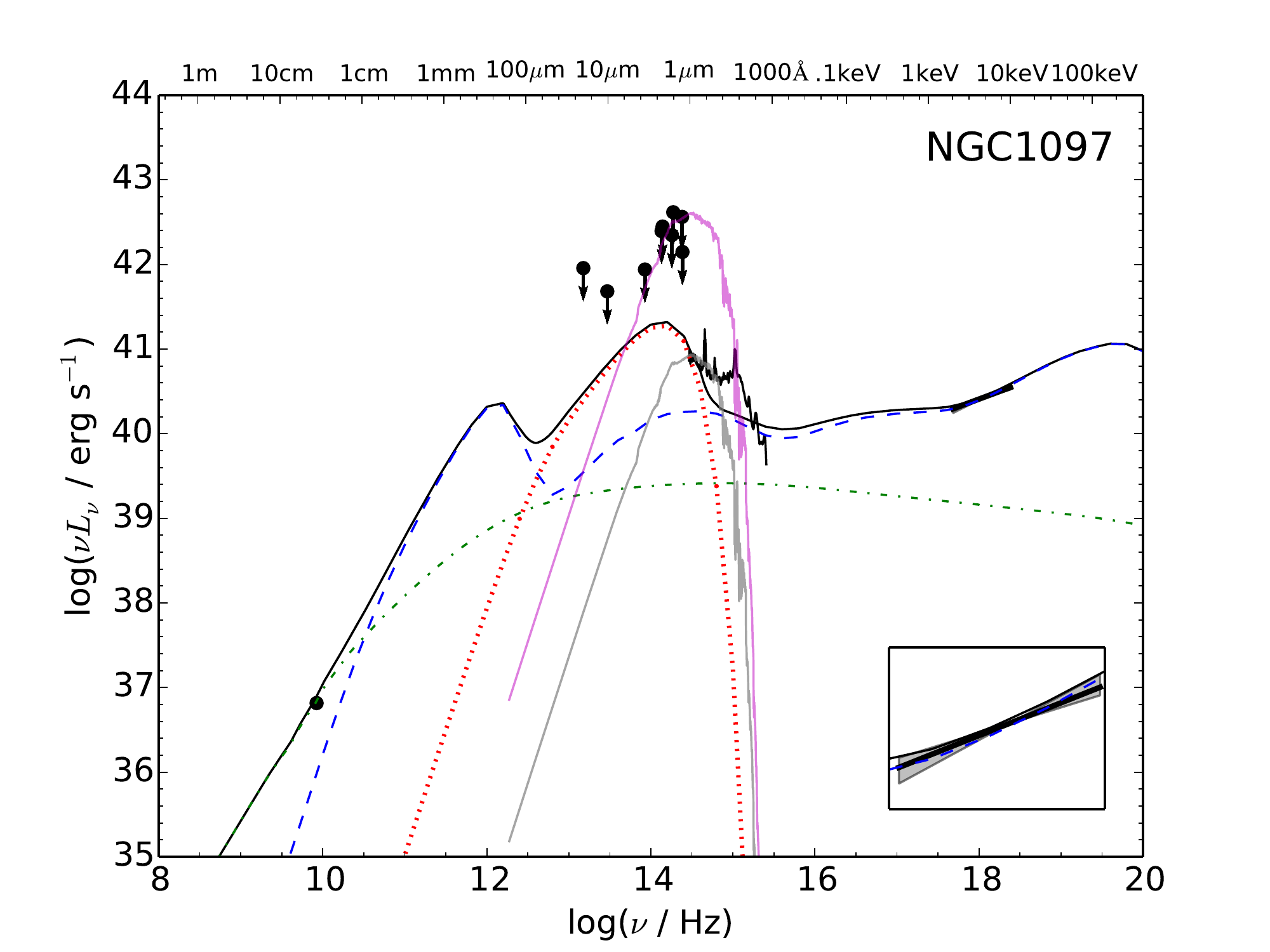}
\caption{The observed SED and ADAF-dominated model for NGC 1097. The dashed blue, dotted red and dot-dashed green lines correspond to the emission from the ADAF, truncated thin disk and jet, respectively. The solid black line represents the sum of the emission from all components. The inset shows the zoomed 2-10 keV spectrum. The upper solid magenta line shows the spectrum of an old stellar population with $\sim 7 \times 10^9 M_\odot$ which accounts for the near-IR upper limits observed at lower spatial resolution. The lower solid gray line displays the spectrum of an old stellar population fitted to the optical observations. }
\label{fig:n1097}
\end{figure}

The SED of this object was previously studied in detail by \cite{Nemmen06} and the model displayed in Figure \ref{fig:n1097} corresponds to the model obtained by \citet{Nemmen06}. 
The transition radius is independently constrained from the modeling of the broad double-peaked H$\alpha$ emission line ($r_{\rm tr} \sim 225$). Once the transition radius is fixed and requiring that the accretion rate in the outer thin disk is the same as in the outer parts of the ADAF, there is not much flexibility in the resulting thin disk spectrum; as such, the truncated thin disk spectrum self-consistently reproduces the HST optical data \citep{Nemmen06}.
The jet emission accounts for the radio data point while the hard X-ray spectrum from the ADAF accounts for the \emph{Chandra} observations. The jet emission is not able to reproduce the X-ray spectrum since it would require $p<2$ (see \textsection \ref{sec:fit}). Our models do not take into account the contribution of the nuclear starburst in the UV. Therefore, they do not reproduce the ``UV bump'' between $\approx 1000$ \AA\ and $\approx 7000$ \AA\ (see Fig. 3 in \citealt{Storchi-Bergmann05}).

Figure \ref{fig:n1097} also illustrates that we cannot rule out the possibility that the optical emission is produced by a unresolved, old stellar population located within $0.''1$ with a mass $1.5 \times 10^8 M_\odot$ (cf. \textsection \ref{sec:stars}). 

The IR upper limits, observed with a lower spatial resolution, are consistent with an old stellar population with a mass $\sim 7 \times 10^9 M_\odot$. As discussed in the previous section, this is presumably the case for the other galaxies in our sample which also have lower spatial resolution IR observations, although the stellar mass will vary from source to source.

\subsection{NGC 3031}	\label{sec:m81}

This source is among the brightest LLAGNs known since it is the nearest AGN besides Centaurus A and has been the subject of a broadband multiwavelength monitoring campaign \citep{Markoff08, Miller10}. 

\citet{Devereux07} modeled the profile of the broad double-peaked H$\alpha$ line with a relativistic thin disk model. They were able to explain the profile with an inclination angle of $50^\circ$. Such a high inclination angle is supported by radio observations of the jet \citep{Bietenholz00}. The H$\alpha$ line profile of M81 is consistent with an inner radius of $\approx 280 - 360 R_S$ for the line-emitting thin disk \citep{Devereux07}. In our models for the SED, we therefore adopt $i=50^\circ$ and $r_{\rm tr} = 360$.
The resulting IR-luminous thin disk spectrum is not strongly constrained by the available IR upper limits and optical-UV observations. On the other hand, the optical data points at $6000$ \AA\ and $4700$ \AA\ are consistent with a stellar population with mass $6.2 \times 10^7 M_\odot$.

\begin{figure*}
\centerline{
\includegraphics[scale=0.5]{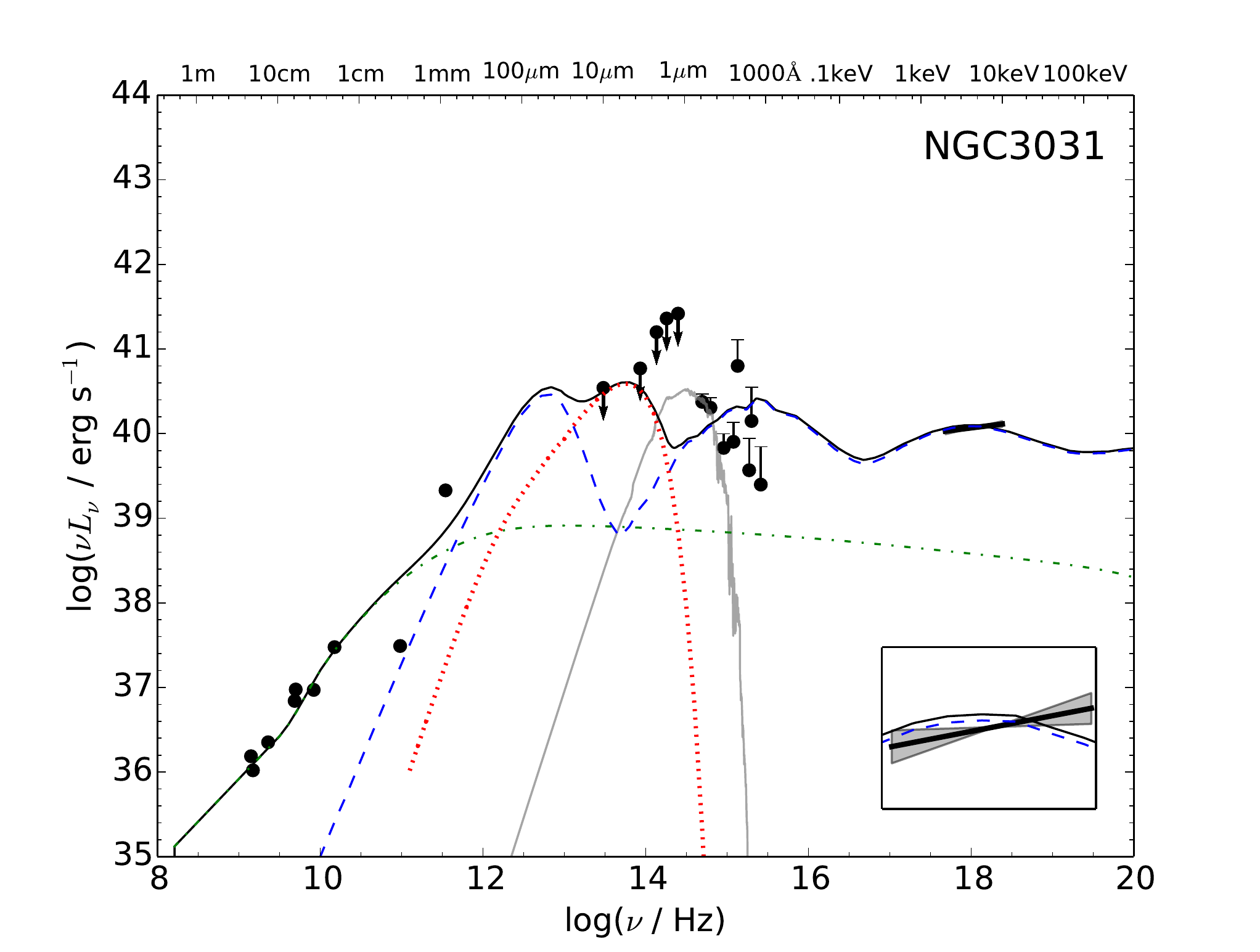}
\hskip -0.3truein
\includegraphics[scale=0.5]{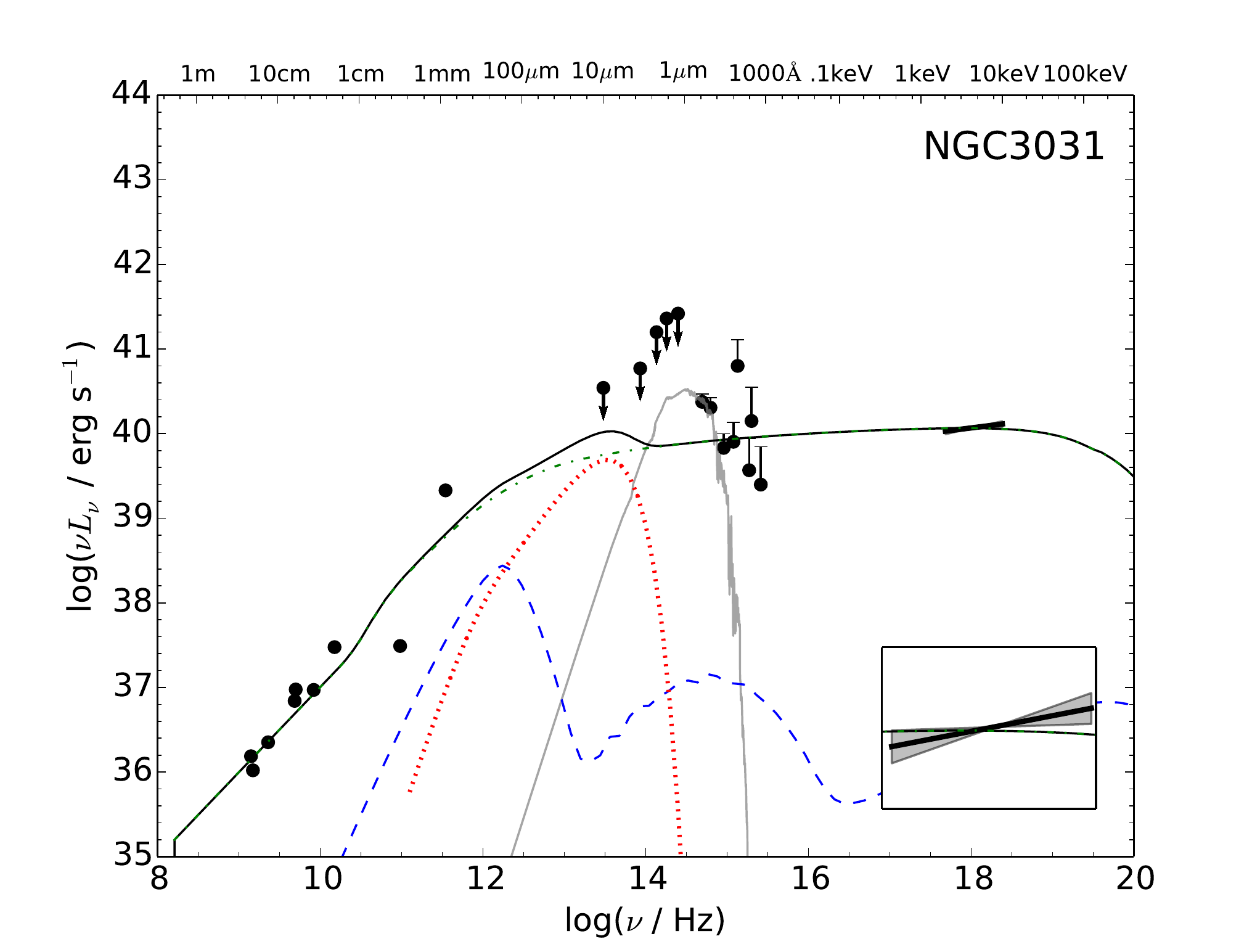}
}
\centerline{
\includegraphics[scale=0.5]{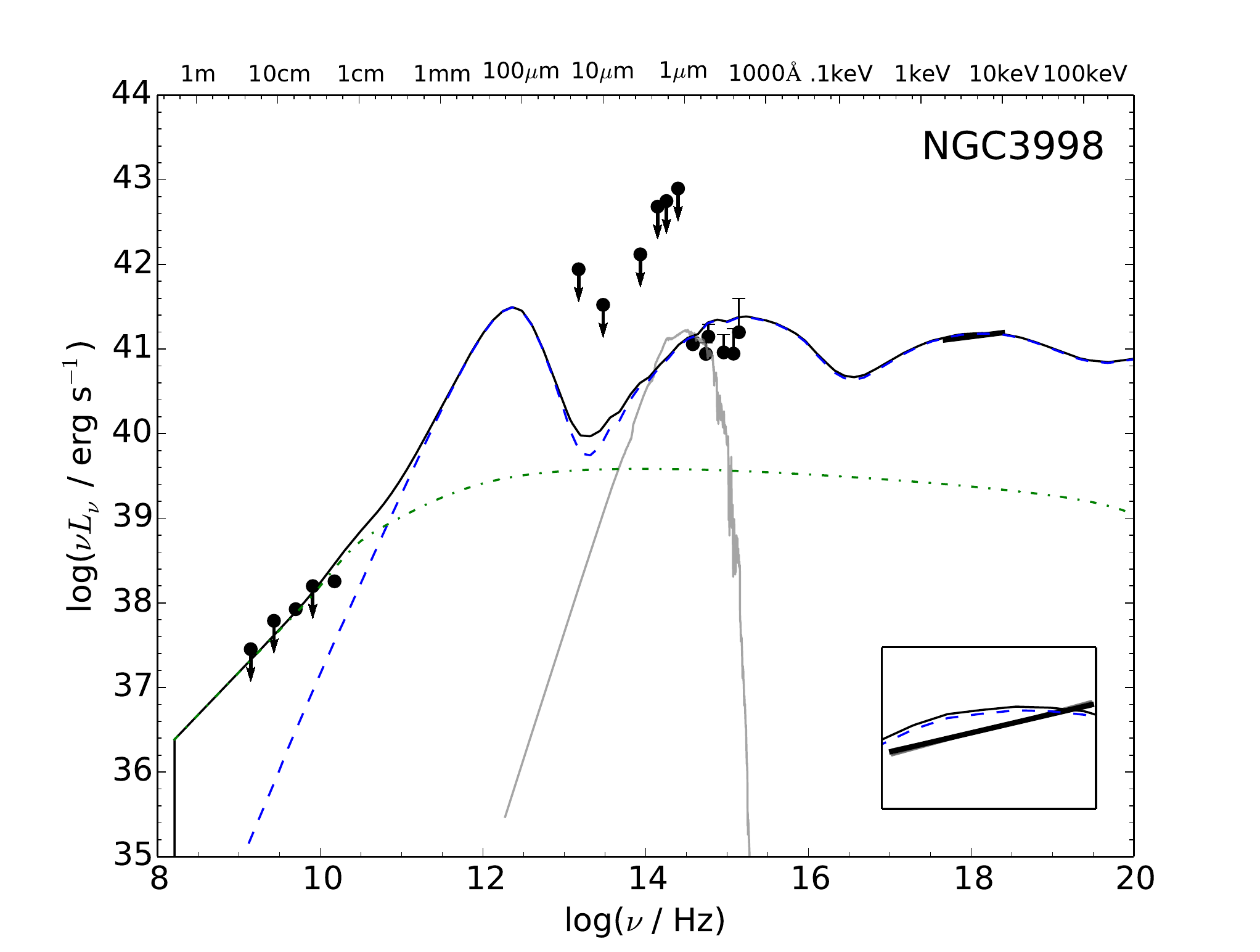}
\hskip -0.3truein
\includegraphics[scale=0.5]{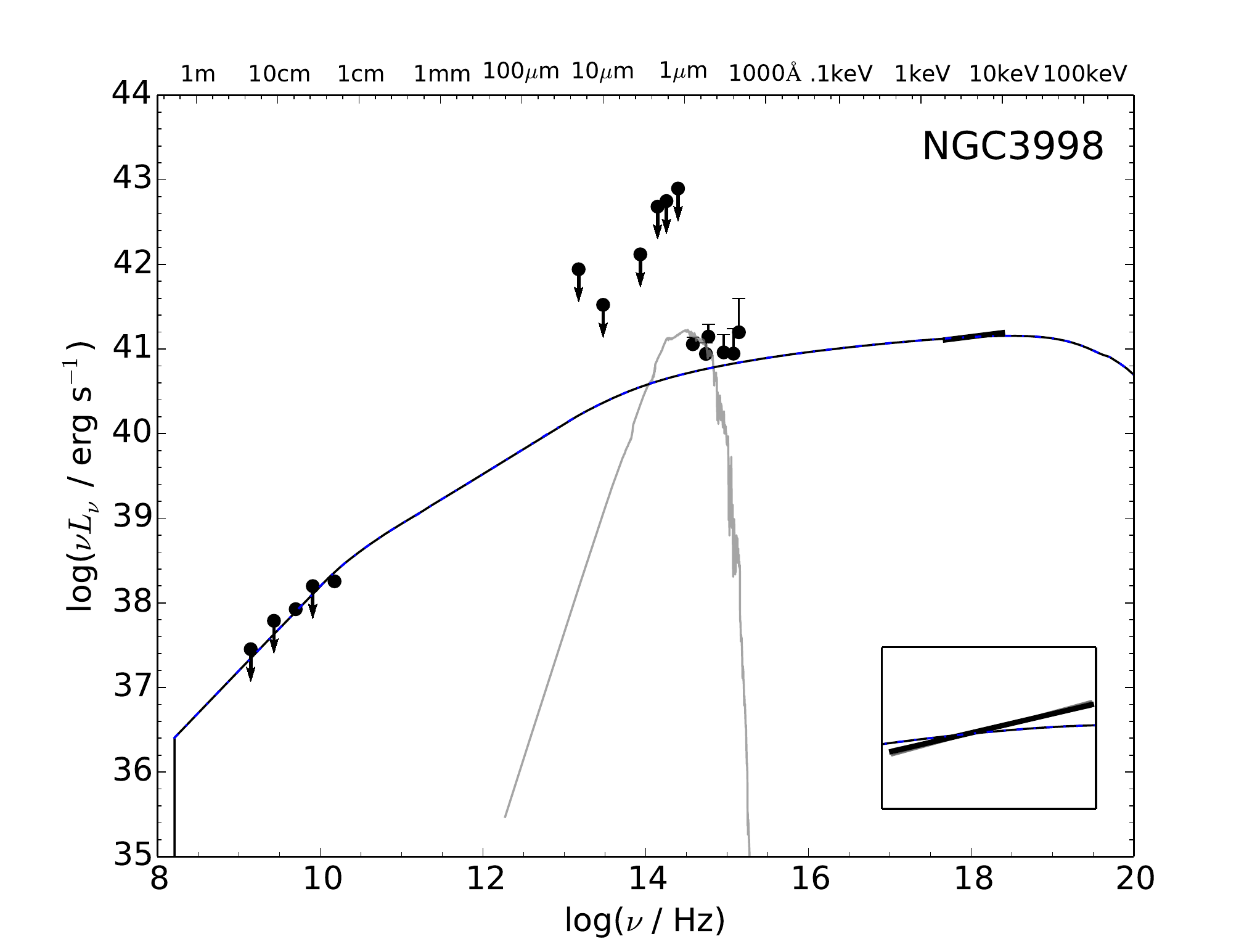}
}
\caption{SEDs and coupled accretion-jet models for NGC 3031 and NGC 3998. The left panel shows the AD models for each object while the right panel displays the JD models. The dashed, dotted and dot-dashed lines correspond to the emission from the ADAF, truncated thin disk and jet, respectively. The solid line when present represents the sum of the emission from all components. The inset shows the zoomed 2-10 keV spectrum. The old stellar population fitted to the optical data is displayed as the gray solid line.}
\label{fig:seds01}
\end{figure*}

Figure \ref{fig:seds01} shows two models for the SED of M81 consistent with the above constraints: in the left-hand side, an ADAF with $\delta=0.01$ dominates the X-ray emission whereas in the model to the right a jet with $p=2$ is the dominant X-ray emitter.

\subsection{NGC 3998}	\label{sec:n3998}

The shortest wavelength (1750 \AA) UV data point in the SED of NGC 3998 presented in EHF10 is anomalously high because of variability \citep{Devereux11n3998}. As discussed by Devereux this data point was obtained many years before all the other observations, when the source was much brighter. For this reason, we exclude this UV measurement from our analysis and plots.

In our models we adopted a large outer radius for the ADAF, but smaller radii are not ruled out by the data. For instance, we are also able to obtain a reasonable fit to the SED with $r_{\rm tr}=500$ and $\dot{m}_{\rm o}=10^{-3}$. This model is consistent with the IR upper limits and accounts for the X-ray data. The transition radius cannot be much smaller than this value, otherwise the emission of the truncated thin disk would exceed the IR upper limits. Furthermore, as noted by \citet{Ptak04}, the lack of Fe K$\alpha$ line emission also suggests that the value of $r_{\rm tr}$ is not so small. 

A JD model accounts quite well for the X-rays but somewhat underestimates the optical-UV data. 

\begin{figure}
\centerline{
\includegraphics[scale=0.5]{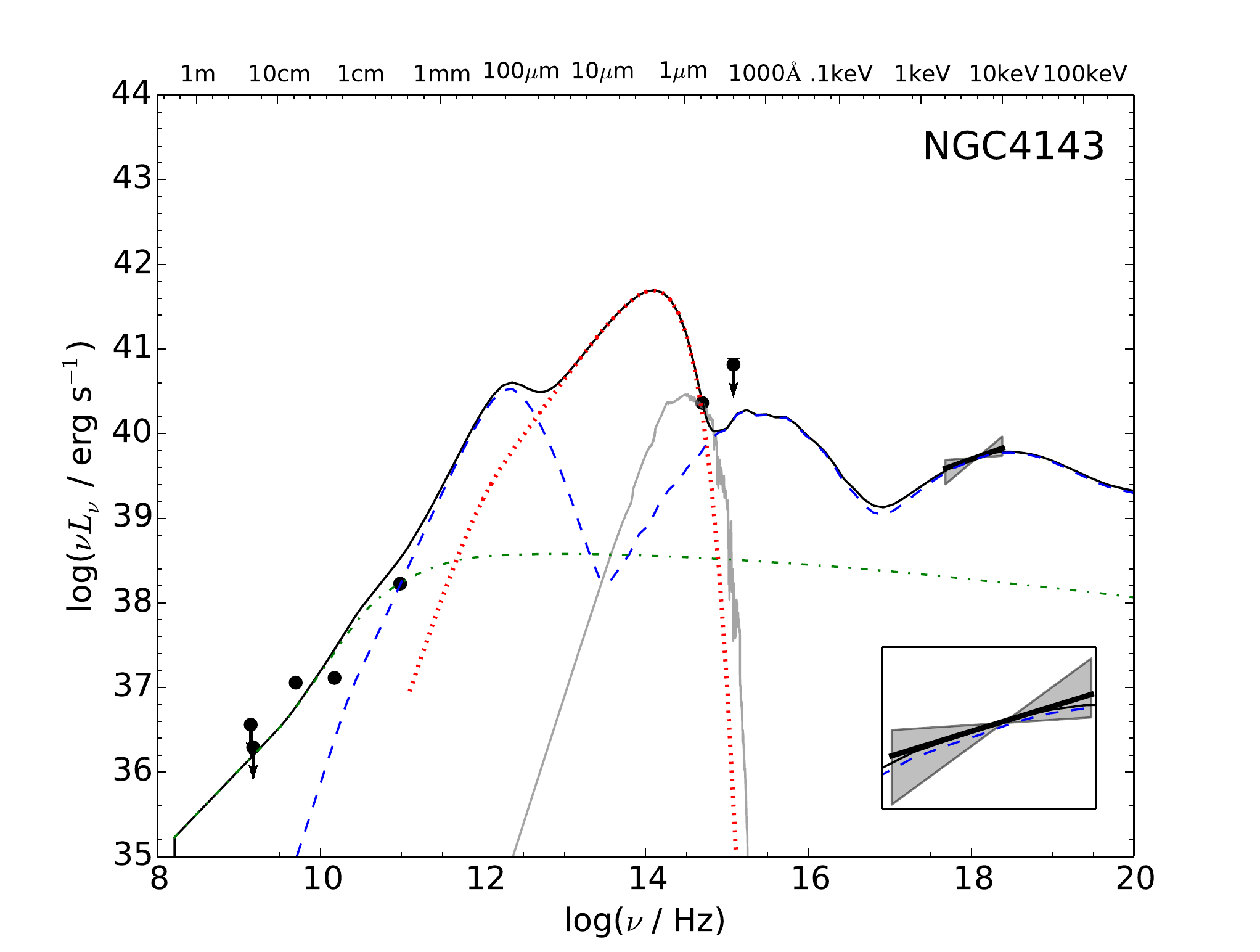}
\hskip -0.3truein
}
\centerline{
\includegraphics[scale=0.5]{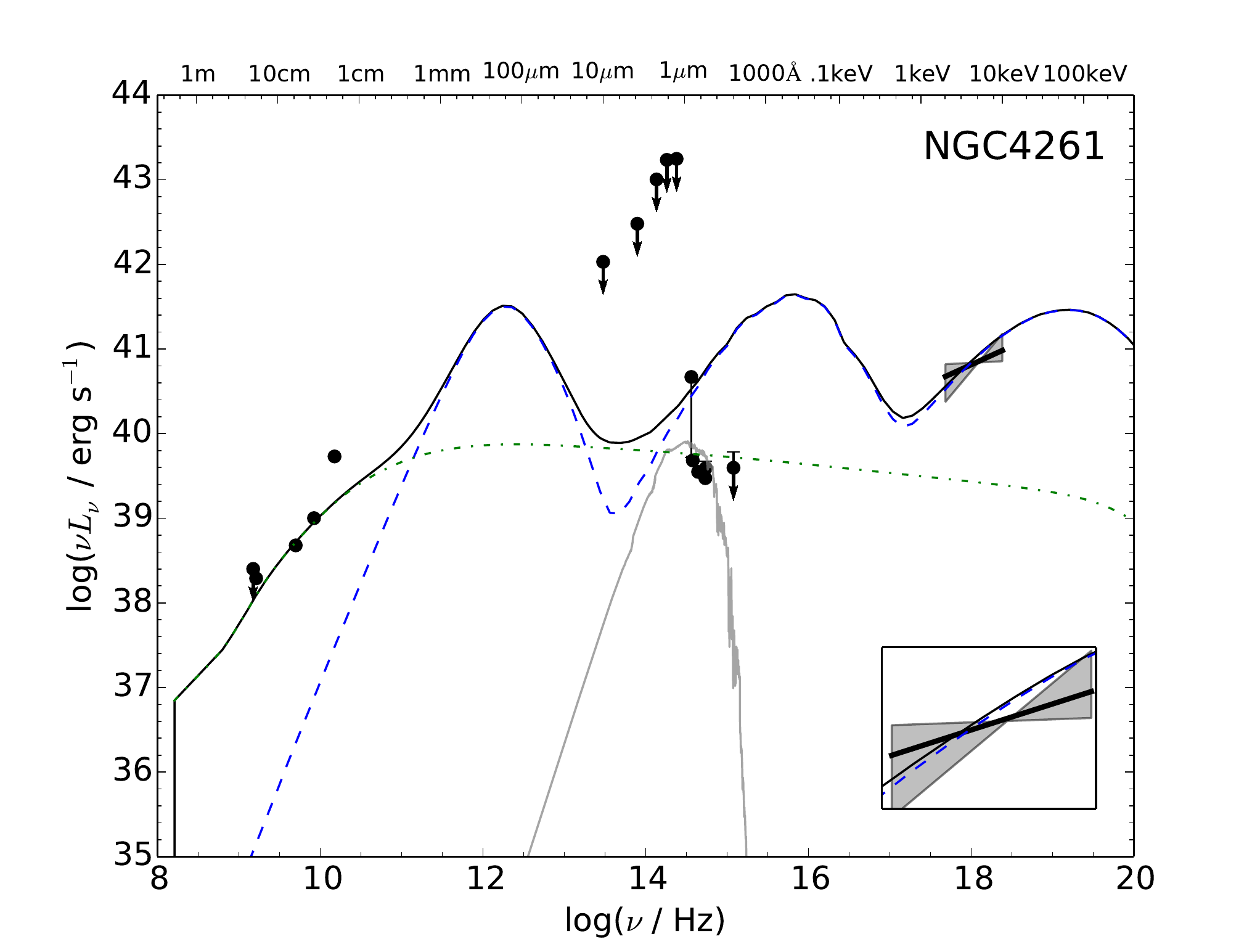}
\hskip -0.3truein
}
\centerline{
\includegraphics[scale=0.5]{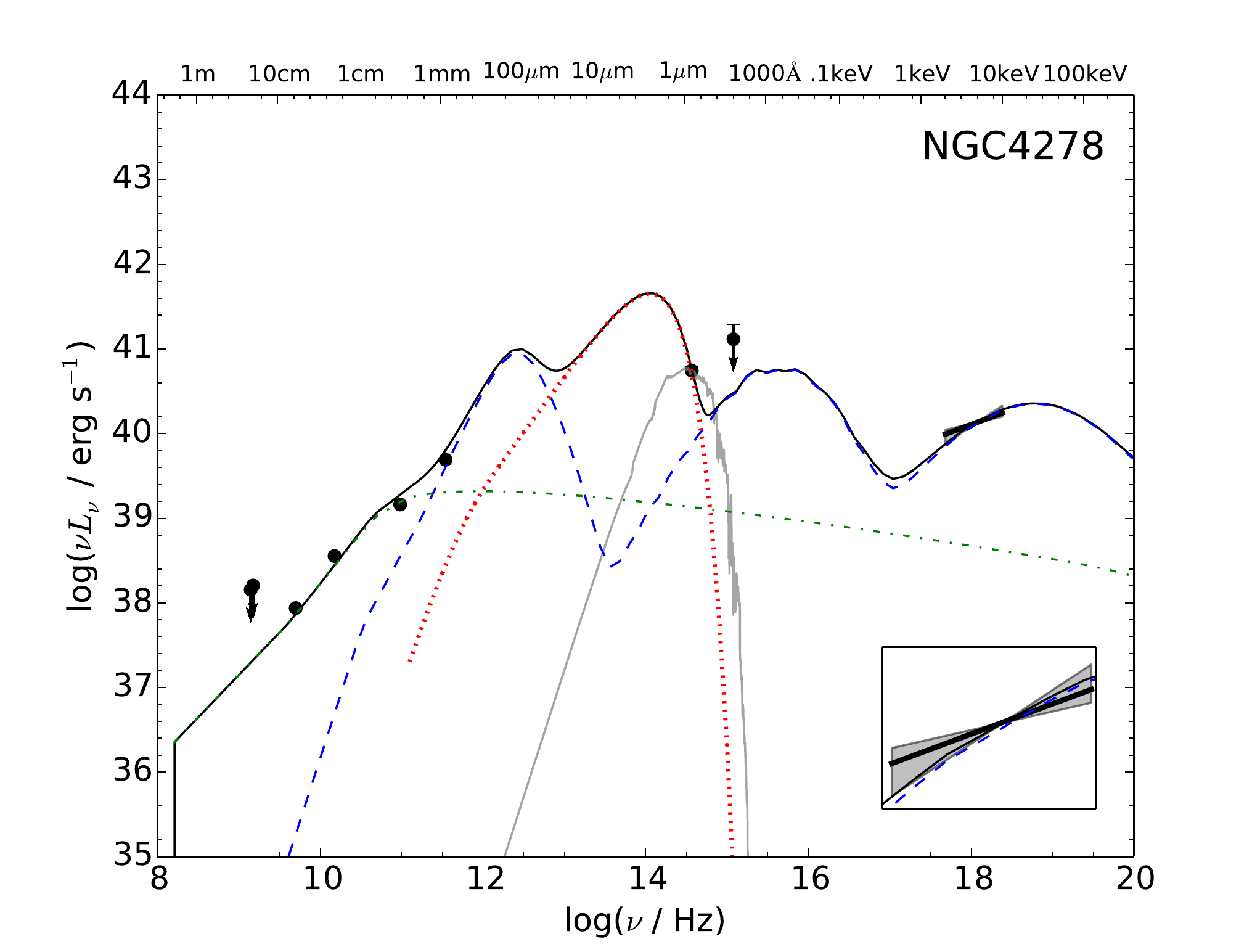}
\hskip -0.3truein
}
\caption{Same as in Figure \ref{fig:seds01} for NGC 4143, NGC 4261 and NGC 4278, displaying only AD models.}
\label{fig:seds02}
\end{figure}

\subsection{NGC 4143}

The optical data point can be reproduced by emission from the truncated thin disk emission with the somewhat small transition radius of $r_{\rm tr} = 70$. Alternatively, the optical observation can be fitted with an old stellar population with mass  $5 \times 10^7 M_\odot$.
Since the jet model requires $p<2$ to reproduce the X-ray spectrum, a jet origin for the X-ray emission in disfavored.

\subsection{NGC 4261}

Estimates of the Bondi rate, inclination angle and jet power are available \citep{Gliozzi03, Merloni07} and there are two independent estimates of the jet power. We adopt $i=63^\circ$ in our models \citep{Gliozzi03}. 
The accretion rate we obtain from the ADAF fit is somewhat smaller than $\dot{m}_{\rm Bondi}$. A jet origin for the X-rays is disfavored since it would require $p<2$.

This AGN is likely to be strongly affected by extinction in the optical-UV (OUV) band (EHF10). Hence, the OUV measurements probably do not capture the emission of the central engine. For this reason, it is not surprising that the models overpredict the OUV emission by almost one order of magnitude in Fig. \ref{fig:seds02}. 
 
\subsection{NGC 4278}

\citet{Di-Matteo01} also estimated that the Bondi rate is $\dot{m}_{\rm Bondi} \sim 0.001-0.01$. 
\citet{Giroletti05} found a two-sided radio structure for the jet using VLA data, and estimated that the jet is oriented close to the line of sight ($2^\circ \lesssim i \lesssim 4^\circ$) and mildly relativistic ($\Gamma_j \sim 1.5$). In our jet models we therefore adopt $i=3^\circ$ and $\Gamma_j = 1.5$. 

The optical data point requires a truncated thin disk with the somewhat small transition radii $r_{\rm tr} \sim 30-40$. If we allow $s$ to vary, we can also reproduce the SED with a larger transition radius, $r_{\rm tr}=100$, $\dot{m}_{\rm o}=4 \times 10^{-3}$, $\delta=0.1$ and $s=0.77$. Alternatively, the optical observation can be fitted with a $10^8 M_\odot$ stellar population.
A JD model is disfavored for this source.

\begin{figure*}
\centerline{
\includegraphics[scale=0.5]{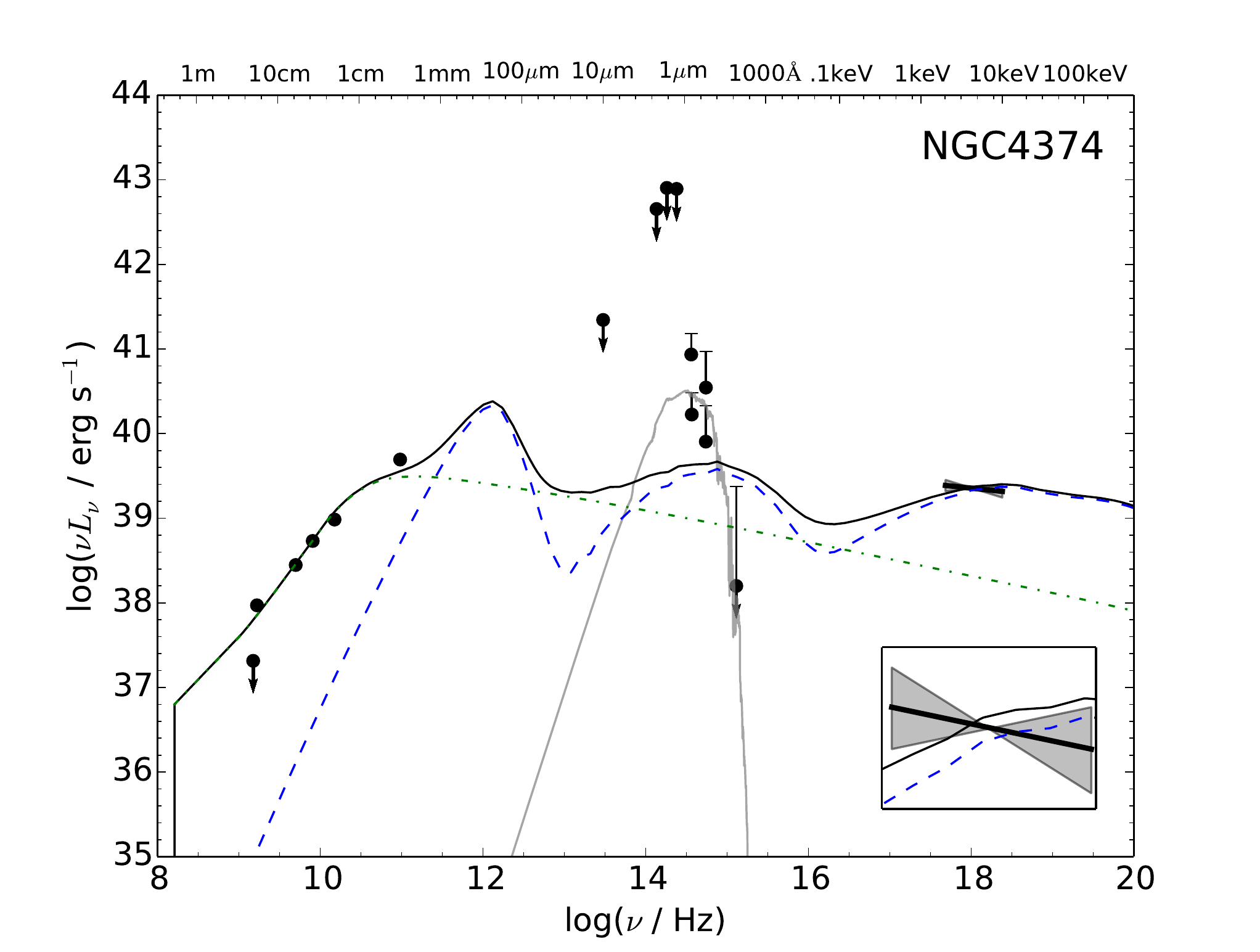}
\hskip -0.3truein
\includegraphics[scale=0.5]{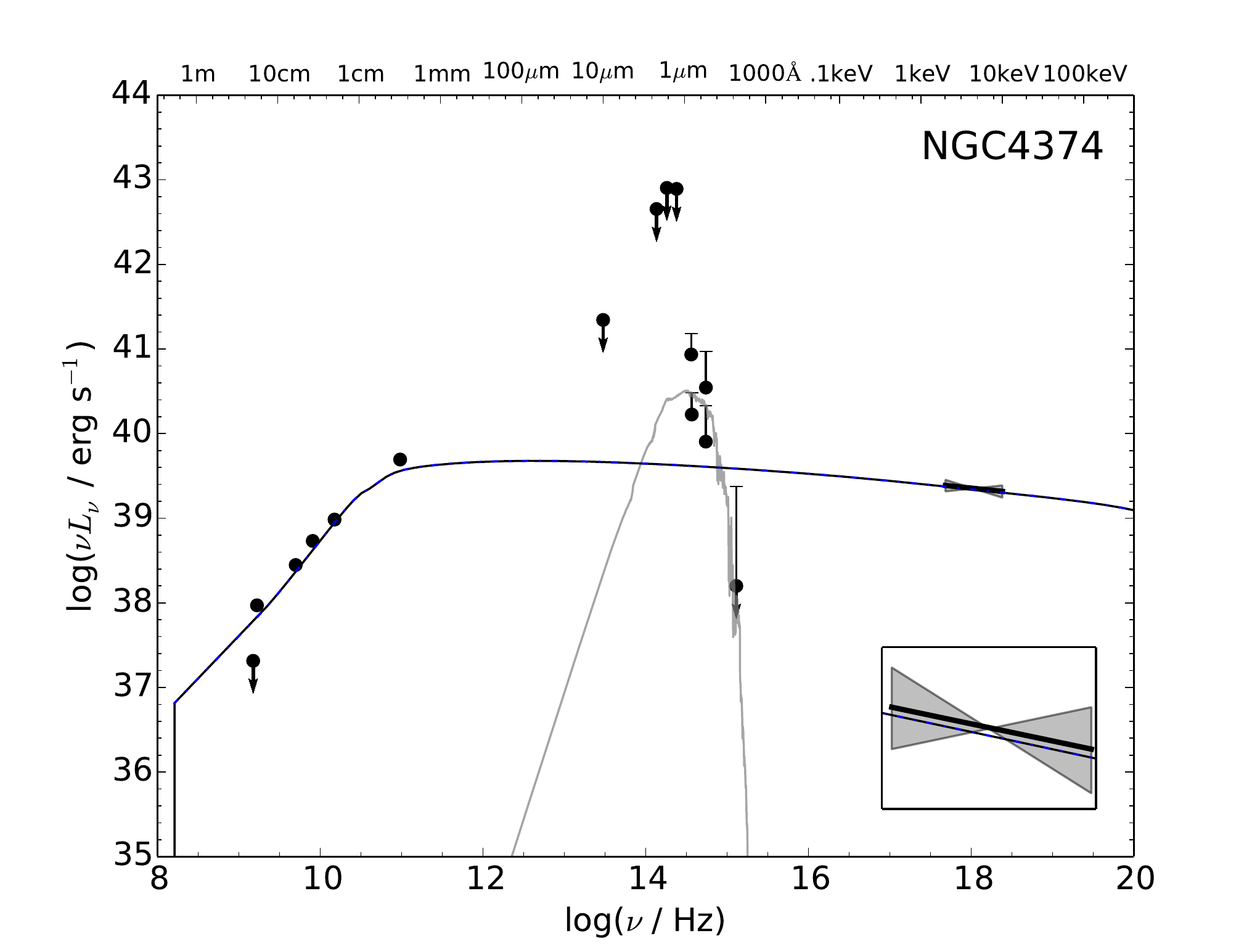}
}
\centerline{
\includegraphics[scale=0.5]{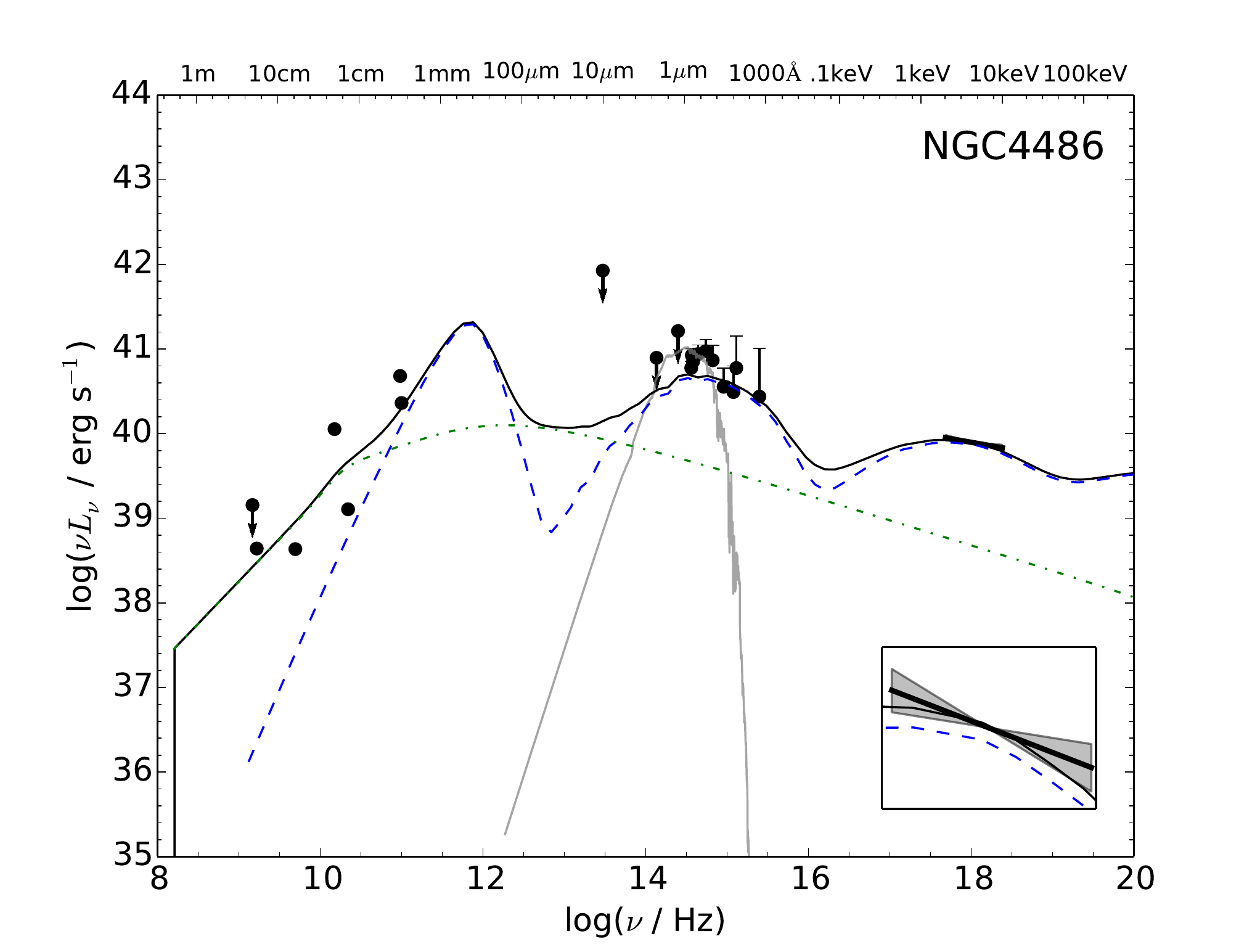}
\hskip -0.3truein
\includegraphics[scale=0.5]{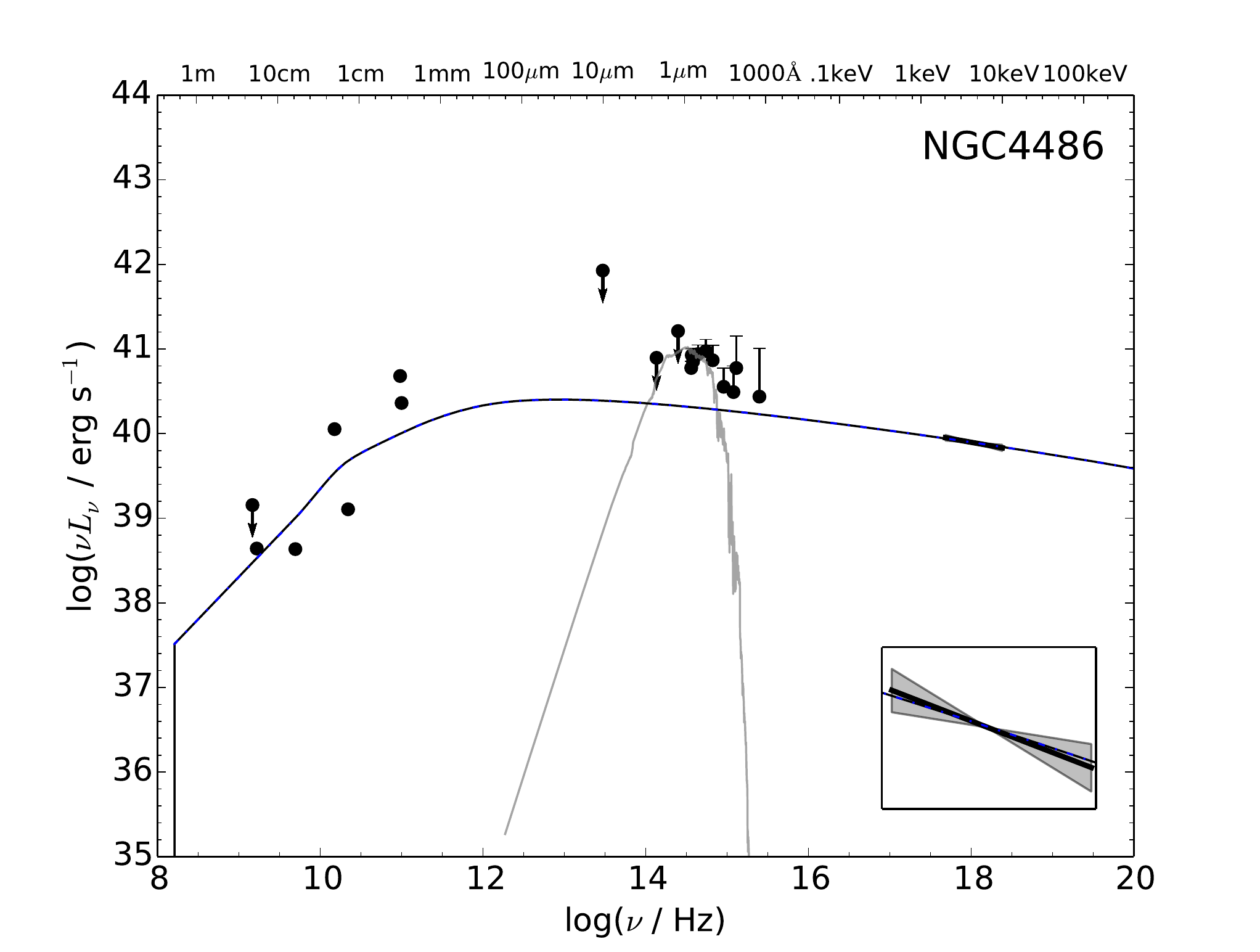}
}
\caption{Same as Figure \ref{fig:seds01} for NGC 4374 and NGC 4486.}
\label{fig:seds03}
\end{figure*}

\subsection{NGC 4374}	\label{sec:m84}

A prominent jet resolved with the VLA is observed to create cavities in the X-ray emitting gas \citep{Allen06, Finoguenov08}. 

We find adequate AD and JD models for the data. As was the case for NGC 4594, the mid to near-IR data are insufficient to make the case of a truncated thin disk compelling and similar models with much larger values of $r_{\rm tr}$ are not ruled out by the available IR data. The jet power in both models is in good agreement with the power estimated by \citet{Merloni07} based on the calorimetry of the X-ray cavities observed with \emph{Chandra}. A $5 \times 10^7 M_\odot$ stellar population nicely reproduces the optical excess above the models.

\subsection{NGC 4486} \label{sec:m87}

\citet{Di-Matteo03} previously estimated the Bondi accretion rate using \emph{Chandra} X-ray data ($\dot{M}_{\rm Bondi} \sim 0.1 \ M_\odot \ \rm{yr}^{-1}$; $\dot{m}_{\rm Bondi} \approx 7 \times 10^{-4}$). 
\citet{Biretta99} analyzed HST observations of the jet in M87 and estimated that $\Gamma_j \geq 6$ and $10^\circ<i<19^\circ$. Based on the results of \citet{Biretta99}, we adopt in our jet modelling the parameters $\Gamma_j=6$ and $i=10^\circ$. 
The jet kinetic power is estimated to be in the range $10^{43} - 10^{44} \ {\rm erg \ s}^{-1}$ (e.g., \citealt{Bicknell99, Owen00, Allen06, Merloni07}).

We find adequate AD and JD models for the data, although the JD model underpredicts the OUV data by a factor of a few. The SED of M87 was previously fitted using ADAF and/or jet models \citep{Di-Matteo03, Yuan09, Li09}. \citet{Di-Matteo03} modelled the SED of M87 with an ADAF model using different values of $\delta$ but not including mass-loss (i.e., $s=0$). The model adopted by \citet{Li09} is quite similar to that of \citet{Di-Matteo03} although the former incorporate general relativistic corrections. \citet{Di-Matteo03} and \cite{Li09} obtained that the ADAF emission with no mass-loss approximately reproduces the SED and results in accretion rates consistent with the Bondi rate. Our AD model is similar to that of \citet{Di-Matteo03} but it also incorporates mass-loss as suggested by numerical simulations. 

\citet{Yuan09} tried to model the SED using an ADAF model with $\delta=0.5$ but failed to fit to data with an AD model. The reason is that for high values of $\delta$ ($\delta > 0.1$) and small accretion rates ($\dot{m}_{\rm o} \ll 0.01$) the ADAF X-ray spectrum is harder than the data. \citet{Yuan09} instead successfully fit the data with a JD model using $p=2.5$ similarly to our JD model.

\subsection{NGC 4552}	\label{sec:4552}

\begin{figure}
\centering
\includegraphics[scale=0.5,trim=30 0 10 10,clip=true]{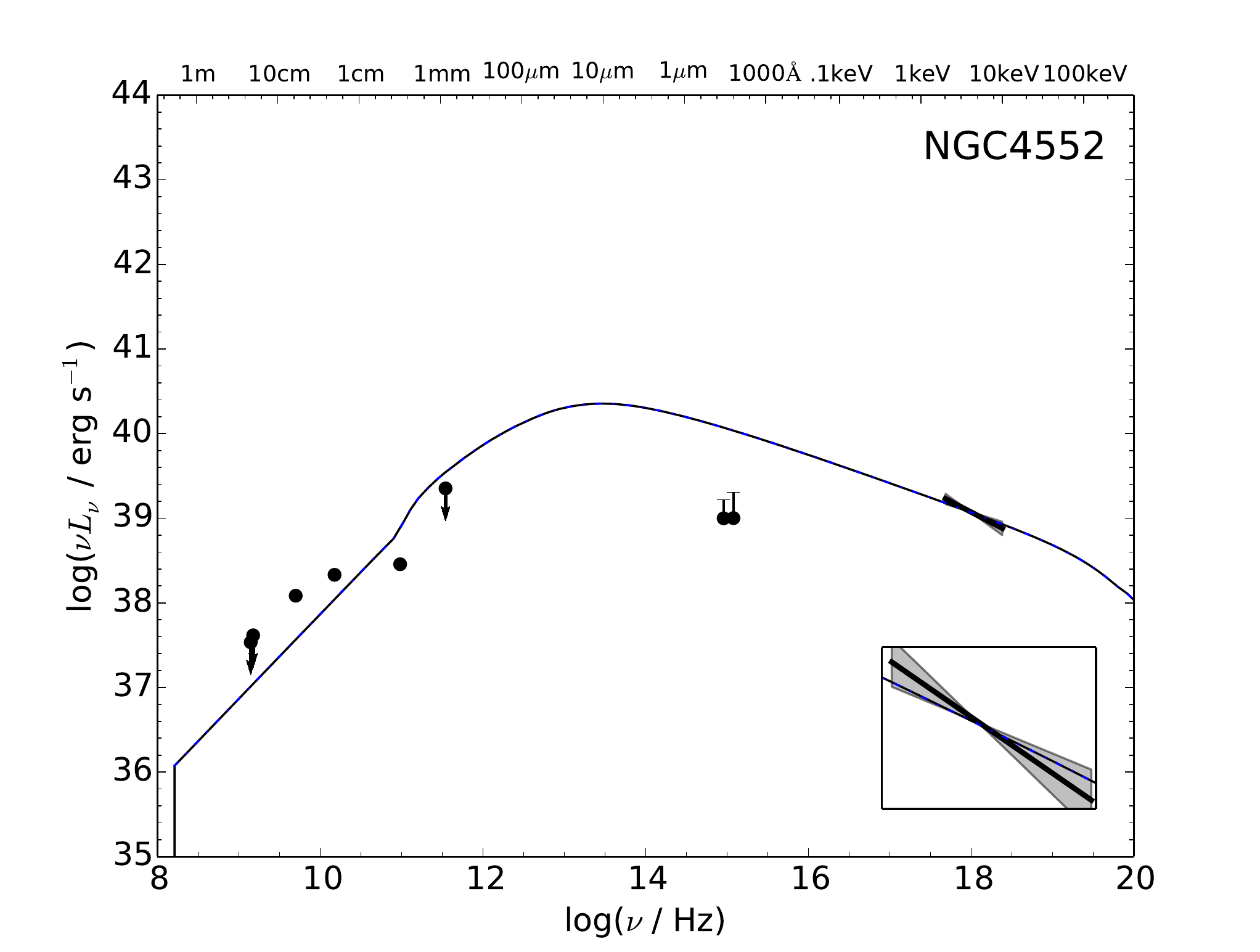}
\caption{The SED and jet-dominated model for NGC 4552. The inset shows the zoomed 2-10 keV spectrum.}
\label{fig:n4552}
\end{figure}

\begin{figure}
\centering
\includegraphics[scale=0.5,trim=30 0 10 10,clip=true]{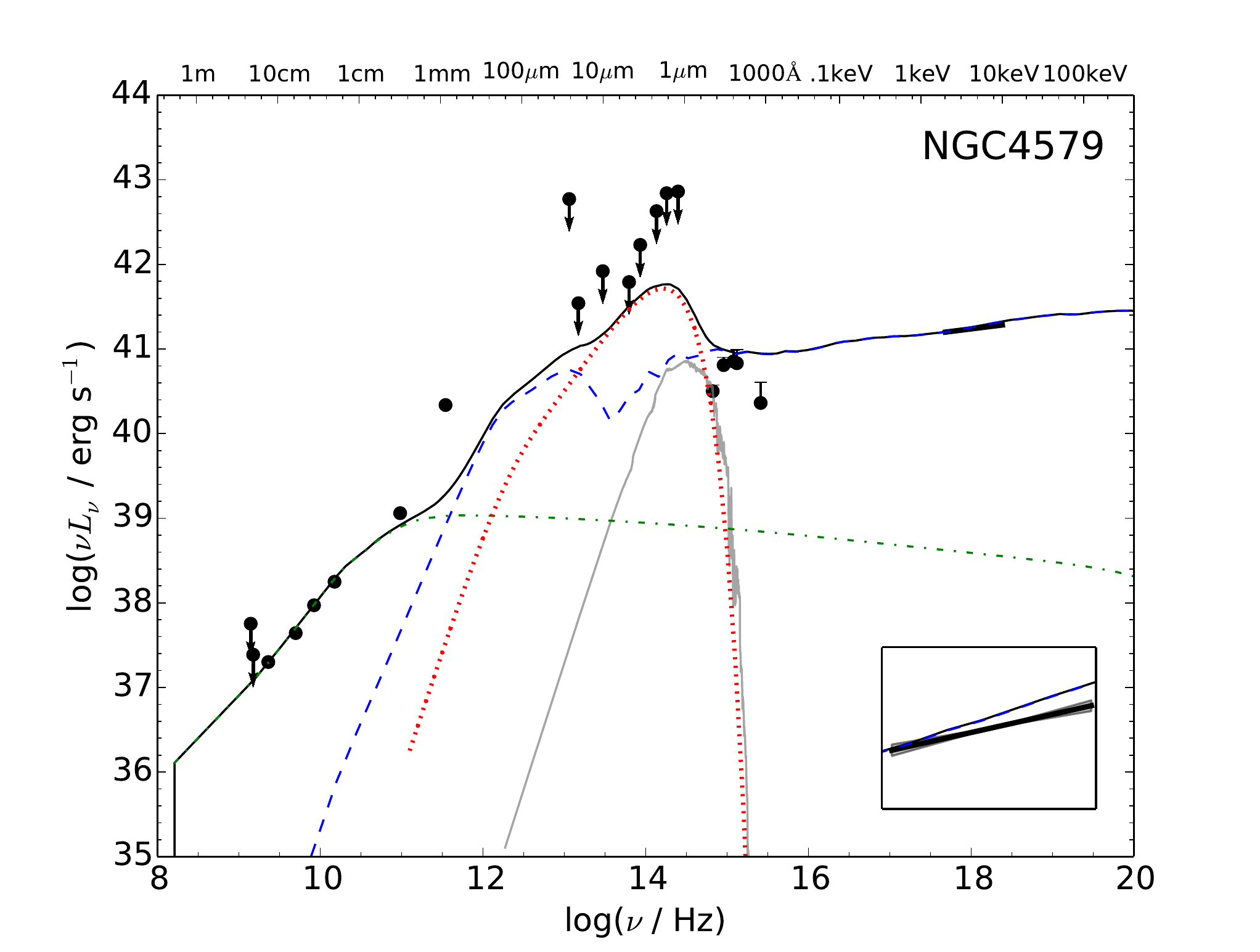}
\caption{Same as Fig. \ref{fig:n1097} for NGC 4579 (AD model).}
\label{fig:n4579}
\end{figure}

The Bondi accretion rate was estimated by \citet{Merloni07} (see also \citealt{Allen06}) from the X-ray profiles of density and temperature. The kinetic power carried by the jet was estimated by \citet{Merloni07} (see also \citealt{Allen06}) from the energy deposited in the X-ray cavities. Since the X-ray spectrum is quite soft, AD models are unable to account for the X-ray emission. The JD model in Fig. \ref{fig:n4552} is roughly consistent with the radio observations and explains quite well the X-ray data, but overpredicts the optical data. The resulting jet power is roughly consistent with the value estimated by \citet{Allen06,Merloni07}. A likely explanation for the unusually faint UV emission compared to the X-rays is source variability \citep{Maoz05} and hence we did not attempt to fit the $2500 \, {\rm \AA}$ and $3300 \, {\rm \AA}$ measurements with a stellar population as in the other LLAGNs.

\subsection{NGC 4579}	\label{sec:4579}

NGC 4579 shares many characteristics with NGC 3031 (M81) and NGC 1097: its nucleus features broad double-peaked Balmer emission lines \citep{Barth01} and a lack of the iron K$\alpha$ line emission \citep{Eracleous02}.
From the width of the broad H$\alpha$ line, \citet{Barth01} obtained a rough estimate of the inner radius of the line-emitting portion of the accretion disk $r_{\rm tr} \sim 160$. Hence, we adopt in our SED models $r_{\rm tr} = 150$ and $i=45^\circ$.
The resulting thin disk spectrum peaks in the IR and its high-frequency tail extends into the optical-UV; however, the available IR upper limits and HST optical-UV data do not allow us to set strong constraints on the thin disk emission.
A jet origin for the X-rays is disfavored by the hardness of the spectrum. 

\subsection{NGC 4594} \label{sec:n4594}

\begin{figure*}
\centerline{
\includegraphics[scale=0.5]{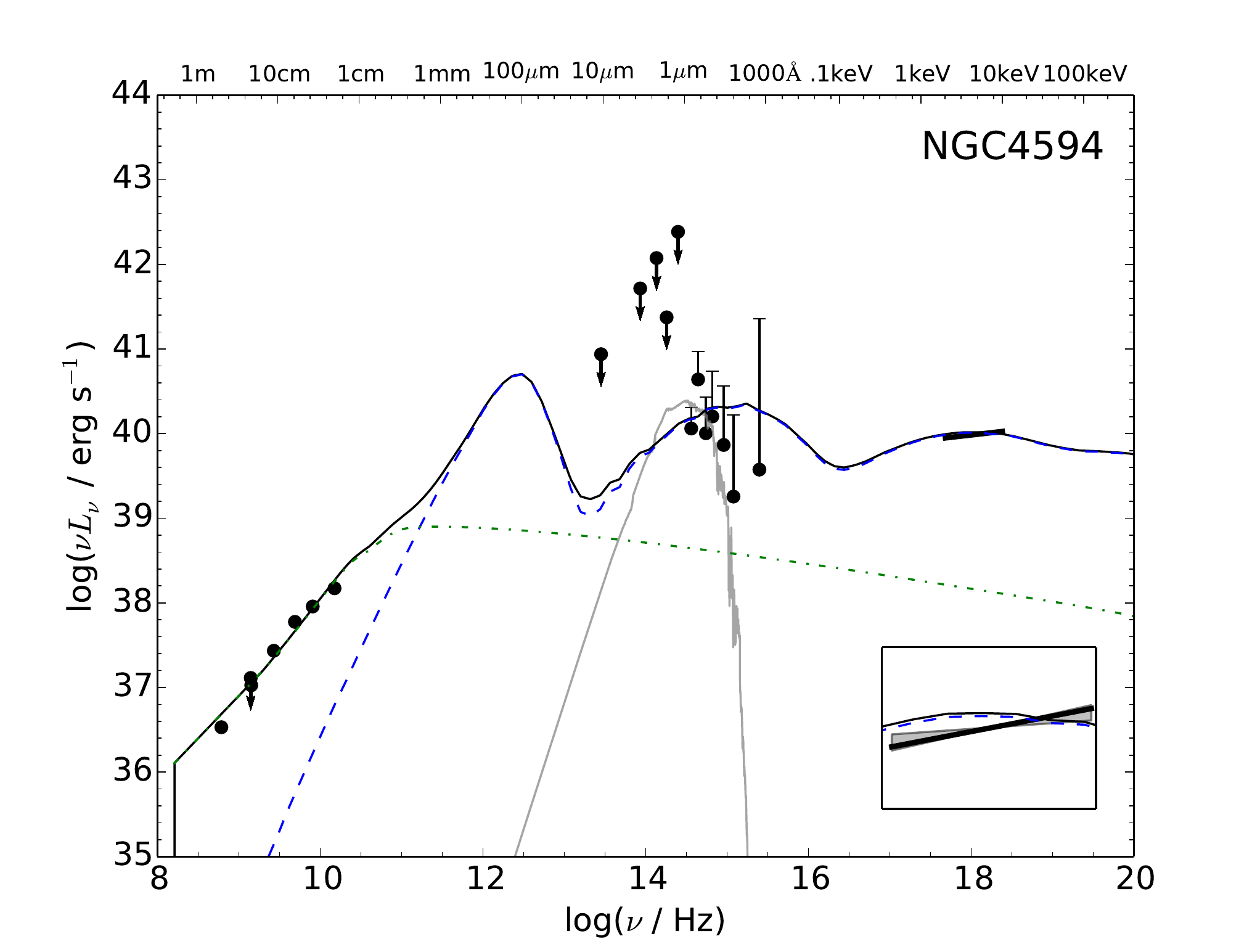}
\hskip -0.3truein
\includegraphics[scale=0.5]{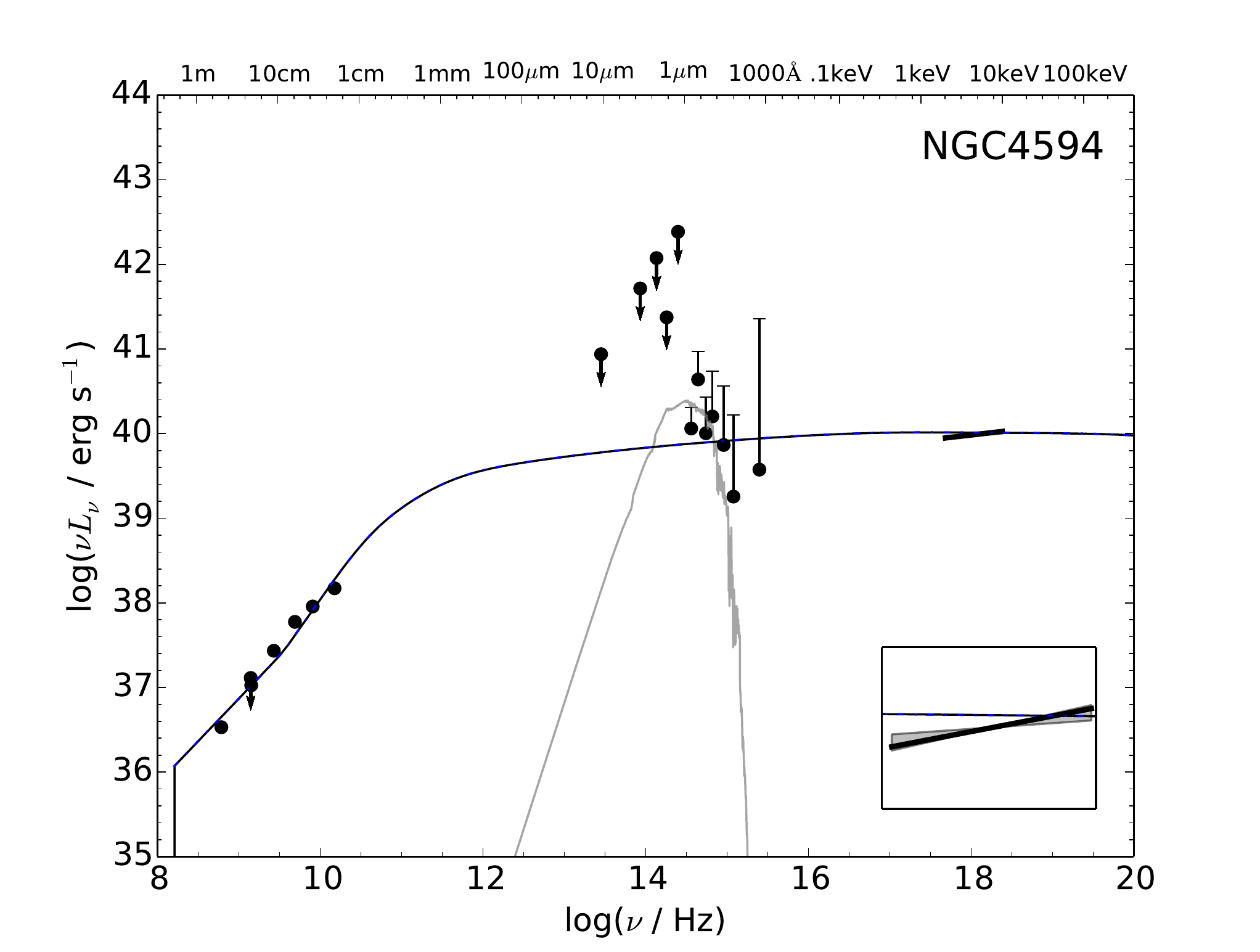}
}
\caption{Same as Figure \ref{fig:seds01} for NGC 4594.}
\label{fig:n4594}
\end{figure*}

The left panel of Figure \ref{fig:n4594} shows an AD model which is consistent with the available optical-UV and X-ray data. In this model, the jet dominates the radio emission and contributes  only weakly to the X-ray flux. The inferred accretion rate is consistent with the Bondi accretion rate estimated by \cite{Pellegrini05}. The available data are not sufficient to constrain the presence of a truncated thin disk in this source, hence we adopt $r_{\rm tr}=10^4$.
The right panel of Fig. \ref{fig:n4594} shows a jet model for the SED of NGC 4594 which is consistent with the data for $p=2$. 

\subsection{NGC 4736}

\begin{figure}
\centering
\includegraphics[scale=0.5,trim=30 0 10 10,clip=true]{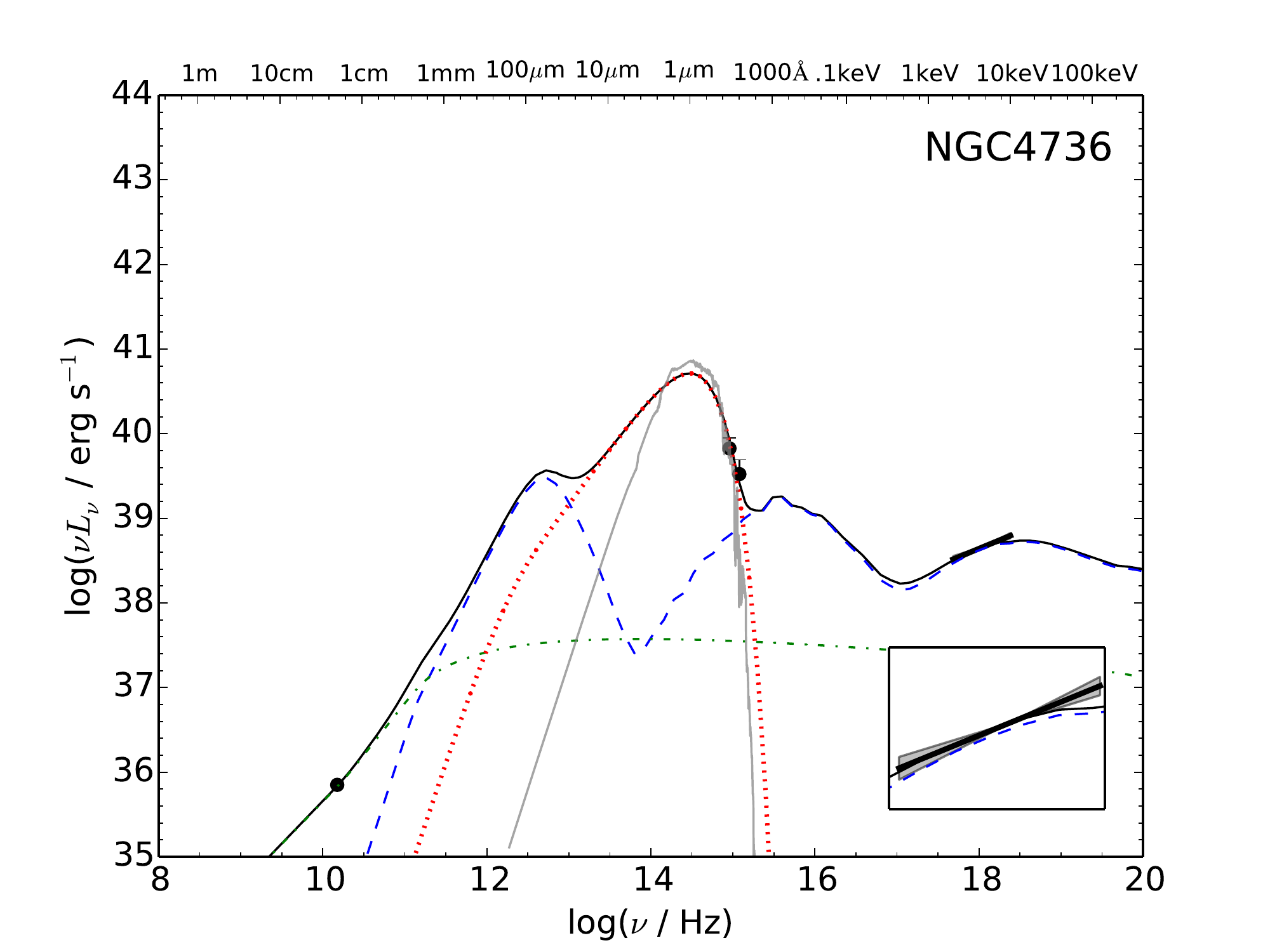}
\caption{Same as Fig. \ref{fig:n1097} for NGC 4736 (AD model).}
\label{fig:n4736}
\end{figure}

As was the case of NGC 1097, NGC 4143 and NGC 4278, this LINER requires a relatively small transition radius in order to explain its UV data. The 3300 \AA\ observation can also be reproduced by a $\approx 1.2 \times 10^8 M_\odot$ stellar population. The X-ray spectrum is too hard to be explained by a jet model (Figure \ref{fig:n4736}).

\begin{table*}
\centering
\caption{Model parameters resulting from the SED fits discussed in Section \ref{sec:seds}. The meaning of the parameters is described in Section \ref{sec:models}.}
\tiny
\begin{tabular}{@{}ccccccccccccccc@{}}
\hline
Galaxy & Model & $\dot{m}_{\rm o}$ & $r_{\rm tr}$ & $\delta$ & $s$ & $\dot{m}_{\rm jet}$ & $p$ & $\epsilon_e$ & $\epsilon_B$ & $i$ ($^\circ$) & $P_{\rm jet}^{\rm mod}$ & $P_{\rm jet}^{\rm obs}$ & $\dot{m}_{\rm Bondi}$ & Refs. \\
& &  &   &  &  &  &  &  &  &  &  &  &  & and notes \\
\hline
NGC 1097 & AD & $6.4 \times 10^{-3}$ & 225 & 0.1 & 0.8 & $7 \times 10^{-7}$ & 2.2 & 0.06 & 0.02 & 34 & $10^{43}$ &  - & - &  \\
\hline
NGC 3031 & JD & $5 \times 10^{-4}$ & 360 & 0.01 & 0.3 & $1.2 \times 10^{-5}$ & 2.05 & 0.6 & $10^{-4}$ & 50 & $4.8 \times 10^{42}$ &  $7.1 \times 10^{41}$ & - &  b \\ 
NGC 3031 & AD & $4 \times 10^{-3}$ & 360 & 0.035 & 0.3 & $2 \times 10^{-6}$ & 2.2 & 0.1 & 0.01 & 50 & $8 \times 10^{41}$ &   & - &   \\ 
\hline
NGC 3998 & JD & - & - & - & - & $3.5 \times 10^{-6}$ & 2.01 & 0.75 & $3 \times 10^{-5}$ & 30 & $1.8 \times 10^{43}$ & $4.6 \times 10^{42}$ & - &  b \\ 
NGC 3998  & AD & $6 \times 10^{-3}$ & $10^{4}$ & 0.05 & 0.3 & $1.7 \times 10^{-6}$ & 2.2 & 0.01 & $10^{-3}$ & 30 & $9 \times 10^{42}$ &  & - &  \\ 
\hline
NGC 4143  & AD & $2.5 \times 10^{-3}$ & 70 & 0.1 & 0.76 & $1.2 \times 10^{-7}$ & 2.2 & 0.03 & 0.1 & 30 & $1.6 \times 10^{41}$ &  & - &  \\ 
\hline
NGC 4261  & AD & $1.9 \times 10^{-3}$ & $10^4$ & 0.3 & 0.3 & $6 \times 10^{-6}$ & 2.2 & 0.01 & 0.1 & 63 & $2 \times 10^{43}$ &  & &  \\ 
\hline
NGC 4278 & AD & $7 \times 10^{-4}$ & 40 & 0.07 & 0.3 & $3.5 \times 10^{-6}$ & 2.3 & 0.001 & 0.01 & 3 & $4 \times 10^{42}$ & & &  \\ 
\hline
NGC 4374 & JD & - & - & - & - & $1.6 \times 10^{-6}$ & 2.2 & 0.009 & 0.008 & 30 & $8.4 \times 10^{42}$ & $3.9 \times 10^{42}$ & $4 \times 10^{-4}$ &  1,6,7 \\ 
NGC 4374 & AD & $2.9 \times 10^{-3}$ & $10^4$ & 0.01 & 0.3 & $4 \times 10^{-7}$ & 2.4 & 0.01 & 0.1 & 30 & $2 \times 10^{42}$ &  & &  \\ 
\hline
NGC 4486 & JD & - & - & - & - & $6 \times 10^{-8}$ & 2.3 & 0.001 & 0.008 & 10 & $8.2 \times 10^{42}$ &  $10^{43}-10^{44}$ & $7 \times 10^{-4}$ &  6,7,8,9 \\ 
NGC 4486 & AD & $5.5 \times 10^{-4}$ & $10^{4}$ & 0.01 & 0.1 & $5 \times 10^{-8}$ & 2.6 & 0.001 & 0.001 & 10 & $6.8 \times 10^{42}$ &  & &  \\ 
\hline
NGC 4552 &  JD & - & - & - & - & $1.8 \times 10^{-6}$ & 2.1 & 0.01 & 0.01 & 30 & $2.2 \times 10^{42}$ &  $1.6 \times 10^{42}$ & 0.001 &  7  \\
\hline
NGC 4579 &  AD & 0.02 & 150 & 0.06 & 0.3 & $1.4 \times 10^{-5}$ & 2.2 & 0.011 & 0.008 & 45 & $5.9 \times 10^{42}$ &   & - &   \\
\hline
NGC 4594 & JD & $2 \times 10^{-3}$ & $10^4$ & 0.01 & 0.3 & $4.5 \times 10^{-7}$ & 2 & 0.8 & 0.03 & 30 & $10^{42}$ &  $3 \times 10^{42}$ & 0.002 &  b,1  \\
NGC 4594 &  AD & 0.006 & $10^4$ & 0.01 & 0.3 & $9 \times 10^{-7}$ & 2.3 & 0.005 & 0.003 & 30 & $2 \times 10^{42}$ &  & &   \\
\hline
NGC 4736 &  AD & 0.0035 & 60 & 0.1 & 0.89 & $6 \times 10^{-7}$ & 2.2 & 0.01 & 0.01 & 30 & $6.5 \times 10^{40}$ &  & &   \\
\hline
\end{tabular}
\begin{flushleft}
\footnotesize 
\textbf{Notes:} \\
(a) Black hole mass estimated from the fundamental plane of black holes \citep{Merloni03}, using the X-ray and radio luminosities. \\
(b) Observed jet power estimated using the radio data and the \citet{Merloni07} correlation. \\
(c) For NGC 1097, $\Gamma=10$ (for consistency with the work of \citealt{Nemmen06}). \\
(d) 5 GHz luminosity estimated using the \citet{Merloni03} correlation. \\
(e) The lower limit on $P_{\rm jet}^{\rm obs}$ was derived by \citet{Gliozzi03} and the upper limit results from using the radio data and the \citet{Merloni07} correlation. The Bondi rate and inclination angle is from \citet{Gliozzi03}. \\
\textbf{References:} 1. \citet{Pellegrini05}, 2. \citet{David05}, 3. \citet{Gliozzi03}, 4. \citet{Di-Matteo01}, 5. \citet{Giroletti05}, 6. \citet{Allen06}, 7. \citet{Merloni07}, 8. \citet{Biretta99}, 9. \citet{Di-Matteo03}, 10. \citet{Barth01}
\\
\end{flushleft}
\label{tab:models}
\end{table*}

\section{Constraints on the accretion-jet parameters} \label{sec:pars}

Our detailed SED modeling allows us to place constraints \emph{-- within the framework of ADAF and jet models --} on the mass accretion rate onto the black hole and kinetic power for the LLAGNs in our sample. In this section, we evaluate possible correlations among the parameters describing the accretion flow and jet resulting from our SED models. In particular, when discussing correlations between the mass accretion rate and other parameters, we restrict ourselves only to AD models. The reason is that in most JD fits the estimate of $\dot{m}$ that we obtain from the fits should be regarded as a rough lower limit to the accretion rate. In fact, for most of the JD models, the contribution of an underlying ADAF to the emission is not necessarily required (cf. the JD models for NGC 4594, M87 and NGC 4579). 

\begin{figure}
\centerline{
\includegraphics[width=\linewidth,trim=15 0 10 10,clip=true]{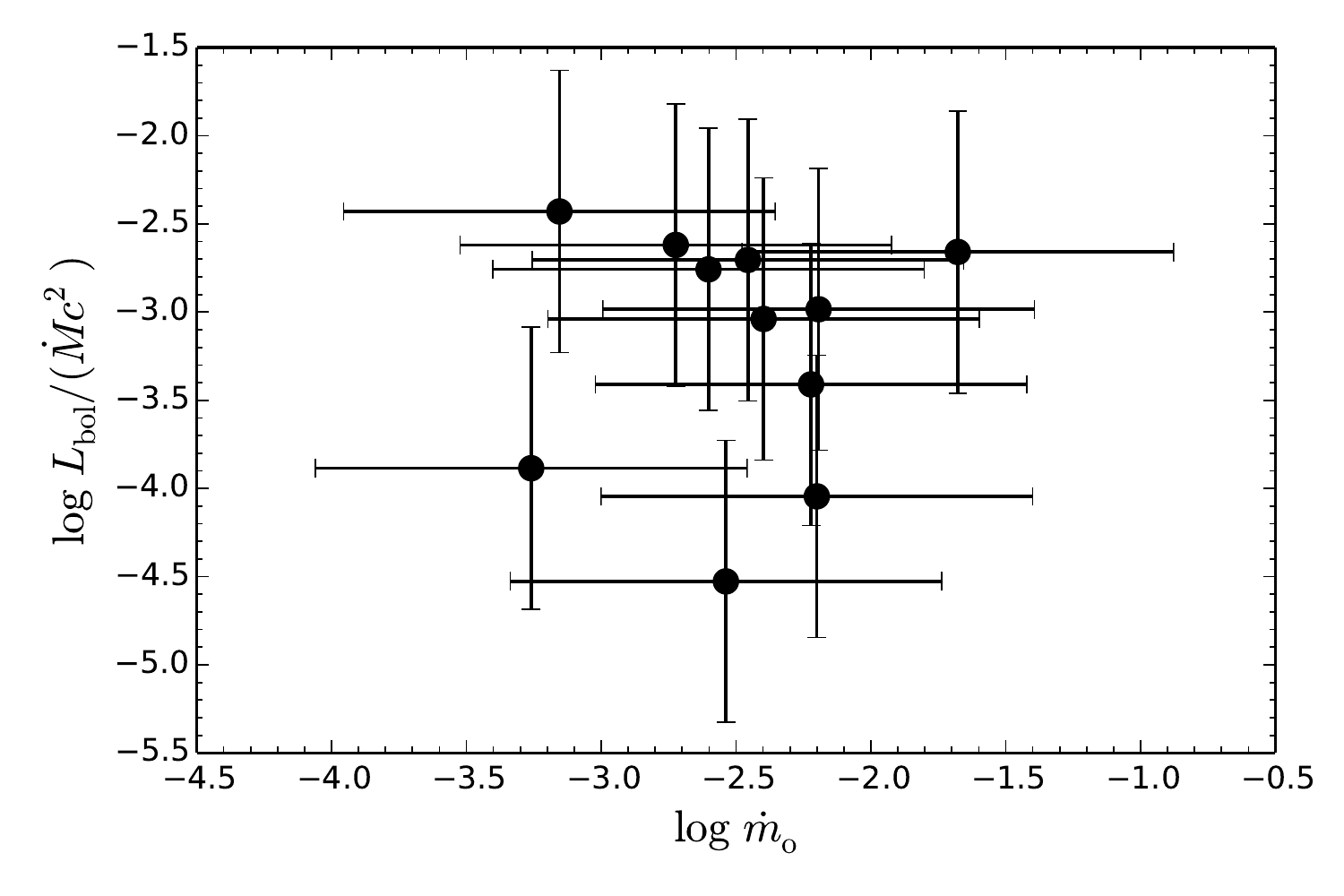}
\hskip -0.3truein
}
\centerline{
\includegraphics[width=\linewidth,trim=15 0 10 10,clip=true]{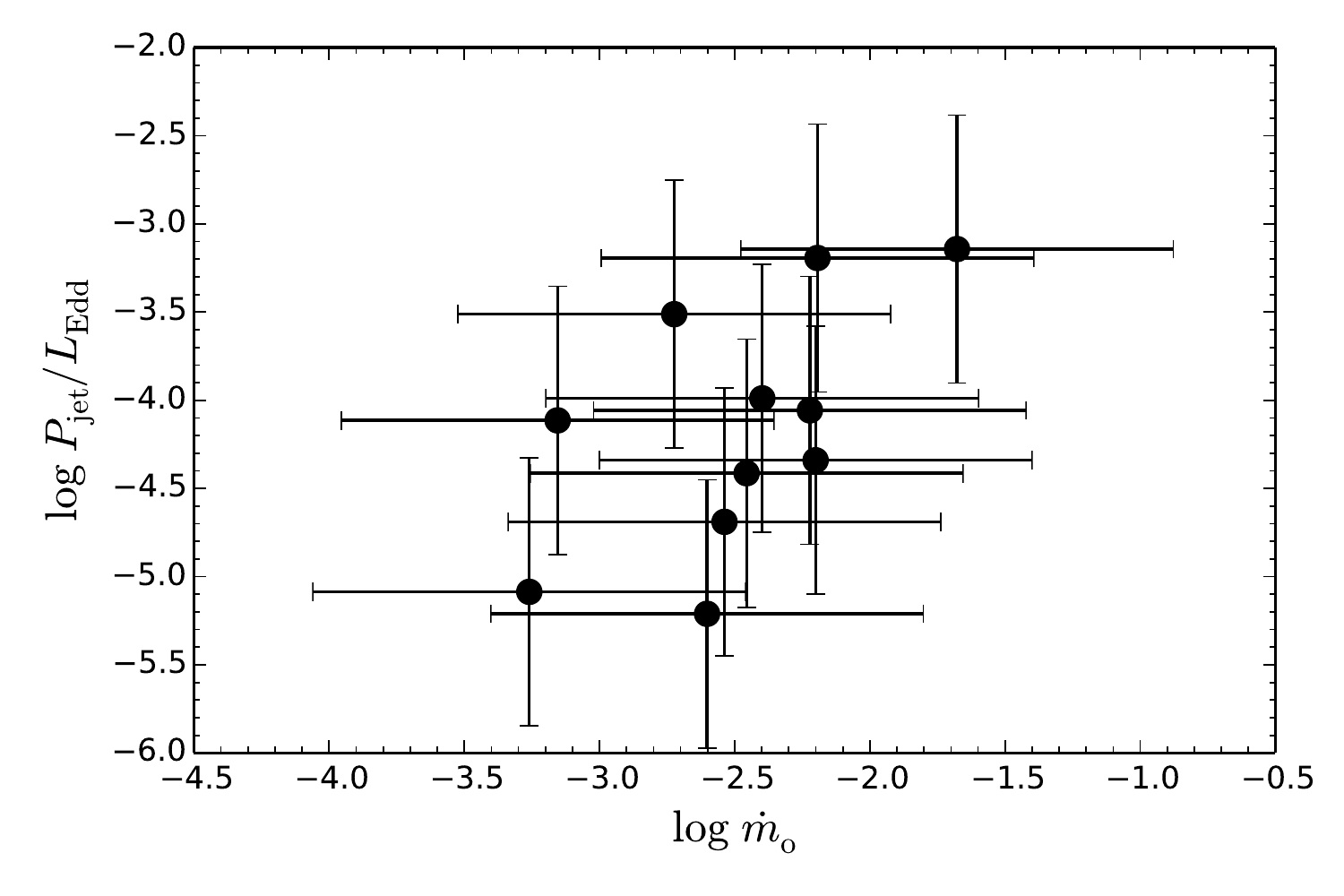}
\hskip -0.3truein
}
\centerline{
\includegraphics[width=\linewidth,trim=15 0 10 10,clip=true]{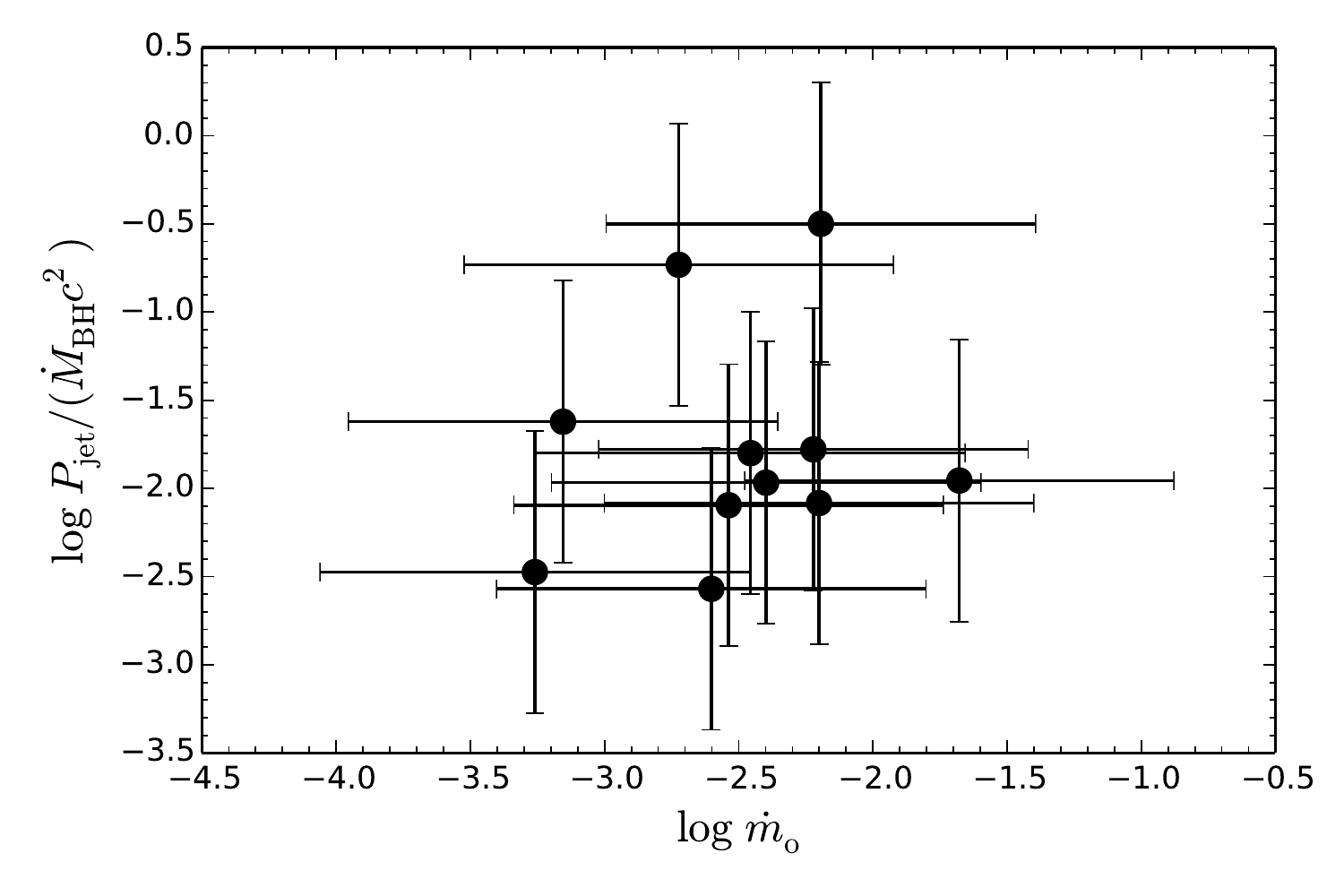}
\hskip -0.3truein
}
\caption{The relation between the mass accretion rate $\dot{m}_{\rm o}$ at the outer radius of the ADAF (in Eddington units) and the global radiative efficiency [$L_{\rm bol}/(\dot{M}_{\rm o} c^2)$] (panel a, top), jet power normalized in Eddington units (panel b, center) and the jet kinetic efficiency defined in terms of the accretion rate at $3 R_S$ (panel c, bottom). The error bars are estimated as described in the Appendix.}
\label{fig:pjet}
\end{figure}

We show in the panel a of Figure \ref{fig:pjet} the relation between the accretion rate at the outer radius of the ADAF and the global radiative efficiency of the systems, defined as $\eta_{\rm rad} \equiv L_{\rm bol}/(\dot{M}_{\rm o} c^2)$ where we obtain $L_{\rm bol}$ by integrating the synthetic spectra obtained in the SED fits including all components (thin disk, ADAF and jet). The values of $\eta_{\rm rad}$ for our sample are in the range $\sim (10^{-3} - 1) \%$ while the accretion rates are in the range $\dot{m} \approx 10^{-4}-0.01$. There is no apparent correlation between the global efficiency and $\dot{m}_{\rm o}$, keeping in mind the significant uncertainties in $\dot{m}_{\rm o}$ and $L_{\rm bol}$. The error bars are estimated as described in the Appendix.

Panel b of Figure \ref{fig:pjet} shows the relation between the accretion rate and the jet power in Eddington units, while panel c displays the relation between the accretion rate and jet kinetic efficiency $P_{\rm jet}/(\dot{M}_{\rm BH} c^2)$ defined in terms of the accretion rate at $3 R_S$, $\dot{M}_{\rm BH} \equiv \dot{M}(3 R_S)$. Panel b shows that there is a hint of a correlation which is significant at the $\approx 2\sigma$ level. However, panel c shows no clear evidence for correlation between the accretion rate and the kinetic efficiency.

\begin{figure}
\centering
\includegraphics[width=\linewidth,trim=70 0 80 10,clip=true]{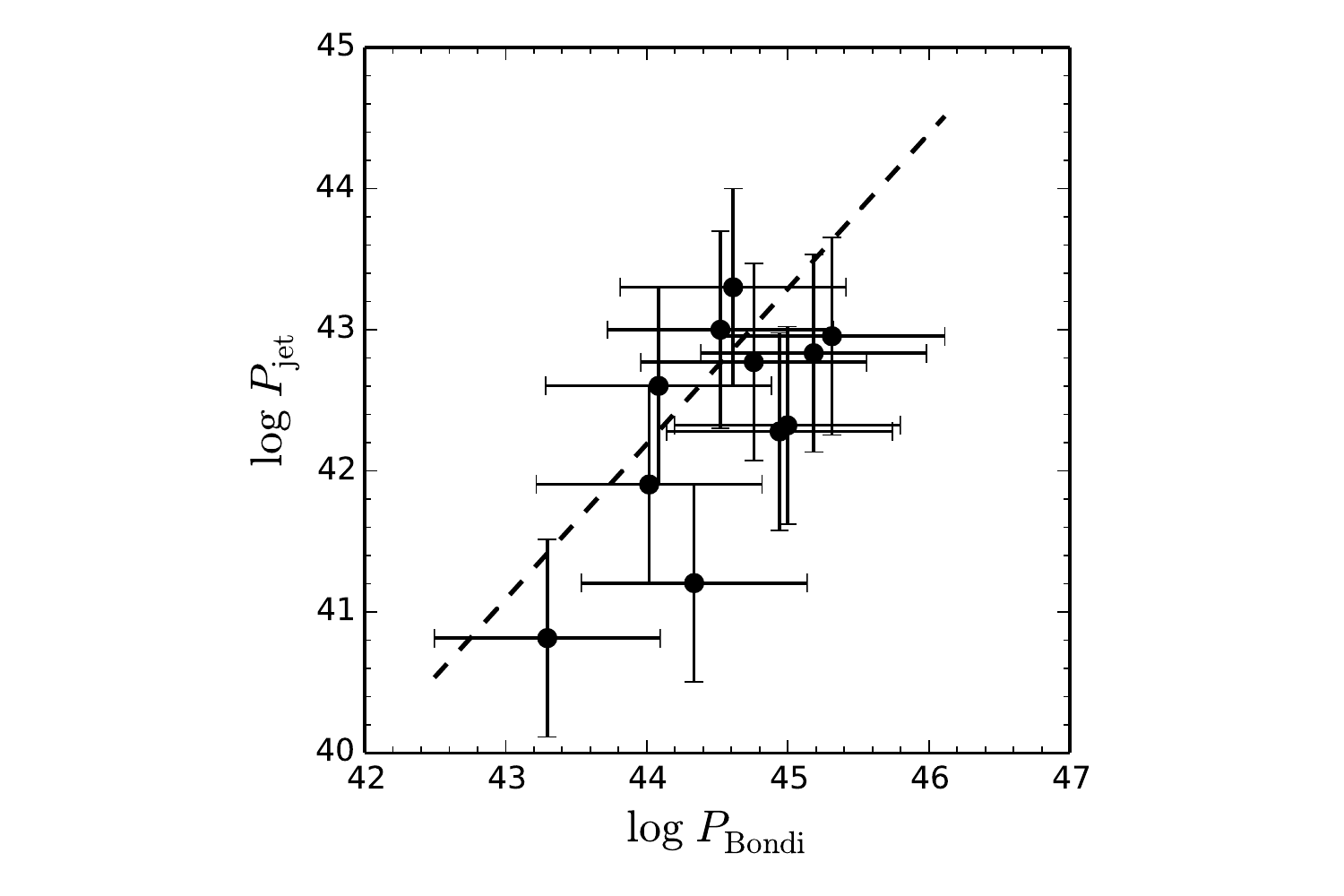}
\caption{The relation between the Bondi power $P_{\rm Bondi}$ as defined in equation \ref{eq:pbondi} and the jet power, using the parameters inferred from the SED models in our sample. The dashed line corresponds to the $P_{\rm Bondi}-P_{\rm jet}$ relation obtained by \citet{Balmaverde08}. }
\label{fig:pbondi}
\end{figure}

Low power radio galaxies exhibit a correlation between the jet power -- derived from X-rays cavities which were presumably created by the AGN outflows -- and the Bondi accretion rates  $\dot{M}_{\rm Bondi}$ -- inferred from the density and temperature profiles that are obtained from the X-ray observations (\citealt{Allen06,Balmaverde08}; but see \citealt{Russell13}). The Bondi rates are usually parametrized in the literature as a ``Bondi power'' assuming a $10\%$ efficiency ($P_{\rm Bondi} \equiv 0.1 \dot{M}_{\rm Bondi} c^2$). Therefore, in order to compare our results with these independent estimates of jet powers and $\dot{M}_{\rm Bondi}$, we use the accretion rates that we derived to define the ``Bondi power'' 
\begin{equation}	\label{eq:pbondi}
P_{\rm Bondi} \equiv 0.1 \dot{M}_{\rm o} c^2/\alpha,
\end{equation}	
taking into account that $\dot{M}_{\rm o} = \alpha \dot{M}_{\rm Bondi}$ in the ADAF model \citep{Narayan08}.
Figure \ref{fig:pbondi} displays the relation between $P_{\rm Bondi}$ defined using the equation above and the jet powers we derived from our SED models. The uncertainties in $P_{\rm Bondi}$ and $P_{\rm jet}$ are both $\sim 0.8$ dex (see Appendix). 
For illustration, we also show in Figure \ref{fig:pbondi} the relation between $P_{\rm Bondi}$ and $P_{\rm jet}$ derived by \cite{Balmaverde08}. Within the uncertainties, we can see that our SED fitting results are consistent with the Balmaverde et al. correlation.

\begin{figure}
\centering
\includegraphics[width=\linewidth,trim=10 0 10 10,clip=true]{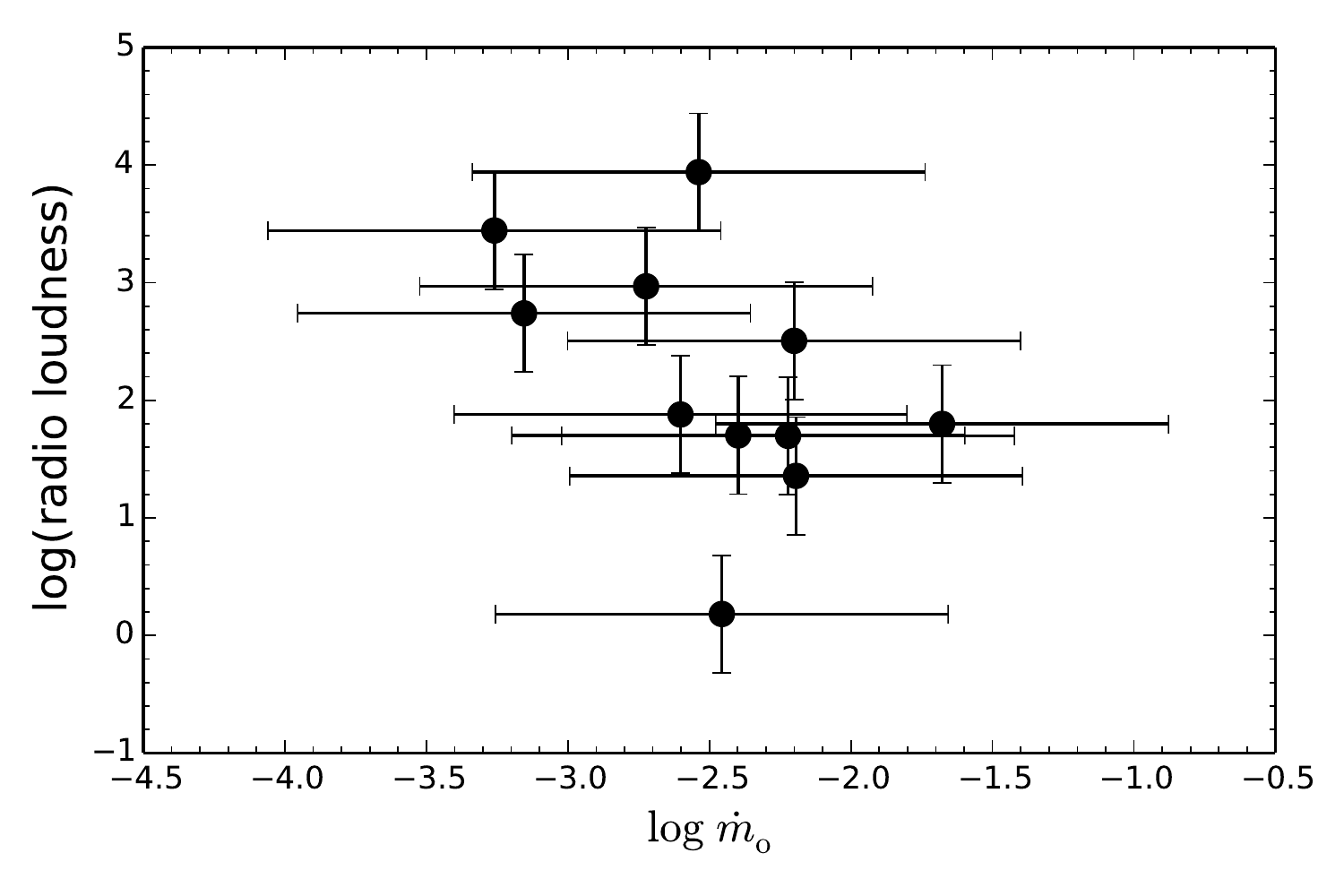}
\caption{Relation between the mass accretion rate $\dot{m}_{\rm o}$ and the radio loudness $R_o$. The anti-correlation is only significant at the $1.6 \sigma$ level.}
\label{fig:rl}
\end{figure}

We then proceed to estimate the radio loudness parameter from the SED fits. The radio loudness can be quantified as the optical to radio ratio $R_o \equiv L_\nu (6 \; {\rm cm}) / L_\nu ({\rm B})$ \citep{Kellermann89} where radio-quiet objects correspond to $R_o = 10$. Figure \ref{fig:rl} shows the relation between $R_o$ and the mass accretion rate. There is a hint of an anti-correlation between these variables, significant at the $1.6 \sigma$ level (keeping in mind the non-negligible uncertainties). This is in general agreement with previous evidence for an anti-correlation between radio-loudness and the Eddington ratio in radio-loud AGNs \citep{Ho02,Sikora07,Broderick11}. 

\begin{figure}
\centering
\includegraphics[width=\linewidth,trim=70 0 80 10,clip=true]{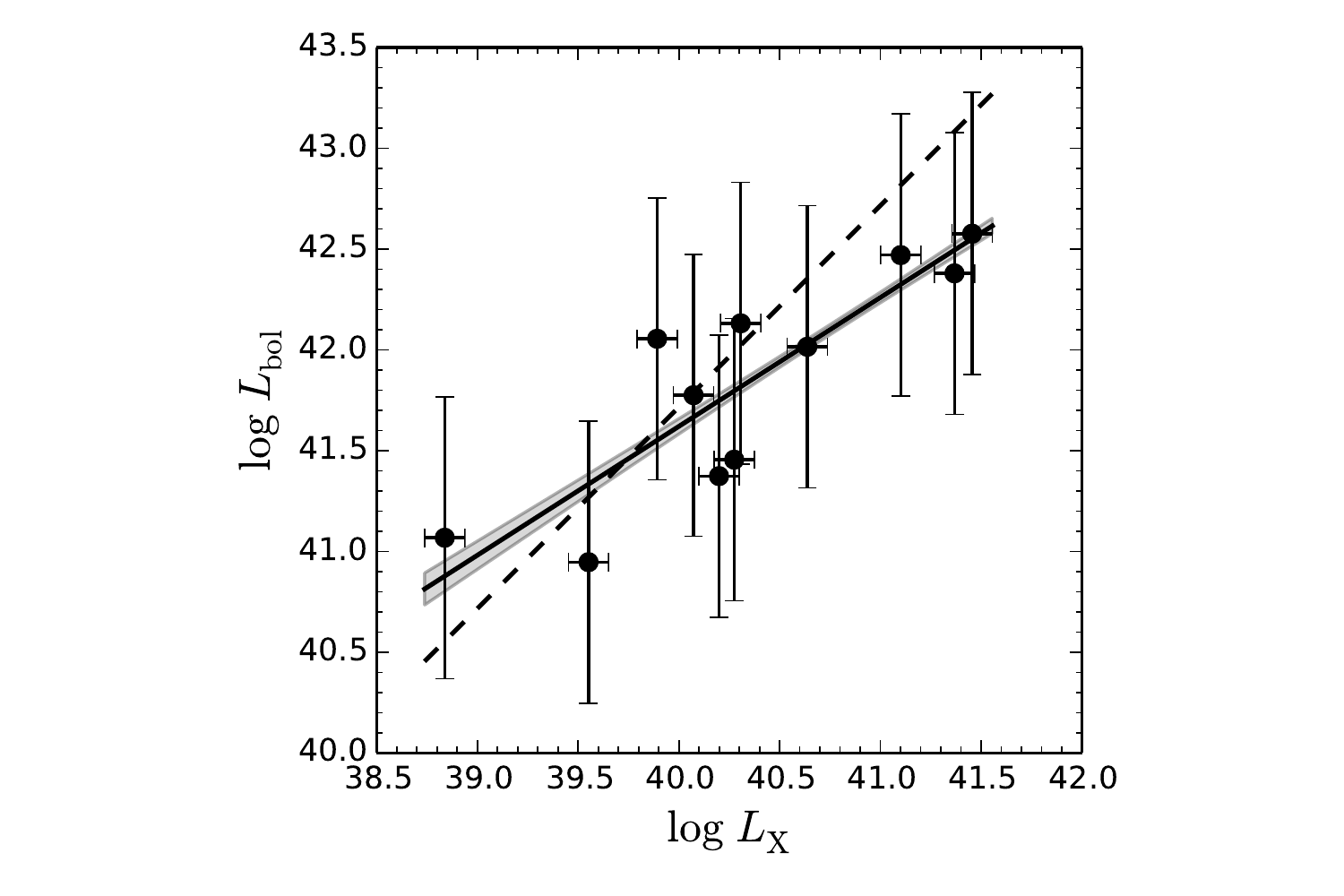}
\caption{The relation between $2-10$ keV X-ray luminosity and the bolometric luminosity. The solid line shows the best-fit power-law model as described in the text (shaded region, $1\sigma$ confidence band). The dashed line corresponds to $L_{\rm bol} = 52 L_X$ as in EHF10.}
\label{fig:bolcorr}
\end{figure}

We consider the relation between the X-ray luminosity and the bolometric luminosity. Figure \ref{fig:bolcorr} displays the correlation between $L_X$ and $L_{\rm bol}$ as well as the corresponding uncertainties. These variables display a correlation significant at the $3.4 \sigma$ level. We perform a linear regression using the least-squares BCES($Y|X$) method with bootstrapping \citep{Akritas96bces}. The best-fit model corresponds to the solid line ($\log L_{\rm bol} = A \log L_X + B$; shaded region corresponds to the $1\sigma$ confidence band). The best-fit parameters are $A=0.63 \pm 0.12$ and $B=16.3 \pm 4.9$. The scatter about the best-fit is 0.3 dex. Hence, this result suggests that the best-fit $2-10$ keV X-ray bolometric correction is given by 
\begin{equation}
\kappa_X \approx 13 \left( \frac{L_X}{10^{41} \ {\rm erg \ s}^{-1}} \right)^{-0.37}.
\end{equation}

EHF10 performed a linear interpolation (in log space) of the SED data points of a control sample of 7 LLAGNs and estimated the median X-ray bolometric correction as 52. Within the uncertainties, this bolometric correction is consistent with our results, as can be seen in Fig. \ref{fig:bolcorr} (dashed line corresponds to $L_{\rm bol} = 52 L_X$). We discuss how our values of $\kappa_X$ compare to those typical of quasars in Section \ref{sec:kbol}.

Finally, we computed the $\alpha_{\rm ox}$ parameter for our sample of AD model SEDs. $\alpha_{\rm ox}$ is a simple parametrization of the UV to X-ray ratio defined in terms of the spectral index of a power law between $L_\nu$ at $2500\, \rm \AA$ and at 2 keV defined as $\alpha_{\rm ox} \equiv 0.384 \log[L_\nu(2 \, {\rm keV})/L_\nu(2500 \, {\rm \AA})]$. We find a median value for $\alpha_{\rm ox}$ of -1.1 with a standard deviation of 0.1. An extrapolation of the $\alpha_{\rm ox} - L_\nu(2500 \, {\rm \AA})$ relation obtained from Seyferts and quasars \citep{Steffen06,Young09} to the low luminosities typical of our sample predict $\alpha_{\rm ox}$ values in the range $-0.9 \lesssim \alpha_{\rm ox} \lesssim -0.6$. Therefore, the typical $\alpha_{\rm ox}$ values in our sample of LLAGNs seem to be lower than what would be expected based on a direct extrapolation of trends obtained from brighter AGNs. We caution the reader, however, that our $\alpha_{\rm ox}$ estimates should be regarded with caution. For instance, given the uncertainty in the degree of extinction affecting the UV luminosities in our sample, we estimate that uncertainties as high as $\sim 0.5$ dex in $L(2500 \, {\rm \AA})$ -- and correspondingly in $\alpha_{\rm ox}$ -- are likely (cf. appendix \ref{sec:error}).

\section{The average SED}	\label{sec:avgsed}

It is of interest to compute the average SEDs resulting separately from the AD and JD scenarios. These SEDs are useful for different purposes. Since the observed SEDs are the main observables that we use to derive the central engine parameters via our fitting, as a consistency check the average model SEDs should reproduce the average SED obtained directly from the observed ones. 

Secondly, SEDs obtained from averaging observed data points are obviously limited by the observed bands. Outside the observed bands different authors adopt \textit{ad-hoc} interpolations between the data, usually a linear interpolation in log-log space (\citealt{Ho99}; EHF10), which might not reflect the actual physical processes involved in the emission. The average from the AD and JD scenarios that we obtained can hence serve as guidelines -- with a physical justification -- to the typical shape of the SEDs in the bands which have not been constrained yet.

\begin{figure*}
\centerline{
\includegraphics[scale=0.5]{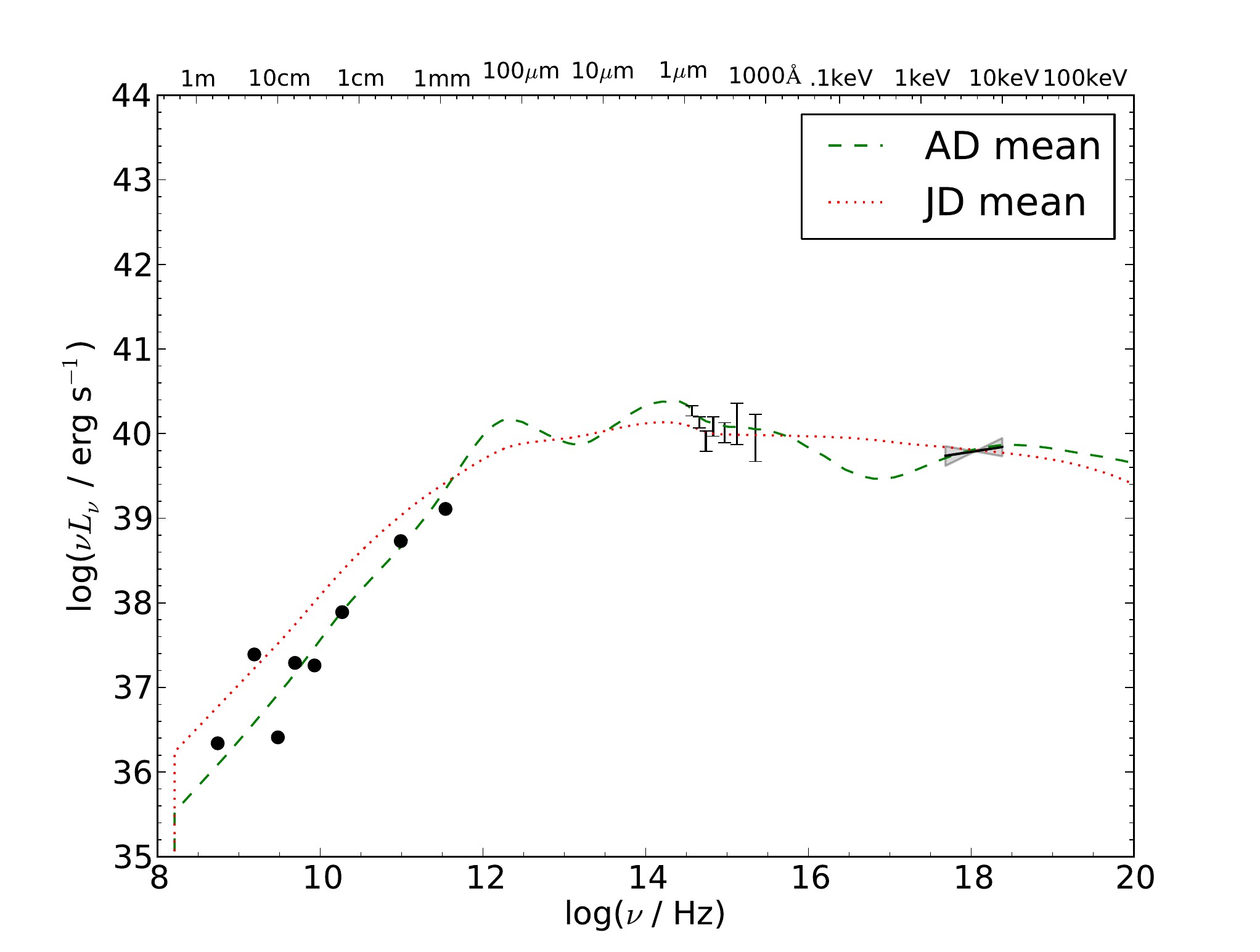}
\hskip -0.3truein
\includegraphics[scale=0.5]{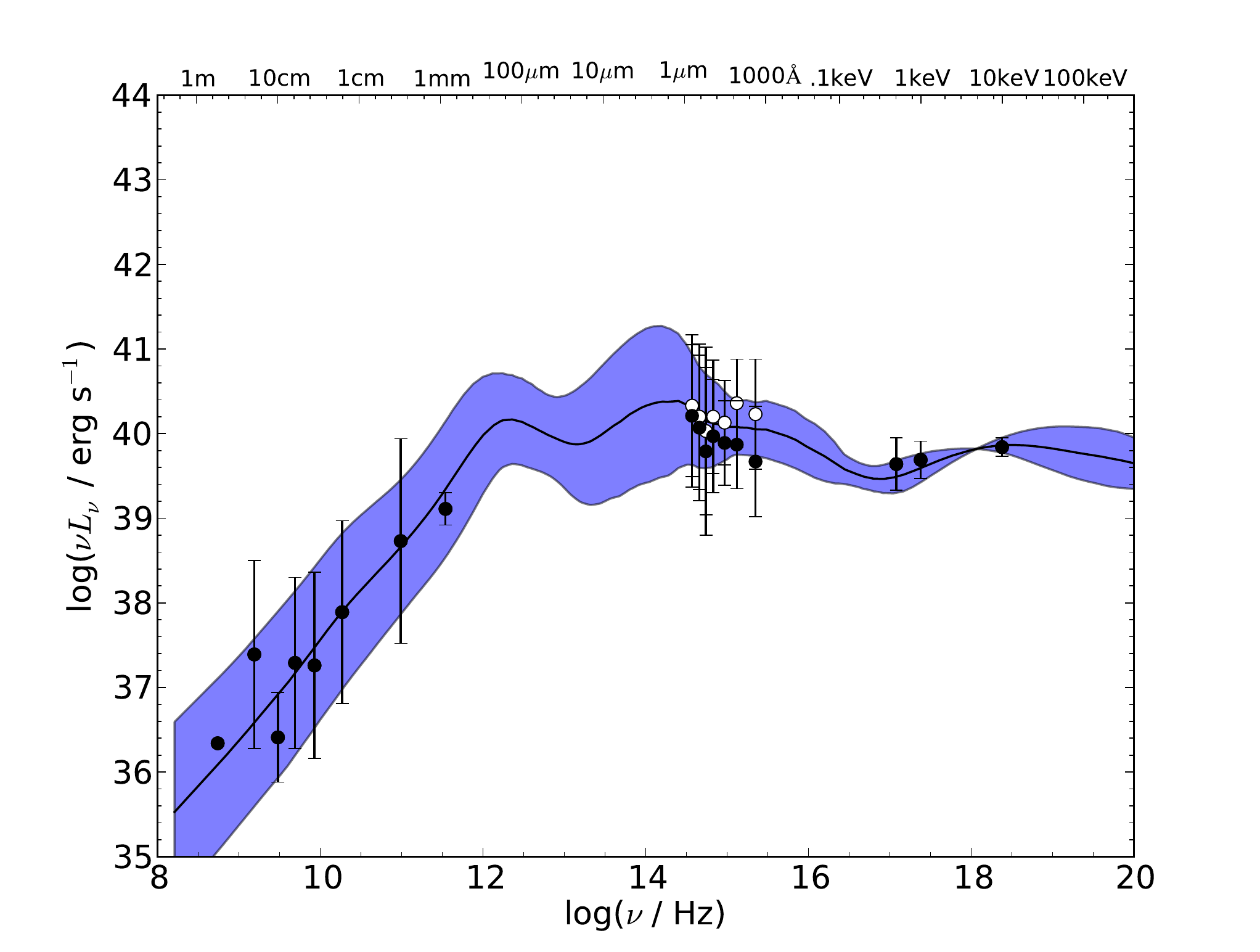}
}
\caption{\textbf{a, left:} The average SEDs (geometric mean) computed separately for the AD and JD models (dashed and dotted lines respectively). The data points correspond to the geometric mean computed by EHF10. The ``error bars'' on the optical-UV data points represent the range of extinction corrections. \textbf{b, right:} $1\sigma$ scatter around the average (model and observed) SEDs illustrating the diversity of individual SEDs. The solid line shows the average AD SED and the shaded region corresponds to the standard deviation from the AD models. The points correspond to the mean computed by EHF10 and the error bars show the scatter in the measurements. In the OUV, the filled circles correspond to measurements without any reddening correction whereas the open circles correspond to the maximal extinction correction.}
\label{fig:avgsed}
\end{figure*}

Motivated by the reasons above, we show in Fig. \ref{fig:avgsed}a the average SEDs computed separately for the AD and JD scenarios. We first normalized the individual SEDs to the same X-ray luminosity of $10^{40} \ {\rm erg \ s}^{-1}$ in the band 2-10 keV. This value is approximately the average X-ray luminosity for the LINERs in our sample. After normalizing the SEDs, for each frequency bin we computed the mean of the model SEDs as $\left\langle \log \nu L_\nu \right \rangle$. Following EHF10 we choose to compute the average of the logarithm of the luminosities instead of using the values of $\nu L_\nu$ themselves in order to reduce the effect of outliers in the resulting average. 

The average JD SED is relatively featureless and has a small bump in the IR between a few $\times10  \mu$m -- a few $\times 1 \mu$m which is due to the truncated thin disk. The average AD SED on the other hand is more complex given the richer variety of radiative processes which take place. Overall, the spectral shapes of the average JD and AD SEDs are similar. The bumps in the average SEDs are much less pronounced than the bumps in the individual SEDs. This occurs because the bumps in the individual SEDs do not peak at the same place and when the SEDs are averaged these bumps are smoothed out. For this reason, the average model SEDs will not resemble any one of the individual SEDs.

Figure \ref{fig:avgsed}a also displays the average data points computed from the observed SEDs in a similar way by EHF10 where the error bars represent the uncertainty in the emission due to the uncertainty affecting the amount of extinction correction involved. The shaded region around the average best-fit X-ray power-law represents the standard deviation in the value of the LINER photon index.
In order to illustrate the wide diversity of the individual SEDs, we show in Figure \ref{fig:avgsed}b the $1\sigma$ scatter affecting the model SEDs (where we show only the AD average SED for simplicity) and the observed ones. 

As expected, the average model SEDs agree well with the observed constraints. There are some details that are worth mentioning. 
The shape of the X-ray spectrum of the average JD SED is slightly softer than the corresponding shape of the average AD SED. Both are within the $1\sigma$ uncertainty in the photon index of the average observed SED.
In the OUV, even though the model SEDs agree with the observed constraints they are quite different from each other. For instance, the red bump is stronger in the average AD SED. The average JD SED predicts a lower UV flux.
In the radio band, the model SEDs are quite similar to each other.

\begin{figure}
\includegraphics[scale=0.5,trim=30 0 0 0,clip=true]{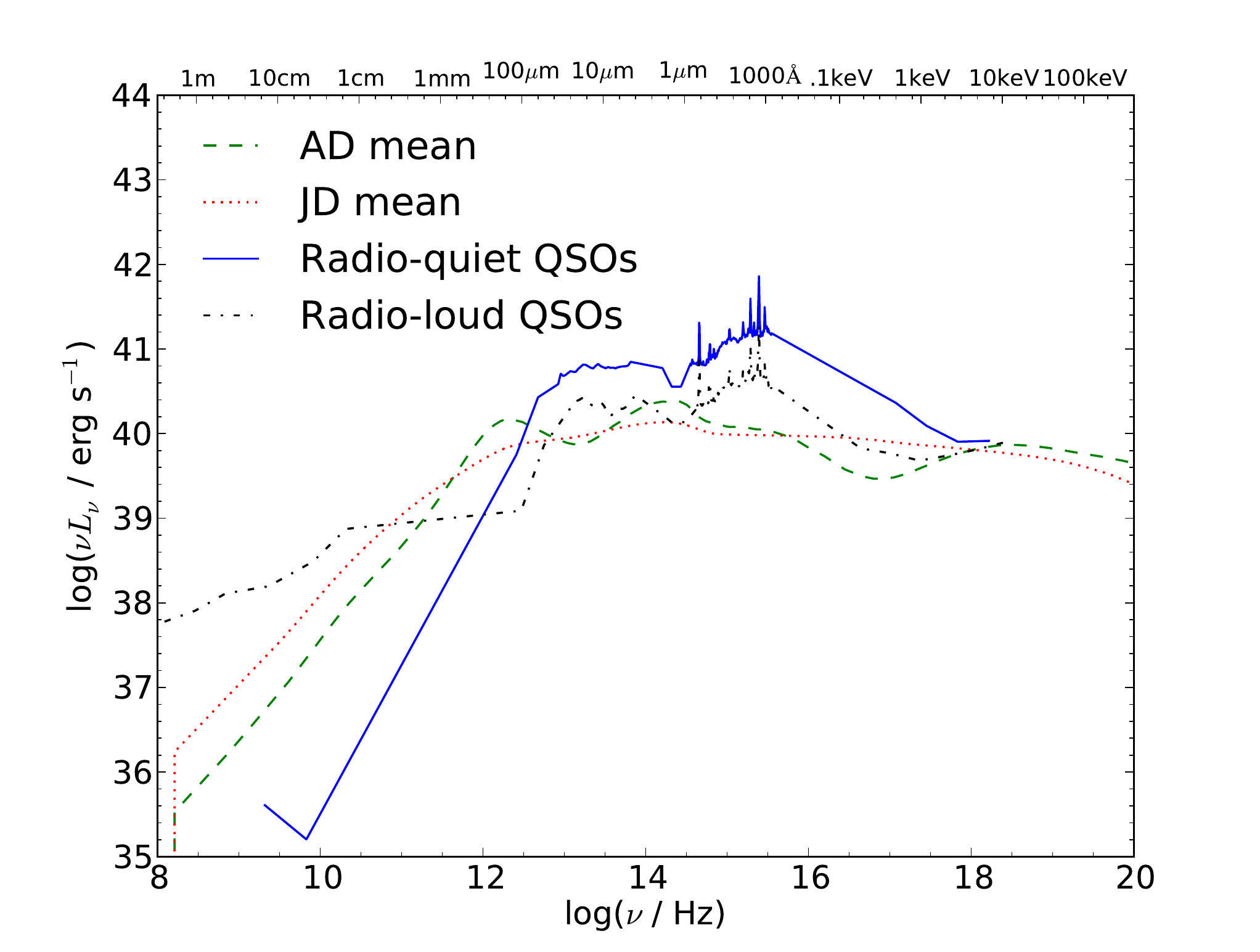}
\caption{The average JD and AD model SEDs compared to the average radio-loud and radio-quiet quasar SEDs computed by \citet{Shang11}. All SEDs are normalized to the same 2-10 keV luminosity.}
\label{fig:qsoavg}
\end{figure}

Figure \ref{fig:qsoavg} shows the average AD and JD SEDs compared to the average ones of radio-loud and radio-quiet quasars computed by \citet{Shang11}. The average quasar SEDs computed by \citet{Shang11} are very similar to the ones of \citet{Elvis94} but the former include more detailed features and are based on more recent data obtained with improved instrumentation.
As in Fig. \ref{fig:avgsed}a, the SEDs were normalized such that they all have the same X-ray luminosity in the 2-10 keV band of $10^{40} \ {\rm erg \ s}^{-1}$. 
The UV excess in the quasar SEDs (the big blue bump) is clearly apparent in comparison to the LLAGN ones. It is also interesting that for $\nu > 10^{17}$ Hz the average AD, average JD and the radio-loud quasar SEDs are quite similar.

\begin{figure}
\includegraphics[width=\linewidth,trim=90 50 60 0,clip=true]{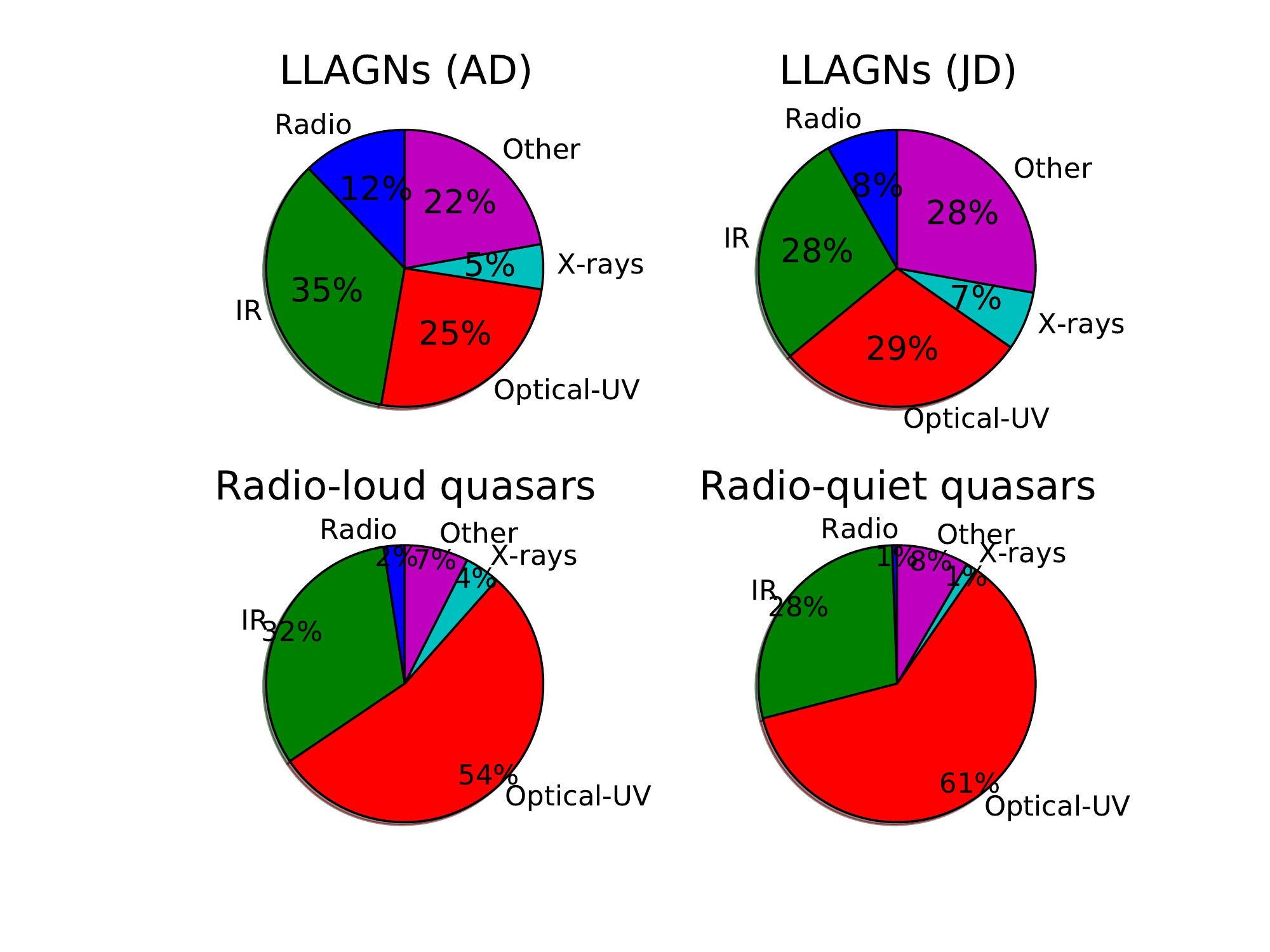}
\caption{$L_{\rm band}/L_{\rm bol}$ ratios for different wavebands, computed for the average AD and JD model SEDs, compared to the corresponding ratios for the average radio-loud and radio-quiet quasar SEDs of \citet{Shang11}. These average SEDs are displayed together in Figure \ref{fig:qsoavg}.}
\label{fig:pies}
\end{figure}

It is worth investigating how the luminosity is partitioned in different wavebands for the SEDs plotted in Fig. \ref{fig:qsoavg}. For this purpose, we computed the luminosity contained in the  radio ($10^8-10^{12.5}$ Hz), IR ($10^{12.5}-10^{14.5}$), optical-UV ($10^{14.5}-10^{16.5}$), X-rays ($2-10$ keV) and ``other'' ($10^{16.5} \ {\rm Hz}-2$ keV and $E>10$ keV) wavebands by integrating the SEDs, with the values in parenthesis denoting the range of frequencies or energies that we adopted for each waveband. The pie charts in Fig. \ref{fig:pies} display the $L_{\rm band}/L_{\rm bol}$ for each waveband defined above, for the average SEDs of LLAGNs and quasars. In the case of quasars, as is well known, the emission is dominated by the UV bump with the IR emission corresponding to the reprocessed emission of the accretion disk by the dust torus. In the case of the average AD and JD SEDs for the LLAGNs in our sample, the IR, optical-UV and ``other'' bands release comparable fractions of the bolometric luminosity contrary to the case of quasars. In particular, the IR dominates the energy budget of the SED in the AD case. It is also notable that a higher fraction of the energy is released in the radio and X-rays compared to the quasar SEDs.

We remind the reader that the shape of the SEDs of individual objects may display a significant variance with respect to the average SEDs (e.g., \citealt{Shang11,Runnoe12}; see also the discussion in section \ref{sec:kbol}). As such, although the energy budget displayed in Fig. \ref{fig:pies} is an average representation of the whole sample of LLAGNs studied in our work and the Shang et al. quasar SEDs, it should be regarded with care when applied to individual sources.

\section{Discussion}	\label{sec:disc}

We discuss the caveats that affect our analysis in Section \ref{sec:caveats}. In Section \ref{sec:xrays} we discuss the nature of X-rays in LLAGNs and in Section \ref{sec:future} we suggest directions for making progress in this issue.

\subsection{Caveats}	\label{sec:caveats}

From an observational point of view, even though we are carrying one of the largest systematic analysis of the SEDs of LLAGNs to date (see also \citealt{Wu07, Yu11}), for most of the sources that we considered we only have a few data points available to fit. The lack of observational constraints, of course, impacts our ability to rule out the different scenarios for the central engine of LLAGNs -- AD vs JD models  -- and to probe the transition between the ADAF and the thin disk. Therefore, it is clear that a \emph{strong case exists for obtaining more measurements and better detections of the sample over a broader range of wavelengths} in order to better constrain the models for the central engine. We discuss specific observational strategies in the Section \ref{sec:xrays}.
Furthermore, the size of our sample is limited. We need to extend our modeling to a larger number of LLAGNs in order to draw further conclusions. 

From a theoretical perspective, there are considerable theoretical uncertainties involved in the the ADAF-jet models. For instance, we cannot expect to constrain the ADAF parameters $s$ and $\delta$ from the SED fitting since there is a degeneracy between these parameters \citep{Quataert99winds}. Previous modeling efforts overcome the particular difficulty of the $s/\delta$ degeneracy by fixing the values of these parameters (e.g., \citealt{Wu07,Yuan09}) to the ones consistent with constraints from Sgr A* \cite{Yuan03}. Despite recent progress (e.g., \citealt{Sharma07}), the value of $\delta$ is essentially unconstrained. For this reason, the values of the parameters inferred in our fits do not correspond necessarily to unique choices but rather should be considered as illustrative of reasonable fits. 
Furthermore, our jet model is basically a phenomenological one: we have more freedom in the jet fitting compared to the ADAF model.

\subsection{The nature of X-ray emission in LLAGNs}	\label{sec:xrays}

The nature of the X-ray emission in LLAGNs has been debated in the last few years by several authors  favoring in some cases the AD or JD models based on the analysis of individual sources (e.g., \citealt{Falcke00,Yuan02,Yuan02n4258,Yuan03,Wu07,Markoff08,Miller10}) or in a statistical sense based on the fundamental plane of black hole activity \citep{Merloni03,Falcke04,Yuan05fp,Yuan09,Plotkin12,Younes12}.

In our AD models, the X-rays are produced predominantly by inverse Compton scattering by the ADAF of seed synchrotron photons produced in the accretion flow itself; in contrast, in our JD models, the X-rays are dominated by the optically thin synchrotron emission at the base of the jet. In this regard, the X-ray spectrum is an important observable that can be used in order to disentangle the dominant component responsible for the LLAGNs. 
Internal shocks in the jet will favor electron energy distributions with power-law indices in the range $p\approx 2-3$ \citep{Bednarz98,Kirk00,Achterberg01,Sironi09}. In the framework of JD models, the above range of values of $p$ results in $\Gamma_X \approx 2-2.5$ (cf. Section \ref{sec:fit}). Therefore, sources characterized by X-ray spectra with $\Gamma_X \leq 2$ (i.e. on the hard side) are unlikely to be jet-dominated in X-rays. 

\begin{figure}
\includegraphics[width=\linewidth,trim=10 0 20 10,clip=true]{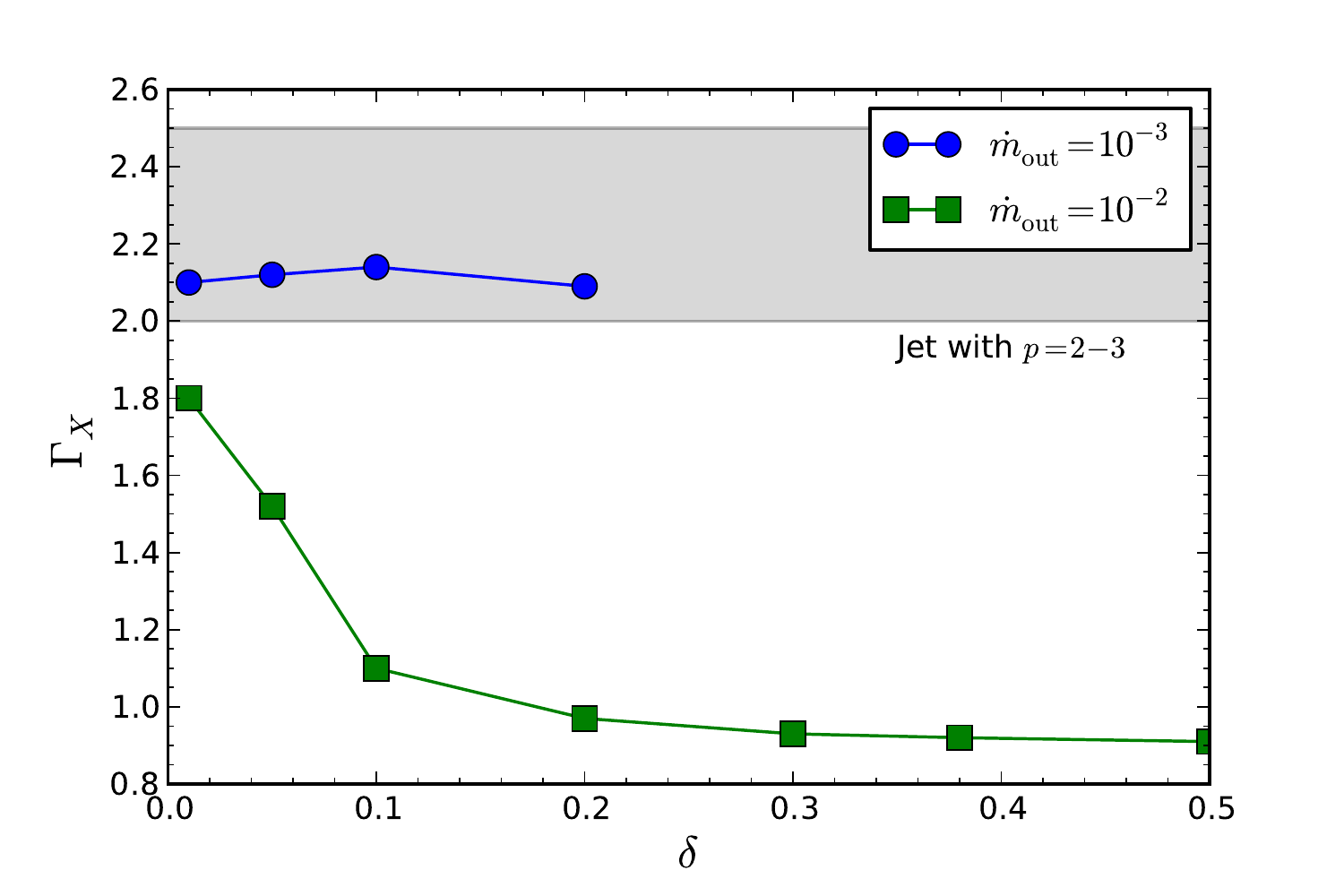}
\caption{The X-ray photon index $\Gamma_X$ predicted by ADAFs as a function of $\delta$ for different accretion rates. We computed $\Gamma_X$ from a grid of ADAF models computed for $R_{\rm o}=500$, $s=0.3$ and $m=10^8$. The shaded region corresponds to the range of photon indexes for jets with spectral index of the electron energy distribution in the range $p = 2-3$.}
\label{fig:hard}
\end{figure}

Figure \ref{fig:hard} illustrates the range of values of the X-ray photon index resulting from ADAFs with typical accretion rate as in the LLAGNs modeled in our sample. This figure demonstrates that  ADAFs naturally accommodate hard X-ray emission with $\Gamma_X<2$. 
In our sample, 6 LINERs have hard X-ray spectra which cannot be easily explained as jet emission whereas for only one source, NGC 4552, we were not able to find an AD model able to account for the soft X-ray SED with $\Gamma_X=2$ (cf. Table \ref{tab:models}).

Based on the observations above, we suggest that the spectral shape of the X-ray spectrum could be used to differentiate between an AD versus a JD origin for the X-rays. Namely, soft X-ray LLAGNs ($\Gamma_X \sim 2-2.5$) would be likely jet-dominated whereas hard X-ray emitters ($\Gamma_X < 2$) would be ADAF-dominated. 

Note that it is still possible that relativistic shocks in jets could give rise to electron energy distributions with $p$ slightly smaller than 2 in a small fraction of sources \citep{Starling08,Curran10} and, as noted by \cite{Yuan09}, reconnection effects could also change the electron energy distribution produced by shocks. 
In addition, ADAFs with relatively high accretion rates ($\dot{m}_{\rm o} \approx 0.01$) can produce X-ray spectra with $\Gamma_X \approx 2.1$ (cf. Fig. \ref{fig:hard}) which has the same hardness as jets with $p=2.2$. At these accretion rates we expect that the thin disk would be truncated at radii $r_{\rm tr} \sim 100$ \citep{Yuan04,Narayan08} and hence the SED would be accompanied by relatively strong thermal emission in the near-IR. In the next section, we outline additional observations which will be crucial to disentangle the high-energy contribution of the jet and ADAF.

\subsection{X-ray bolometric corrections in LLAGNs and quasars}	\label{sec:kbol}

It is instructive to compare the X-ray bolometric corrections we estimate in this work with those typical of quasars. One common way is to integrate the SED using the IR emission as a proxy of the intrinsic nuclear luminosity, assuming that a dust torus reprocesses the intrinsic optical-UV emission into IR radiation (e.g. \citealt{Pozzi07,Vasudevan10,Lusso11}). Another closely related approach is to integrate the SED directly using the optical-UV emission  
(e.g., \citealt{Marconi04,Hopkins07,Vasudevan07,Runnoe12}). For example, \cite{Lusso11} find $\langle \kappa_X \rangle \sim 23$ for type-I quasars whereas \cite{Runnoe12} estimate $\langle \kappa_X \rangle \sim 38$ for their sample. An extrapolation of the results of \cite{Marconi04, Hopkins07} to the luminosities characteristic of LLAGNs predicts values as low as 10 or even lower for $\kappa_X$ (cf. also \citealt{Vasudevan07,Young10,Lusso11}). 

As we described in section \ref{sec:pars}, we estimate $L_{\rm bol}$ by integrating the entire SED from radio up to 100 keV and including the IR emission since it is not clear at all whether the IR photons in LLAGNs should be treated as reprocessed emission from a dusty torus (e.g., \citealt{Mason13}). We find $\kappa_X \approx 13 \left( L_X/10^{41} \ {\rm erg \ s}^{-1} \right)^{-0.37}$ for the LLAGN SEDs in our sample; in other words, our average bolometric correction seems to be in rough agreement with the low values expected from the extrapolations of the quasar results mentioned above. At lower luminosities, our results suggest that $\kappa_X$ increases slightly ($\kappa_X \approx 32$ at $L_X=10^{40} \ {\rm erg \ s}^{-1}$).

We caution that extrapolations of bolometric corrections derived from quasar samples to lower luminosities have to be regarded with caution since, as we previously argued, it is likely that the accretion physics in LLAGNs may be significantly different compared to quasars. We also remind the reader that our $\kappa_X$ estimates are subject to non-negligible uncertainties ($\sim 0.3$ dex; cf. section \ref{sec:pars} and appendix \ref{sec:error}) given the uncertainties in the $L_{\rm bol}$ determinations. In addition, the mean X-ray bolometric corrections can be inaccurate for individual objects due to variation in the X-ray emission and the SED shapes (or equivalently, the different degrees of contributions of the ADAF, thin disk and jet emission) from object to object, i.e. there is a large intrinsic scatter in the values of $\kappa_X$ (cf. the scatter in the SEDs in Figure \ref{fig:avgsed}b). Such a large scatter was also noted in the case of quasar SEDs (e.g., \citealt{Marchese12,Runnoe12}).

We now discuss a few individual sources and their associated X-ray bolometric corrections. Two examples of SEDs with relatively low $\kappa_X$ correspond to the AD model for NGC 3998 (Fig. \ref{fig:seds02}; $\kappa_X \approx 10$) and the JD model for NGC 4594 (Fig. \ref{fig:n4594}; $\kappa_X \approx 11$). More generally, SEDs with $\kappa_X \lesssim 10$ are roughly flat (to first order) in the IR to the X-rays range. 
On the other hand, many of the modeled SEDs in our sample -- in particular, the AD models -- have considerable imbalances in the fraction of the luminosity radiated in the different wavebands. For instance, the integrated luminosity contained in the IR, optical-UV and $> 0.1$ keV wavebands may considerably exceed $L_X$ (see Fig. \ref{fig:pies}). For example, the models for NGC 1097 and NGC 4278 ($\kappa_X$ values of 24 and 67, respectively) with their pronounced IR ``bumps'' due to the truncated thin disk emission, and the extreme case of NGC 4143 with $\kappa_X \approx 146$. 

\subsection{The optical emission}	\label{sec:disk}

The peak of the emission for a thin disk truncated at $r \sim 50-1000$ for $m \sim 10^8$ is typically located in the wavelength range $1-10 \mu$m with a high-frequency tail extending into the optical and UV (e.g., \citealt{Nemmen06,Yu11,Taam12}). 
Indeed, six sources (NGC 1097, NGC 3031, NGC 4143, NGC 4278, NGC 4374 and NGC 4736) display an optical-UV excess with respect to the ADAF and jet models. In four of these sources (NGC 1097, NGC 4143, NGC 4278 and NGC 4736), on the assumption that the optical-UV observations are dominated by the AGN, we were able to reproduce the data with the truncated thin disk emission with reasonable accretion rates and transition radii. For NGC 1097, NGC 3031 and NGC 4579 in particular, the transition radius is fixed by the modeling of the H$\alpha$ double-peaked emission-line (e.g., \citealt{Storchi-Bergmann03,Nemmen06} and references therein); for these three LLAGNs, the thin disk component is significantly luminous (note that 4579 does not display an optical-UV excess with respect to the ADAF emission), given the relatively small transition radii implied by the H$\alpha$ emission line. While the optical spectrum of NGC 1097 is explained by the thin disk emission, the available IR upper limits and optical-UV observations for NGC 3031 and NGC 4579 do not allow us to put useful constraints on the emission of the truncated thin disk for these sources. For the remaining sources in the sample, the lack of appropriate IR constraints -- i.e. observations with high enough spatial resolution to accurately isolate the AGN -- prevents us from better constraining the outer thin disk emission. We discuss observations that will be helpful in order to shed light into this issue in section \ref{sec:future}.

Even though the truncated thin disk is an appealing explanation for the optical-UV observations in many of the LINERs in our sample, we are unable to rule out the possibility that the optical emission in our sources is produced by a unresolved stellar population based solely on the SED modeling. This possibility was already pointed out by \cite{Maoz05,Maoz07}. From the fits of stellar population models to the optical data we obtained stellar masses in the range $\sim {\rm a\ few\ }10^7-10^8 M_\odot$ (Table \ref{tab:star}). In fact, as we argued in section \ref{sec:stars}, such an amount of mass due to unresolved stellar populations in the bulge within the HST aperture is quite feasible. 

\begin{table}
\centering
\caption{Masses of stellar populations required to fit the HST optical emission for the galaxies in our sample.}
\begin{tabular}{@{}llrccccc@{}}
\hline
Galaxy & log mass \\
       & ($M_\odot$)   \\
\hline
NGC 1097                 &  8.2  \\
NGC 3031           &  7.8  \\
NGC 3998                 &  8.5  \\
NGC 4143                 &  7.7  \\
NGC 4261                 &  7.1  \\
NGC 4278                 &  8  \\
NGC 4374  &  7.7  \\
NGC 4486   &  8.3  \\
NGC 4552    & --  \\
NGC 4579        &  8.1  \\
NGC 4594     &  7.6 \\
NGC 4736    &  8.1 \\
\hline
\end{tabular}
\label{tab:star}
\end{table}

In addition, some of the galaxies in our sample could potentially host a compact nuclear star cluster, similarly to the Milky Way \citep{Schodel09,Genzel10}. The compact nuclear star cluster in our Galaxy has a density $\rho_{\rm NC} \sim 10^6 M_\odot \, {\rm pc}^{-3}$ \citep{Schodel09,Genzel10}. If the galaxies in our sample host nuclear star clusters with similar densities as the Milky Way, then the implied star cluster mass enclosed within a radius $R$ is 
\begin{equation}
M_{\rm NC} \sim 4 \times 10^9 \left( \frac{\rho_{\rm NC}}{10^6 M_\odot \, {\rm pc}^{-3}} \right) \left( \frac{R}{10 {\rm pc}} \right)^3 \, M_\odot.
\end{equation}
This mass estimate is likely to be on the high-end for our sample; however we do note that studies of galaxies hosting both nuclear star clusters and massive black holes suggest that the masses of these components are quite similar \citep{Seth08}.

\subsection{Future observational progress}	\label{sec:future}

We suggest different routes to clarify the issue of the nature of X-ray emission in LLAGNs in the future.
The AD and JD models should predict different characteristic radio and X-ray variability timescales (e.g., \citealt{Ptak98}), hence we should be able to test which component is dominant in X-rays by carrying out a simultaneous monitoring of the variability of radio and X-rays. By comparing the variability pattern predicted by jet/ADAF models we could pinpoint the X-ray dominant component. This promising strategy is quite similar to what the observational campaigns carried out for M81 \citep{Markoff08,Miller10} and could be applied to many other LLAGNs. 

The \emph{Fermi} Gamma-ray Space Telescope can be quite helpful in this regard. LLAGNs are potential sources of $\gamma$-rays \citep{Takami11} and in fact one of the source in this sample (NGC 4486) has already been detected by the \emph{Fermi} LAT \citep{m87fermi}. By using the model parameters that we derived in our radio-to-X-rays fits we should be able to predict the $\gamma$-ray spectrum \citep{Mahadevan97,Takami11} and compare with future \emph{Fermi} detections. In this way we should be able to compare AD/JD predictions and through $\gamma$-ray observations place further constraints on the production site responsible for the high-energy emission and the jet-disk connection in LLAGNs.

Observations in the mm and sub-mm are also very helpful because they constrain the ADAF synchrotron emission and hence also the X-ray emission. This follows because the synchrotron photons in the ADAF are inverse-Compton-scattered to X-rays.

Another future direction for a shedding light on the role of the jet in the SEDs of LLAGNs lies in refining and better understanding of the fundamental plane of black hole activity \citep{Merloni03,Falcke04,Gultekin09,Yuan09,Plotkin12} as a tool for constraining the radiative processes shaping the radio and X-ray emission in sub-Eddington black hole sources.

Finally, 
it is crucial to observe the LLAGNs of our sample at high spatial resolution in the IR wavebands in order to probe the presence of the outer thin disk component. There has been recent observational progress on this front \citep{Asmus11, Fernandez-Ontiveros12, Mason12}. Since emission from hot dust is also important in these IR bands, the contribution of the putative dusty torus \citep{Ramos-Almeida11} must also be taken into account. Modeling these recent observations is beyond the scope of this work (cf. Mason et al. 2013, in preparation).

\section{Summary}	\label{sec:end}

We performed a detailed exploratory modeling of the broadband spectral energy distributions of a sample of 12 low-luminosity AGNs in LINERs selected from EHF10 based on the presence of a compact radio cores and nuclear X-ray point sources as well as the availability of high-resolution optical/UV observations. Our coupled accretion-jet model consists of an accretion flow which is radiatively inefficient in the inner parts and becomes a thin disk outside a certain transition radius. The relativistic jet is modeled in the framework of the internal shock scenario. In the framework of our model, our main results are as follows.

(i) We find that there are two broad classes of models that can explain the majority of the observed SEDs. We call the first one AD which stands for ``ADAF-dominated'' since the ADAF dominates most of the broadband emission, particularly in X-rays. In the second class of models, the jet component dominates the majority of the continuum emission and for this reason we call this scenario JD as in ``jet-dominated''. 

(ii) We suggest that the spectral shape of the X-ray spectrum could be used to differentiate between an AD versus a JD origin for the X-rays: soft X-ray LLAGNs (photon spectral indices $\sim 2-2.5$) are likely jet-dominated whereas hard X-ray emitters (photon indices $< 2$) are likely ADAF-dominated. We discuss different observational strategies to make progress in understanding the origin of X-rays in LLAGNs.

(iii) The radio band is almost always dominated by the synchrotron emission from the jet, confirming previous results.

(iv) Six sources in our sample display an optical-UV excess with respect to ADAF and jet models; in four of them (NGC 1097, NGC 4143, NGC 4278 and NGC 4736), this excess can be reproduced by  emission from a truncated thin accretion disk with transition radii in the range $30-225 R_S$. Alternatively, unresolved, old stellar populations with masses $\sim 10^7-10^8 M_\odot$ located within $\approx 2-15$ pc of the nuclei can also reproduce the HST optical observations.

(v) The mass accretion rates that reach the outer radius of the ADAF -- based on the AD models -- lie in the range $\sim 1 \times 10^{-4}-0.02$ in Eddington units. These accretion rates are likely to represent upper limits if the jet is the dominant contributor to the broadband high-energy emission. 

(vi) Typically $\sim 10\%$ of these accretion rates reach $3 R_S$ with the remaining gas being presumably lost due to outflows.

(vii) The efficiency of conversion of rest mass energy associated with gas supplied to the accretion flow into disk+jet radiation is in the range $\sim (10^{-3}-0.1)\%$. 

(viii) Jet kinetic powers from the SED models are in the range $\sim 10^{40} - 10^{44} \ {\rm erg \ s}^{-1}$. The efficiency with which the rest mass energy associated with gas supplied to the black hole is converted into jet kinetic power is in the range $0.1-10\%$. 

(ix) We find hints of correlations between the accretion rate and jet power ($2\sigma$) and anti-correlation ($\approx 1.6\sigma$) between the radio-loudness -- estimated using core radio luminosity -- and the accretion rate.

(x) We compute the average SED for LLAGNs using our model fits. The average SEDs are quite different from the quasar ones and display a slight IR bump. High spatial resolution IR observations are required in order better constrain the physics of IR emission in LLAGNs.

(xi) There is evidence for a nonlinear relation between the $2-10$ keV and the bolometric luminosity for the sources in our sample [$\log L_{\rm bol} = (0.63 \pm 0.12) \log L_X + 16.3 \pm 4.9$].

This work highlights the need for a better multiwavelength coverage of LLAGN SEDs in order to make progress in understanding the role of the different components of the central engine.
The individual and average SED models that we computed can be useful for a number of different applications.

\section*{Acknowledgments}

We are grateful to: Feng Yuan for useful discussions as well as help with the models and allowing us to use some of his codes; Renyi Ma and Hui Zhang for their help with setting up the models; Jo\~ao Steiner, Michael Brotherton, Rog\'erio Riffel, Francesco Tombezi, Judith Racusin, Rafael Eufrasio, Rachel Mason and Roman Shcherbakov for productive discussions; and to the referee for the useful comments and the suggestion of considering the contribution of stellar populations. RSN was supported by an appointment to the NASA Postdoctoral Program at Goddard Space Flight Center, administered by Oak Ridge Associated Universities through a contract with NASA. TSB acknowledges the financial support of the Brazilian institutions CNPq and CAPES. This research has made use of the NASA/IPAC Extragalactic Database (NED) which is operated by the Jet Propulsion Laboratory, California Institute of Technology, under contract with the National Aeronautics and Space Administration.

\bibliographystyle{mn2e}

\appendix

\section{Uncertainties in derived parameters of LLAGNs}	\label{sec:error}

In this section, we estimate the uncertainties in the main parameters obtained from the SED models such as mass accretion rates and jet powers.

\noindent
{\it Accretion rate --}
In order to estimate the uncertainty in $\dot{m}$ from the AD models, we assume that the bolometric luminosity of LLAGNs can be approximated as the ADAF total luminosity, $L_{\rm bol} \sim L_{\rm adaf}$. We note that this may not be a good approximation for systems in which the thin disk is truncated at low radii, since in such situations the truncated disk can contribute a significant part of the luminosity. Correspondingly, for LLAGNs in which the emission is dominated by the jet, this approximation is also not appropriate. We use the fitting formulae for the ADAF radiative efficiencies  computed by \cite{xie12} for different values of $\delta$, assuming $s=0.4$ and $R_{\rm o}=100$ (cf. their Table 1). We evaluate how much $\dot{m}_{\rm o}$ varies by changing the value of $\delta$ between $10^{-3}$ and 0.5, for $L_{\rm bol}= {\rm const}$. We find that $\dot{m}_{\rm o}$ is affected by a systematic uncertainty of 0.2 dex due to our uncertain knowledge of the value of $\delta$. 

The above estimate is appropriate for fixed values of $s$ and $R_{\rm o}$. It is possible that  ADAFs have systematically different amounts of mass-loss due to outflows or convection than what we considered (e.g., \citealt{Shcherbakov12}). If this is the case, then our values of $\dot{m}$ could be biased towards higher rates if $s>0.3$ and lower rates if $s<0.3$ (the fiducial value of $s$ that we adopted in our SED models). In order to estimate the degree to which different values of $s$ could affect the estimated accretion rates, we computed a grid of ADAF SED models for $R_{\rm o}=10^4$, $\delta=0.1$, $m=10^8$ and $\alpha=0.3$ for which we varied the accretion rates and the values of $s$ in the ranges $\dot{m}_{\rm o}=10^{-4}-0.01$ and $s=0.1-0.5$. By considering how much the accretion rate varies in order to reproduce a fixed value of $L_{\rm bol}$, we estimate rough uncertainties in $\dot{m}_{\rm o}$ of a factor of 5 in each direction (0.7 dex; slightly smaller for smaller values of $R_{\rm o}$).

Combining the above uncertainties in quadrature with the systematic scatter of 0.23 dex resulting from the $M-\sigma$ relation \citep{Tremaine02}, we estimate a systematic uncertainty in $\dot{m}_{\rm o}$ of $\sim 0.8$ dex (keeping in mind that this is a rough estimate  since $s$ and $\delta$ should be correlated). 
The uncertainty in the ``Bondi accretion power'' $P_{\rm Bondi}$ as defined in the main text is estimated to be of the same order.

We expect that the uncertainty in the accretion rate due to our lack of understanding in the structure of ADAFs (values of $s$ and $\delta$) and the uncertainty in the black hole mass will dominate over the observational sources of uncertainty (i.e., error in the observed fluxes, paucity of SED sampling in some sources). 

\noindent
{\it Bolometric luminosity --}
The bolometric luminosities and Eddington ratios for the LINERs in our sample are affected by a factor of 5 ($\sim 0.7$ dex) uncertainties according to the estimate of EHF10.  Correspondingly, the typical uncertainty in $\eta_{\rm rad}$ is $\sim 0.8$ dex.
The typical relative uncertainty in the X-ray luminosities was estimated by EHF10 as $\sigma L_X/L_X \approx 0.2$ which corresponds to $\approx 0.1$ dex.

\noindent
{\it Jet power --}
In order to estimate the uncertainty in the jet powers derived from the SED models, we consider for illustration how the relativistic jet model (\textsection \ref{sec:jet}) reproduces the radio spectrum of NGC 4594. By fixing $\epsilon_e=0.1$ and $\epsilon_B=0.01$, we consider two values of the bulk Lorentz factor, 2.3 (the fiducial one) and 7 as suggested by \cite{Merloni07}, and vary $\dot{m}_{\rm jet}$ in order to fit the radio SED. The uncertainty in the value of $\Gamma_j$ results in a factor of 5 systematic uncertainty in $P_{\rm jet}$, similarly to the uncertainty affecting $L_{\rm bol}$. The uncertainty in $P_{\rm jet}/L_{\rm Edd}$ corresponds to $\sim 0.76$ dex. The uncertainty affecting the jet kinetic efficiency should be of the same order of magnitude as the one affecting $P_{\rm jet}/L_{\rm Edd}$.

\noindent
{\it Radio loudness --}
The uncertainty in the radio loudness depends on the corresponding errors affecting the radio and optical luminosities. We estimate that the typical uncertainties in $L_\nu (6 \; {\rm cm})$ and $L_\nu (4400 {\rm \AA})$ correspond to $\sim 0.1$ dex and $\sim 0.5$ dex, respectively. The latter error is due to the uncertain amount of intrinsic extinction. Therefore, our $R_o$ estimates have a $\sim 0.5$ dex uncertainty.

\section{Fits to sparsely sampled SEDs} \label{ap:seds}

We present in this section the model fits to the 9 SEDs that, although displaying nuclear X-ray point sources observed with \emph{Chandra}, did not pass the quality cut discussed in Section \ref{sec:sample} for lacking a radio core or HST optical/UV observations.

 but nevertheless have estimates of the black holes mass and X-ray SEDs available. The SED fits are shown in Figures \ref{fig:n266}-\ref{fig:seds07}. In the subsections below, we briefly describe the details of the fits to the SED of each object. The basic properties of the LLAGNs and the model parameters are summarized in tables \ref{tab:samplep} and \ref{tab:modelsp}, respectively.

\begin{table*}
\centering
\caption{Sample of galaxies with sparsely sampled SEDs and their basic properties. Notation is the same as in Table \ref{tab:sample}.}
\begin{tabular}{@{}llrccccc@{}}
\hline
Galaxy & Hubble & Distance$^{\rm b}$ & $\log$  & LINER & $L_{\rm X}$ & $L_{\rm bol}$ & $L_{\rm bol}/L_{\rm Edd}$ \\
       & Type   & (Mpc)    & $(M_{\rm BH}/M_\odot)$ & Type  & (erg s$^{-1}$)$^{\rm c}$ & (erg s$^{-1}$)$^{\rm d}$ &   \\
\hline
NGC  266                 &  SB(rs)ab   & 62.4 (1) & 7.6$^{\rm e}$ & L1 & $7.4 \times 10^{40}$ & $2.2 \times 10^{42}$ & $4 \times 10^{-4}$  \\
NGC 2681                 &  SBA(rs)0/a & 16.0 (2) & 7.1 & L1      & $6.1 \times 10^{38}$ & $1.8 \times 10^{40}$ & $1 \times 10^{-5}$  \\
NGC 3169                 &  SA(s)a     & 19.7 (1) & 7.8 & L2      & $1.1 \times 10^{41}$ & $3.3 \times 10^{42}$ & $4 \times 10^{-4}$  \\
NGC 3226                 &  E2         & 21.9 (2) & 8.1 & L1      & $5.0 \times 10^{40}$ & $1.5 \times 10^{42}$ & $1 \times 10^{-4}$  \\
NGC 3379         &  E1         &  9.8 (2) & 8.2 & L2/T2   & $1.7 \times 10^{37}$ & $5.1 \times 10^{38}$ & $3 \times 10^{-8}$  \\
NGC 4457                 &  SAB(s)0/a  & 10.7 (4) & 6.9 & L2      & $1.0 \times 10^{39}$ & $3.0 \times 10^{40}$ & $3 \times 10^{-5}$  \\
NGC 4494                 &  E1-2       & 15.8 (2) & 7.6 & L2      & $9.2 \times 10^{38}$ & $2.8 \times 10^{40}$ & $6 \times 10^{-6}$  \\
NGC 4548           &  SBb(rs)    & 15.0 (3) & 7.6 & L2      & $5.4 \times 10^{39}$ & $1.6 \times 10^{41}$ & $3 \times 10^{-5}$  \\
\hline
\end{tabular}
\begin{flushleft}
\footnotesize 
Notes: \\
(e) The mass estimate for NGC 266 is subject to a large uncertainty ($\approx 1$ dex) since it is based on using the fundamental plane of \citet{Merloni03} (see text).
\end{flushleft}
\label{tab:samplep}
\end{table*}

\subsection{NGC 0266}

\begin{figure}
\centering
\includegraphics[scale=0.5,trim=30 0 10 10,clip=true]{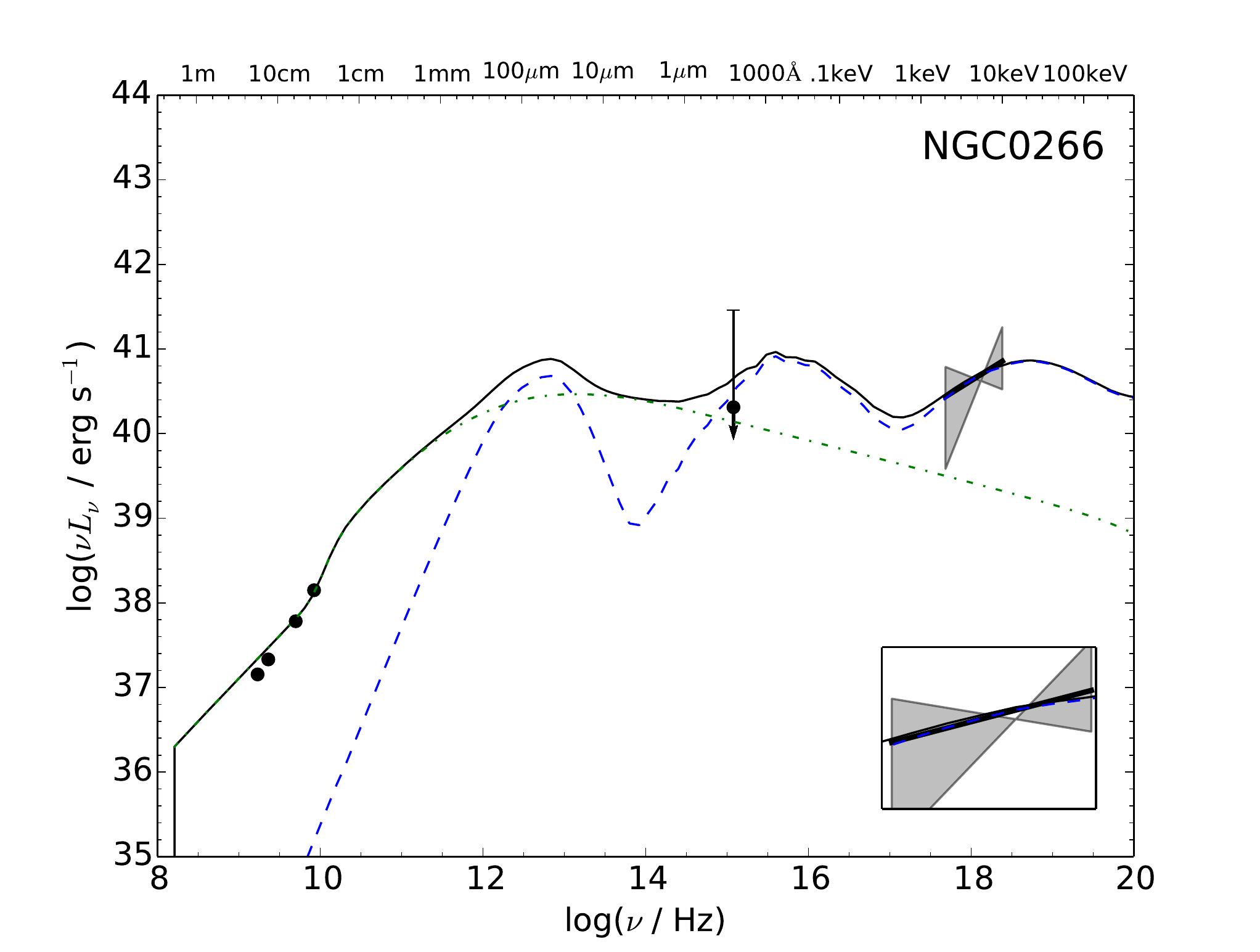}
\caption{The SED and ADAF-dominated model for NGC 266. The dashed, dotted and dot-dashed lines correspond to the emission from the ADAF, truncated thin disk and jet, respectively. The solid line represents the sum of the emission from all components. The inset shows the zoomed 2-10 keV spectrum.}
\label{fig:n266}
\end{figure}

The AD model accounts slightly better for the best-fit X-ray slope, but note the pronounced uncertainty on the value of observed photon index. The JD model requires $p<2$ which is below the usual range $2<p<3$ suggested by relativistic shock theory \citep{Bednarz98, Kirk00} and hence is disfavored (Fig. \ref{fig:n266}).

\subsection{NGC 1553}	\label{sec:1553}

\begin{figure}
\centering
\includegraphics[scale=0.5,trim=30 0 10 10,clip=true]{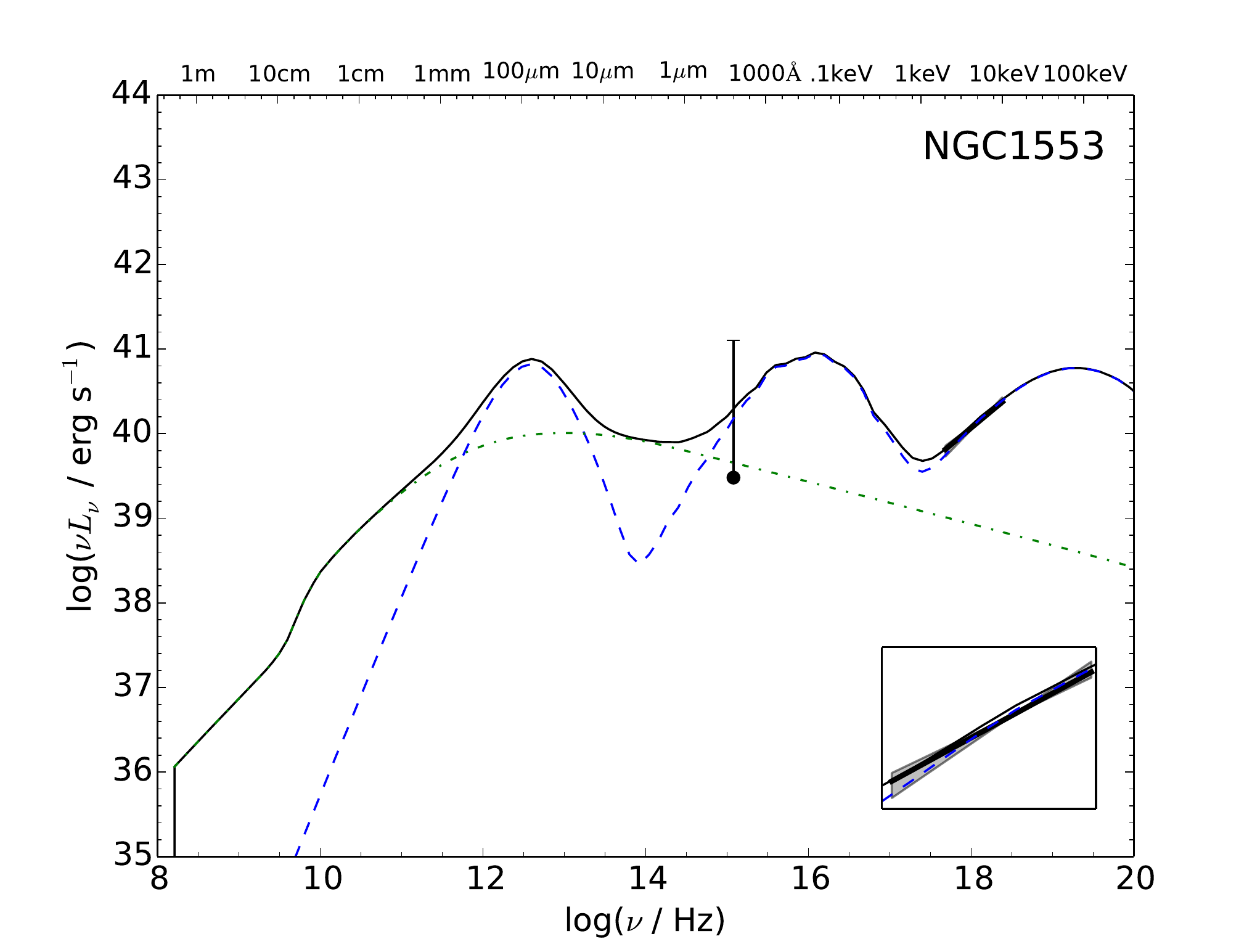}
\caption{Same as Figure \ref{fig:n266} for NGC 1553.}
\label{fig:n1553}
\end{figure}

The Bondi accretion rate was estimated by \citet{Pellegrini05}. 
Even though the JD model roughly accounts for the estimated $L_X$, it fails to fit the slope of the X-ray spectrum even with a small value of $p$. We take this as evidence that this source is unlikely to be JD (Fig. \ref{fig:n1553}). The estimated $\dot{m}_{\rm o}$ is consistent with the lower limit on $\dot{m}_{\rm Bondi}$ obtained by \citet{Pellegrini05}.

\subsection{NGC 2681}

\begin{figure*}
\centerline{
\includegraphics[scale=0.5]{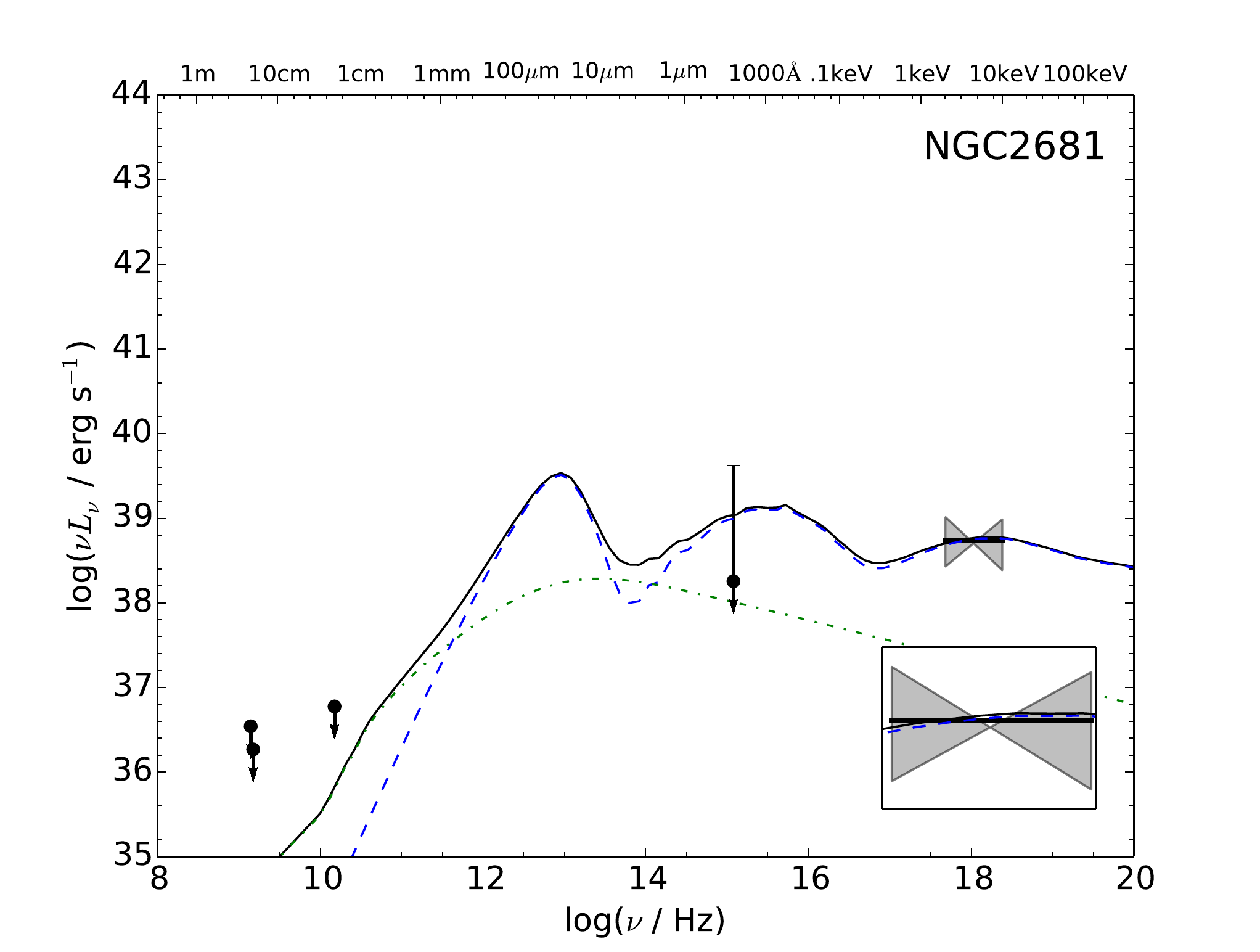}
\hskip -0.3truein
\includegraphics[scale=0.5]{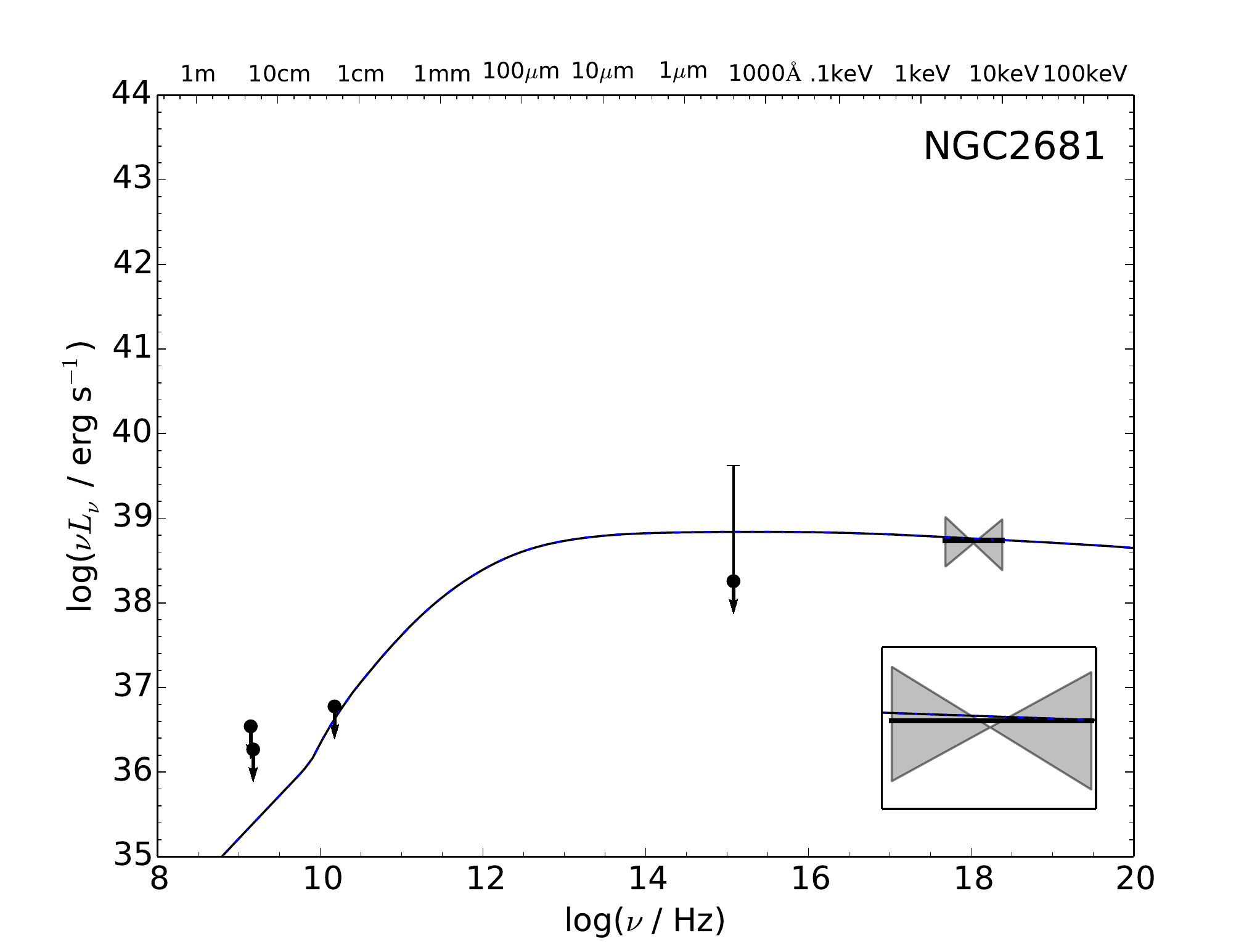}
}
\caption{SEDs and coupled accretion-jet models for NGC 1553. The left panel shows the AD models for each object while the right panel displays the JD models.}
\label{fig:n2681}
\end{figure*}

\begin{figure*}
\centerline{
\includegraphics[scale=0.5]{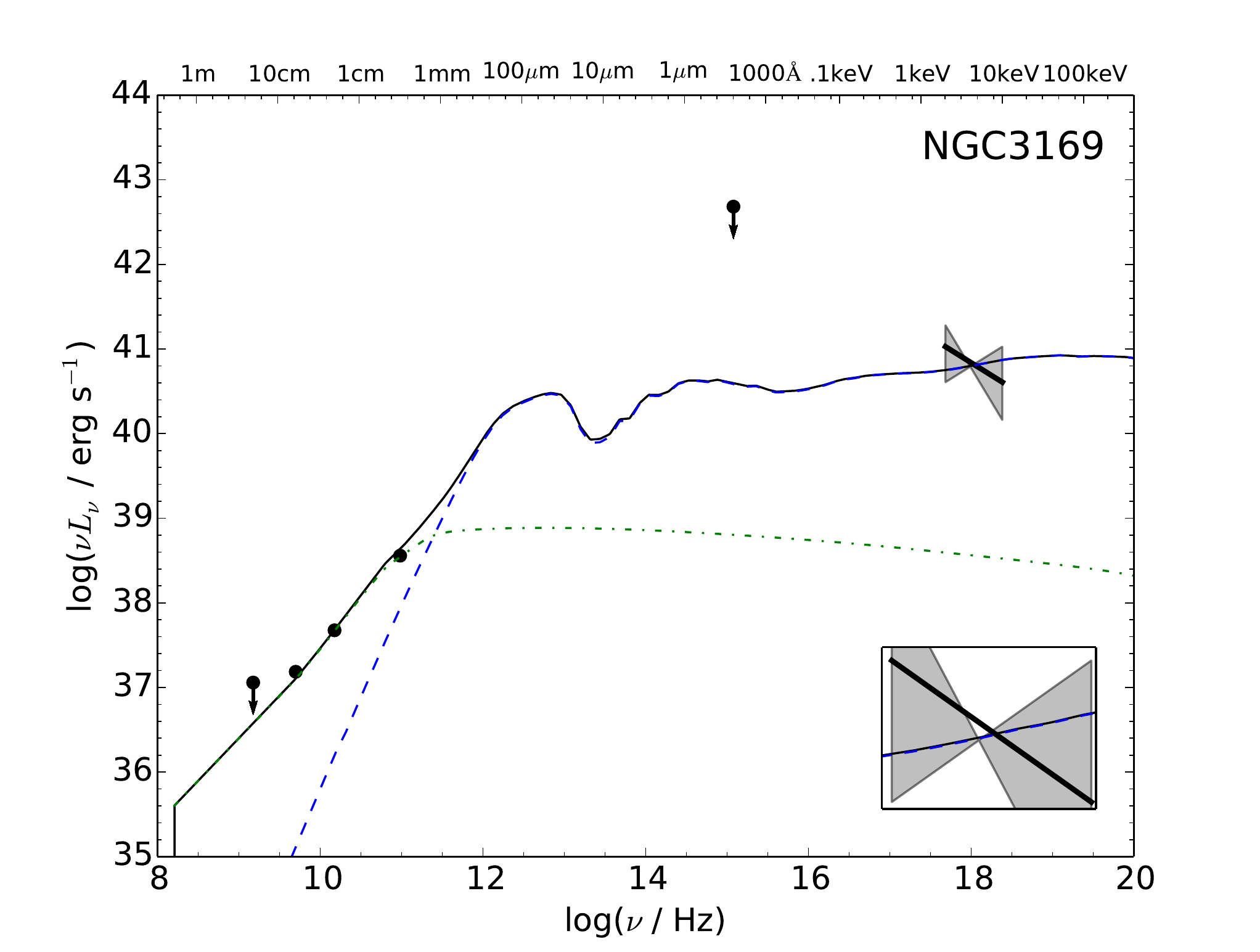}
\hskip -0.3truein
\includegraphics[scale=0.5]{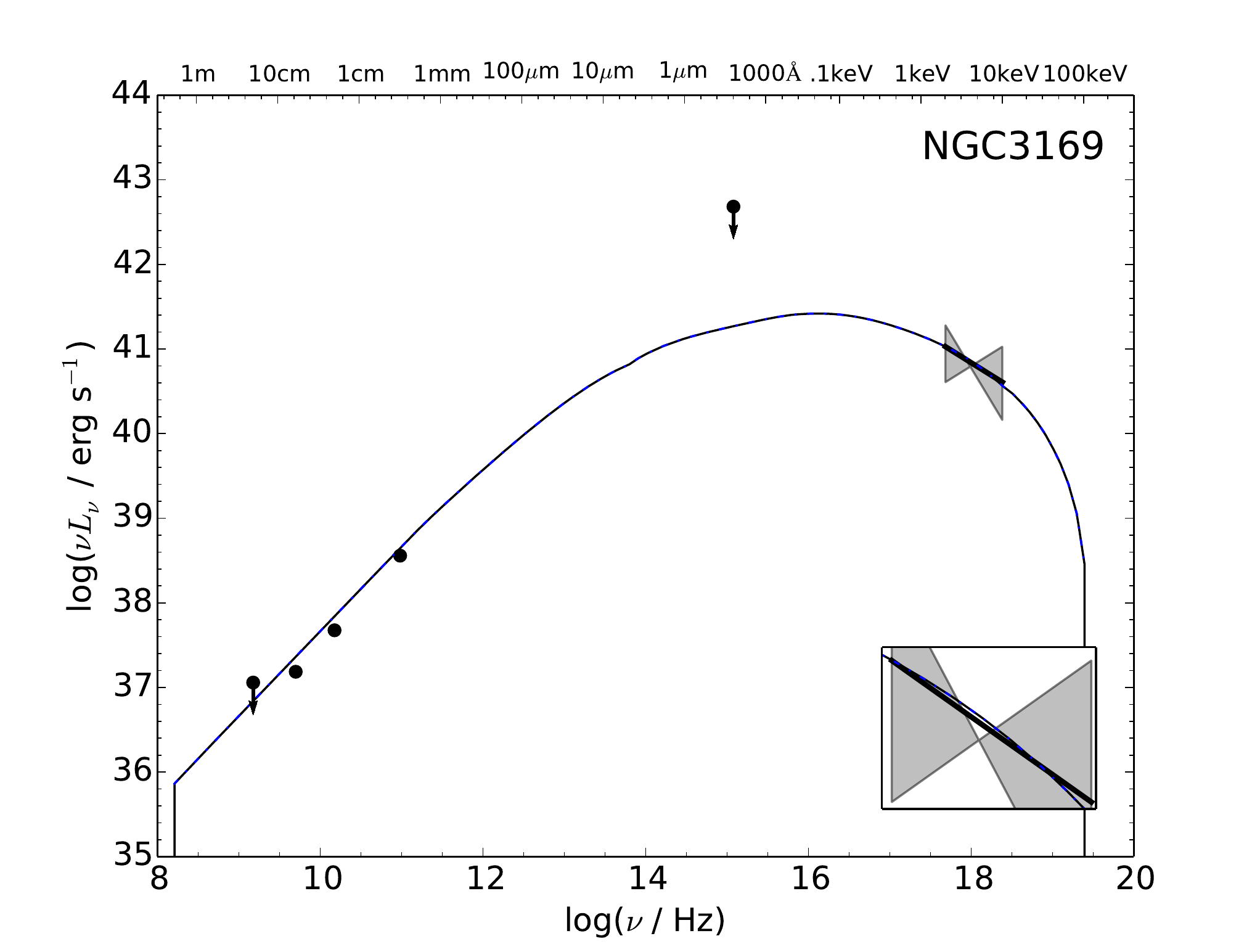}
}
\centerline{
\includegraphics[scale=0.5]{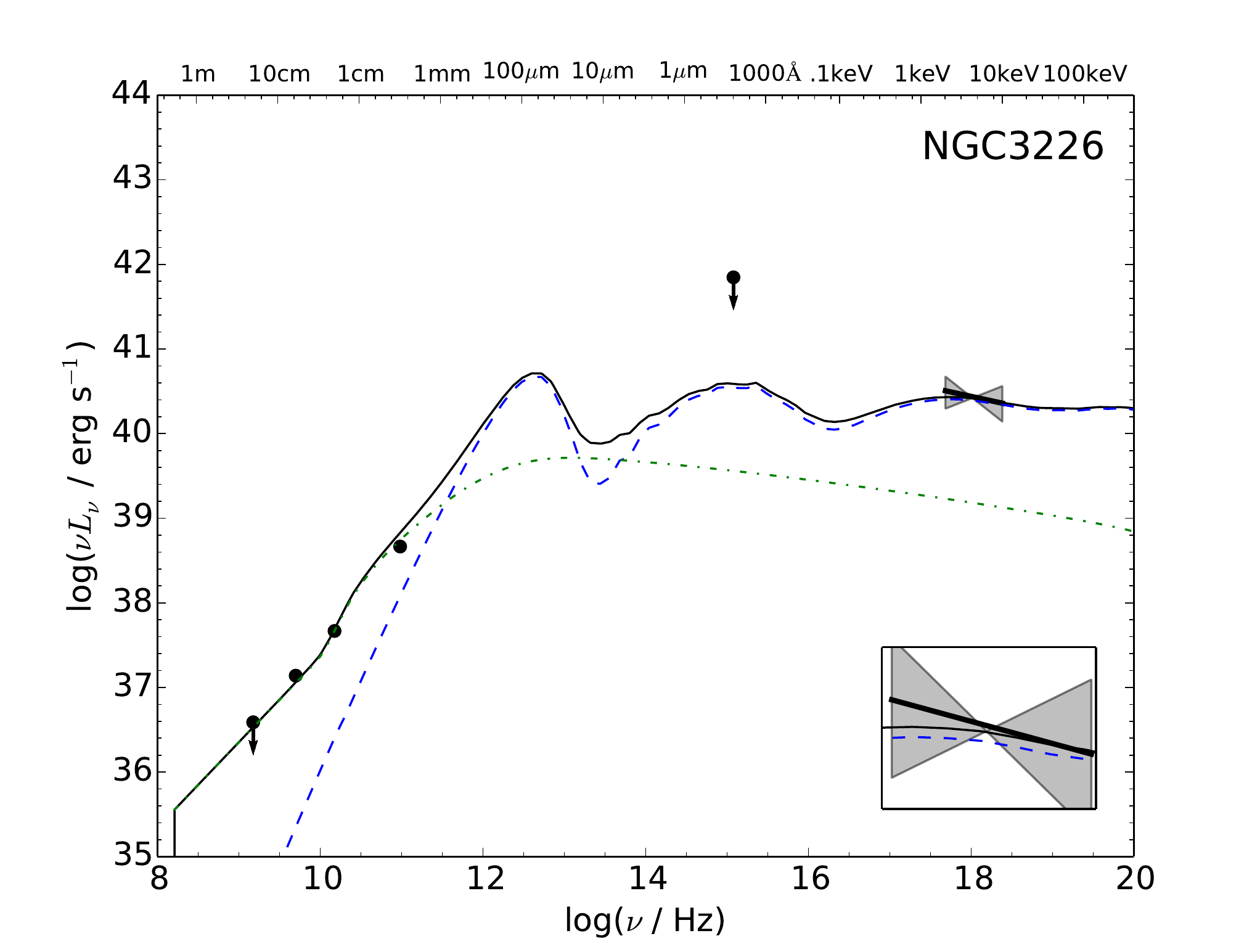}
\hskip -0.3truein
\includegraphics[scale=0.5]{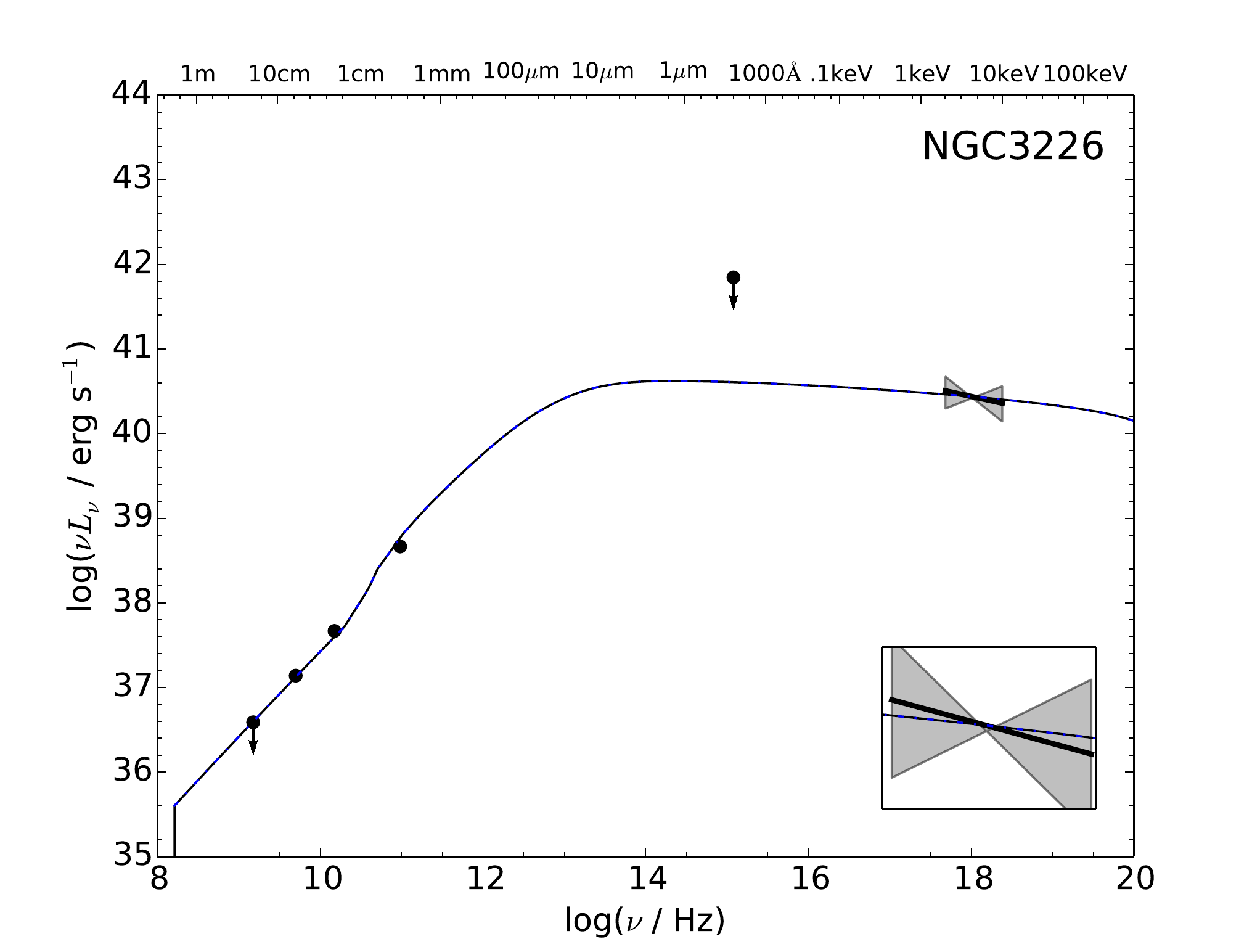}
}
\centerline{
\includegraphics[scale=0.5]{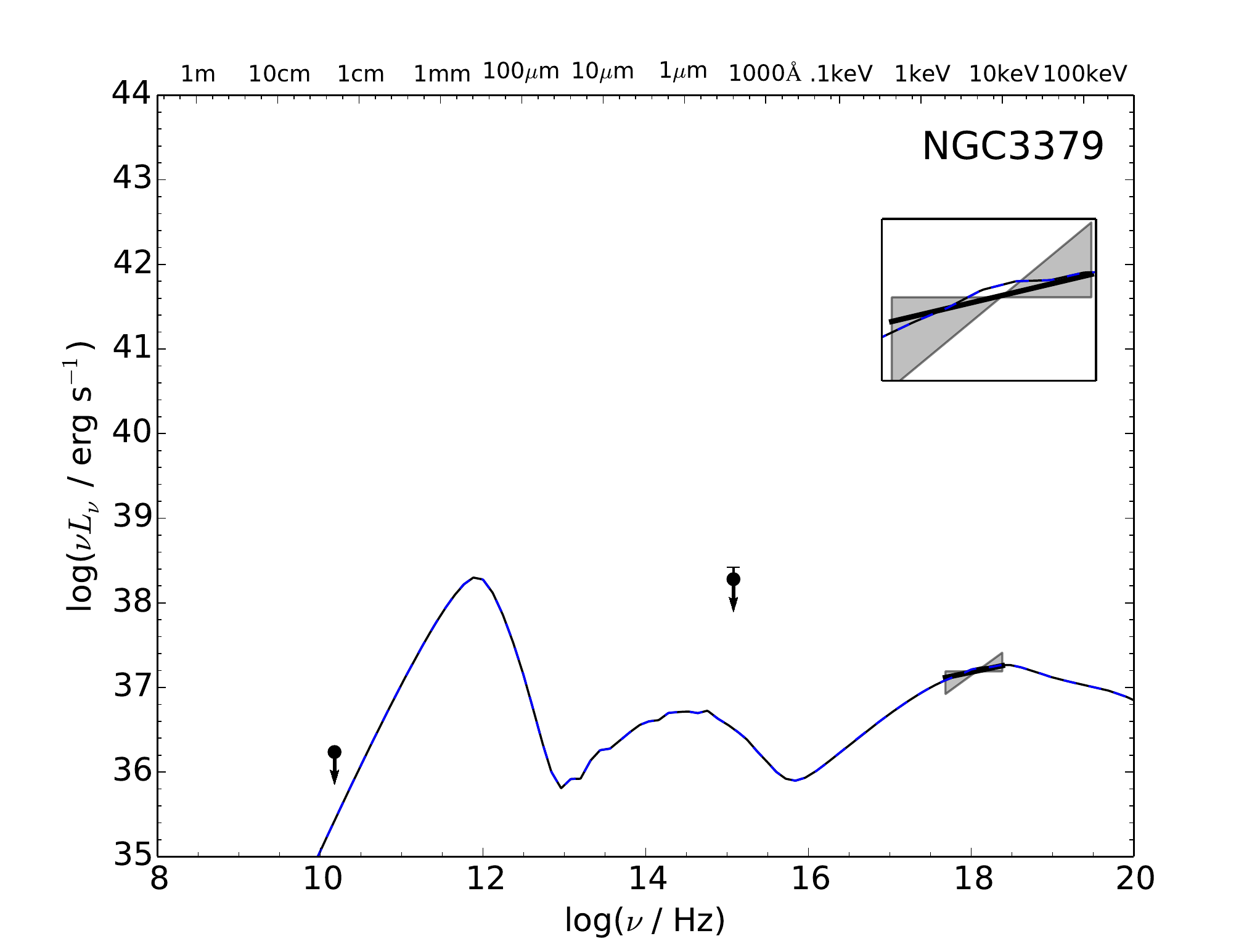}
\hskip -0.3truein
\includegraphics[scale=0.5]{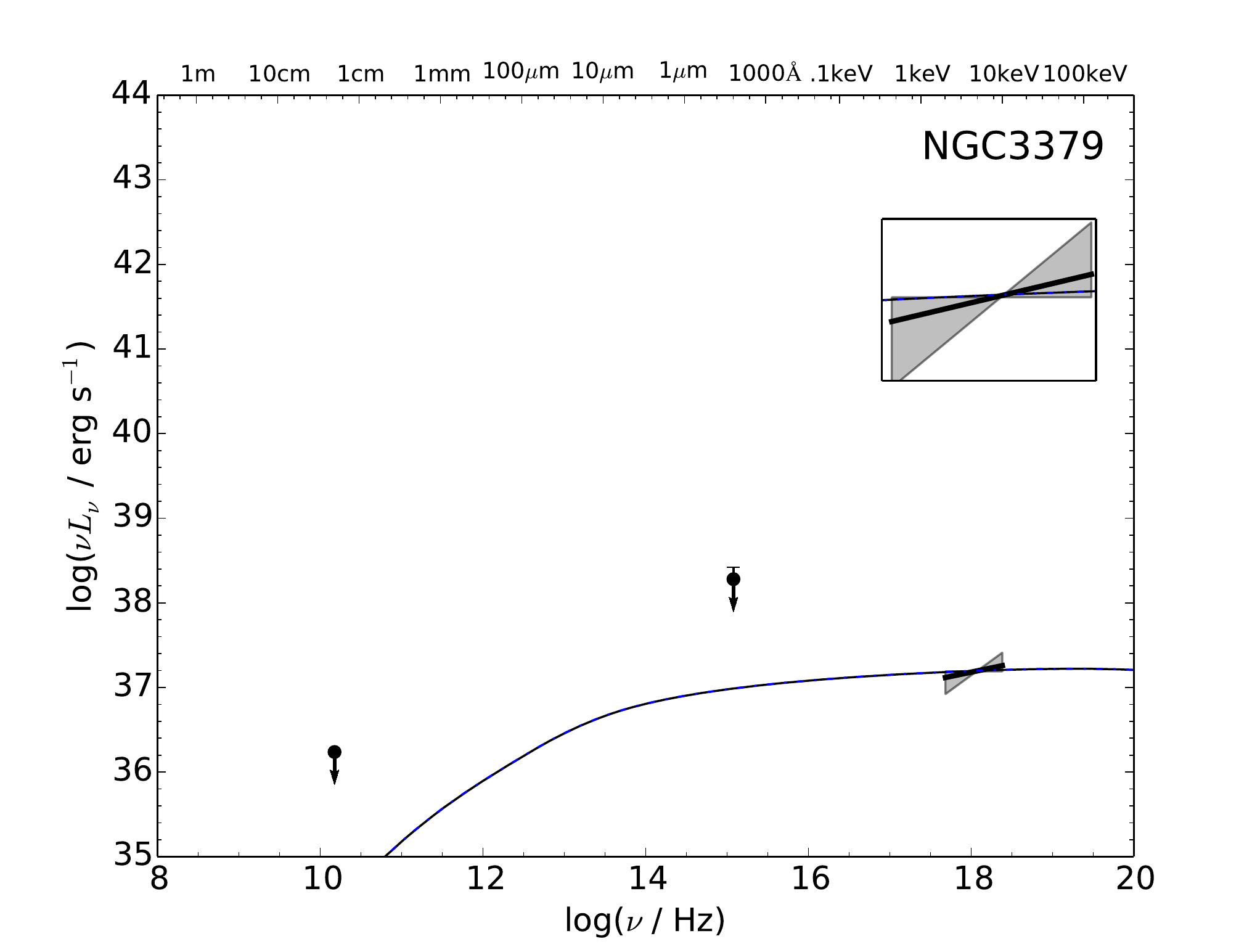}
}
\caption{Same as Figure \ref{fig:seds01} for NGC 2681, NGC 3169 and NGC 3226.}
\label{fig:seds06}
\end{figure*}

\begin{figure*}
\centerline{
\includegraphics[scale=0.5]{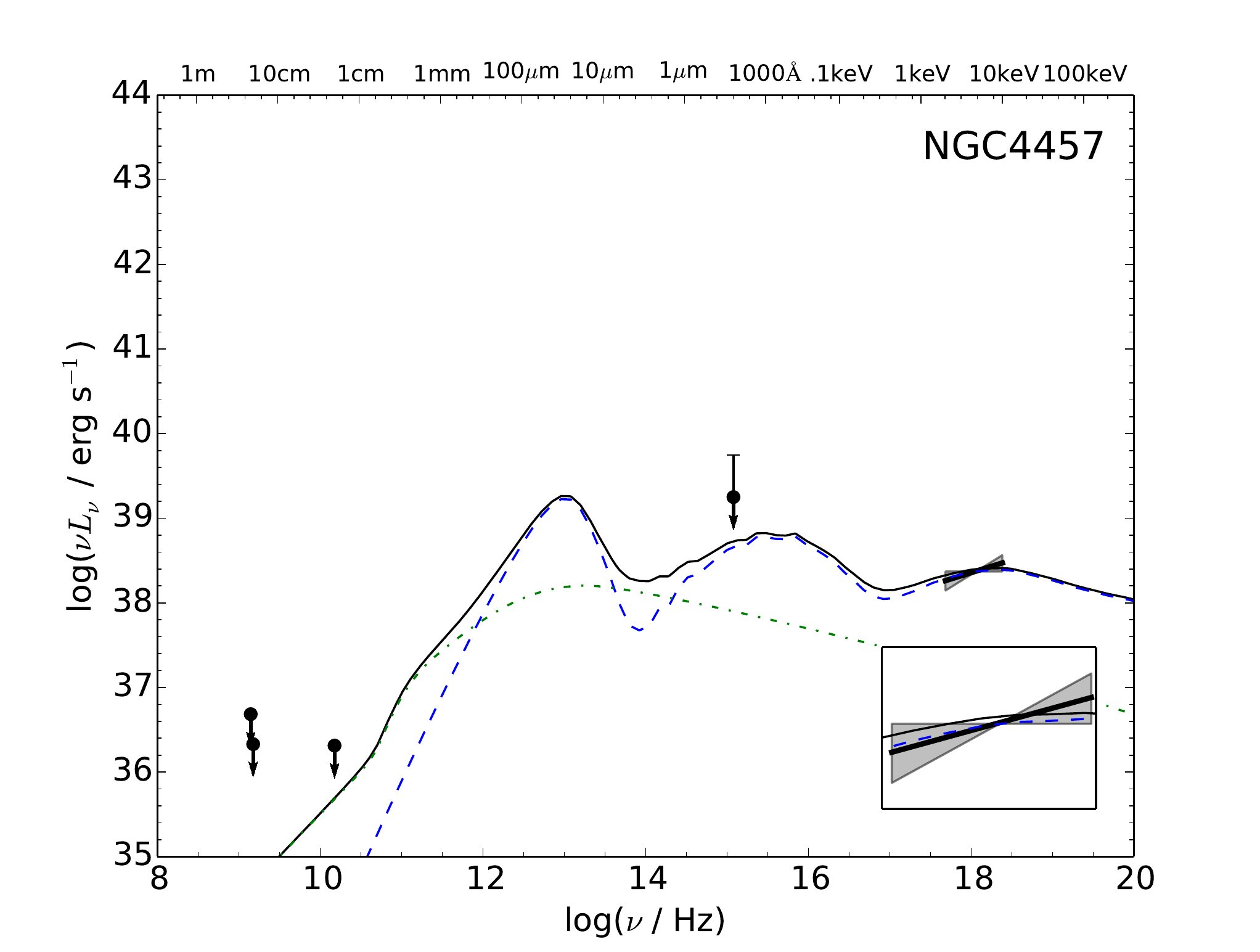}
\hskip -0.3truein
\includegraphics[scale=0.5]{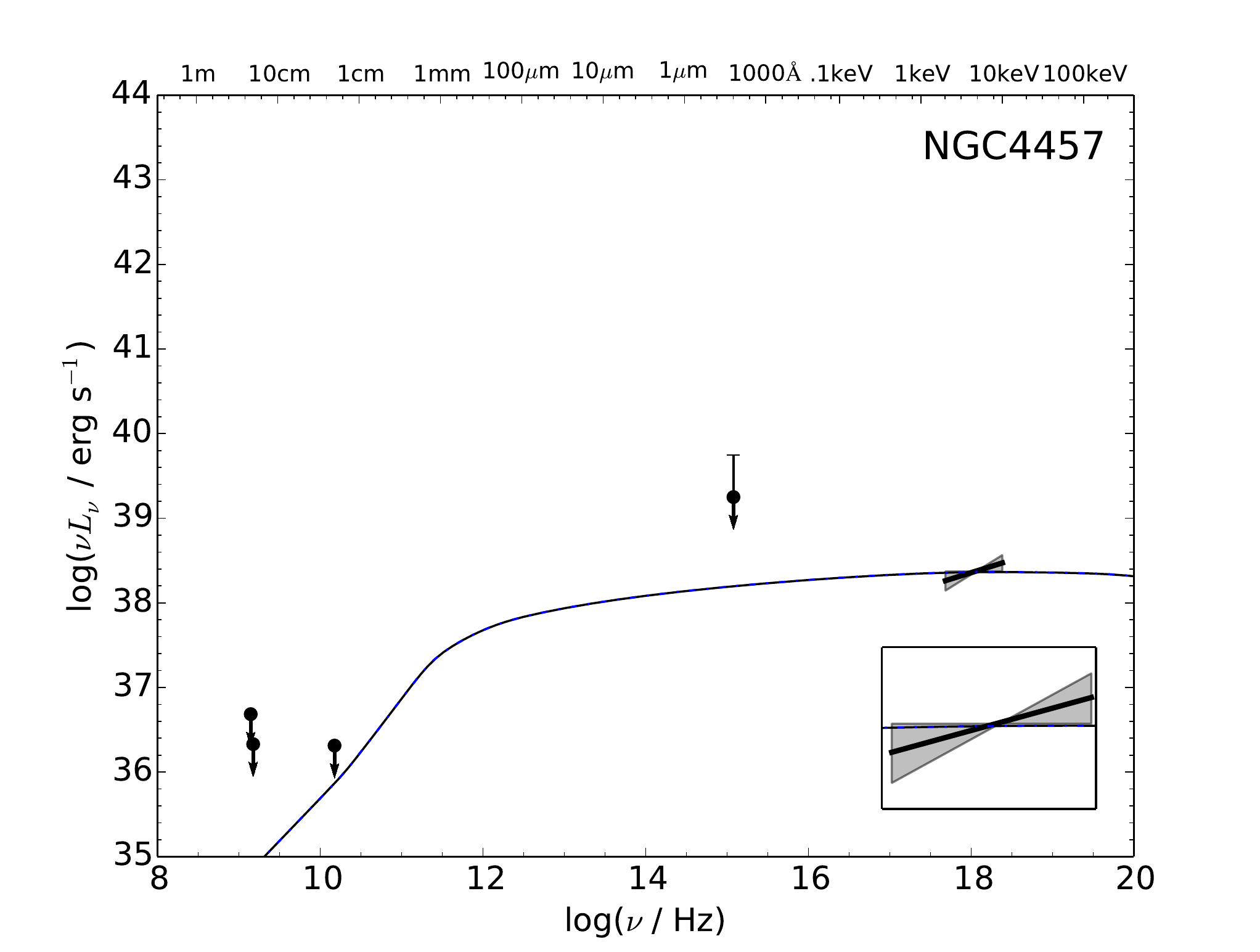}
}
\centerline{
\includegraphics[scale=0.5]{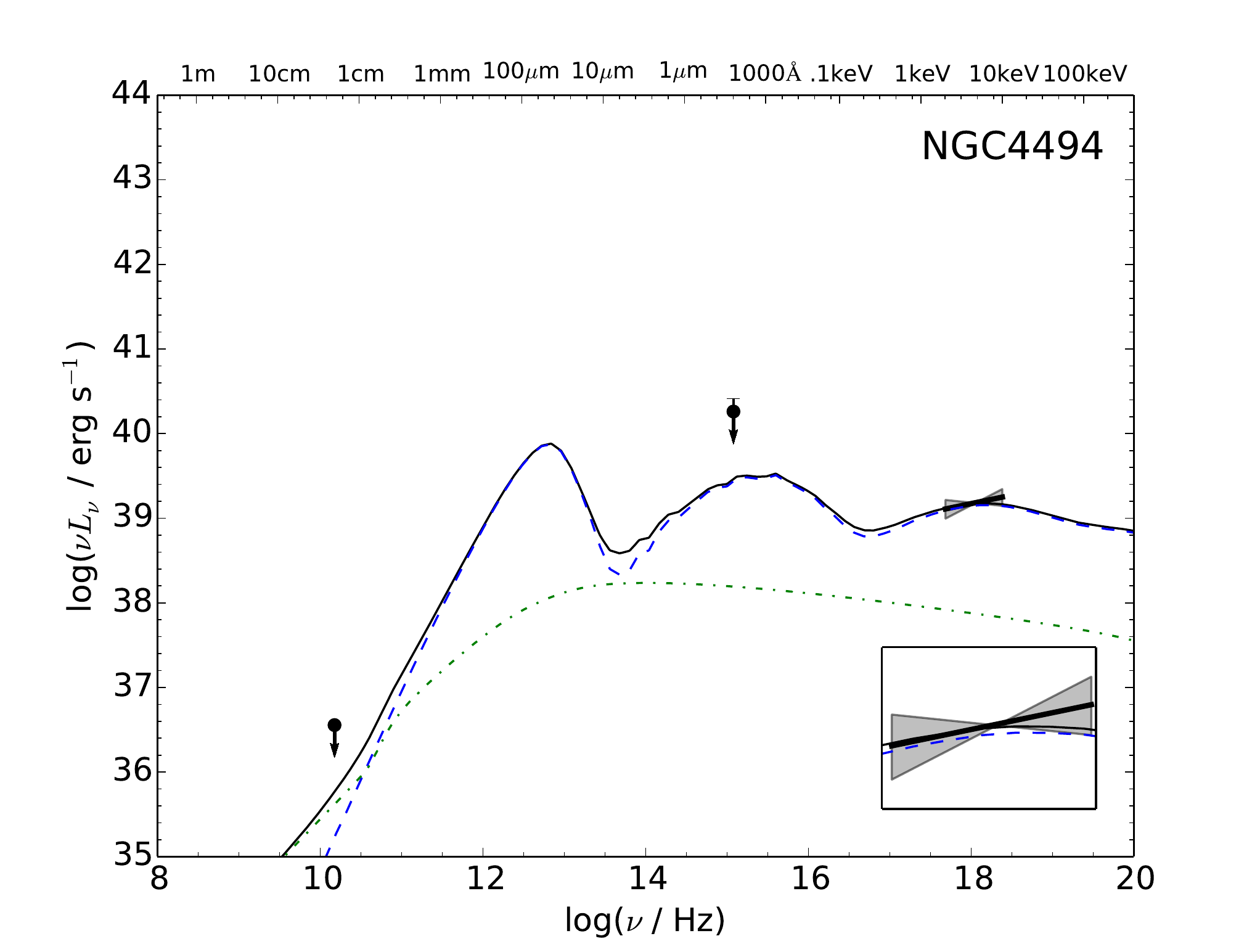}
\hskip -0.3truein
\includegraphics[scale=0.5]{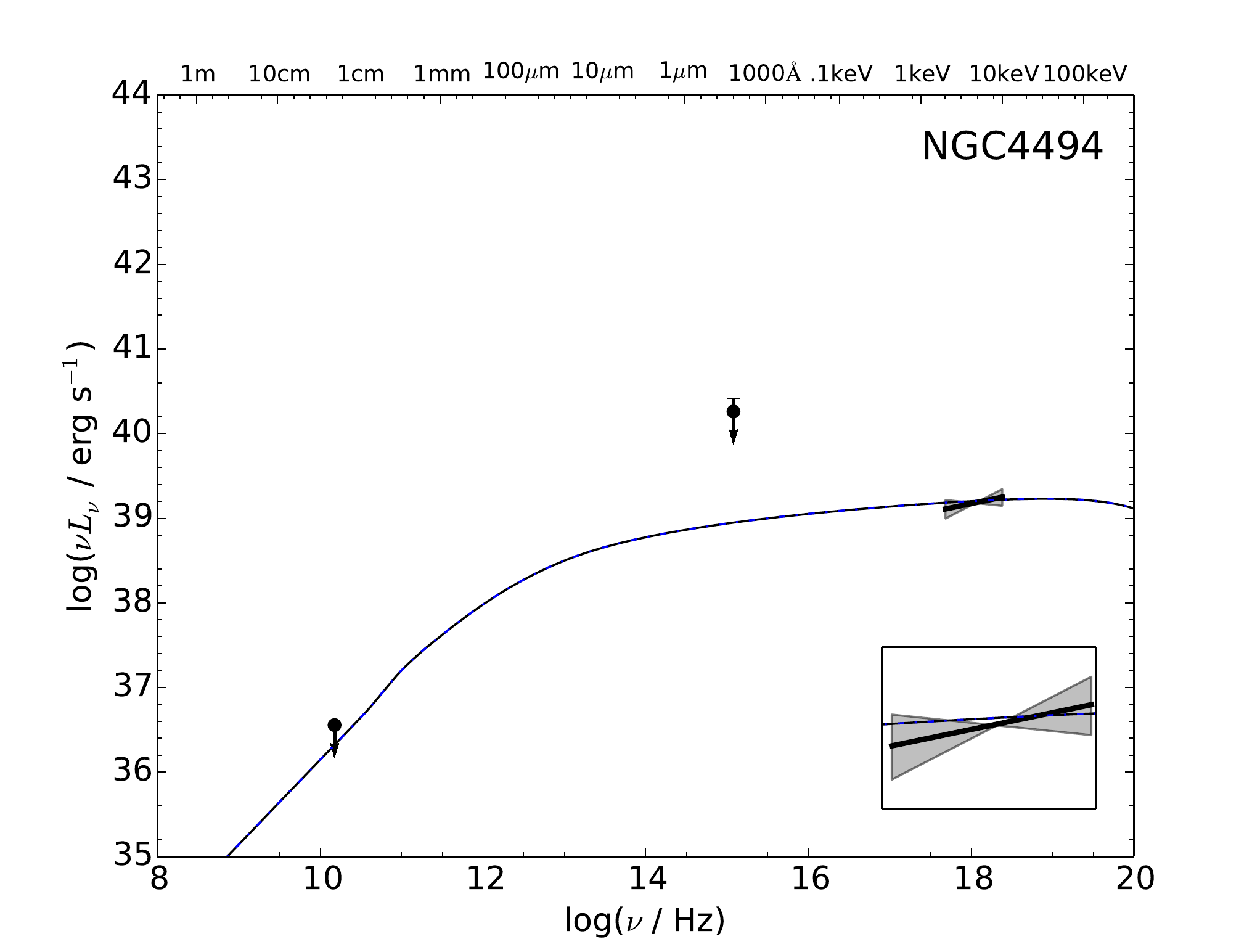}
}
\centerline{
\includegraphics[scale=0.5]{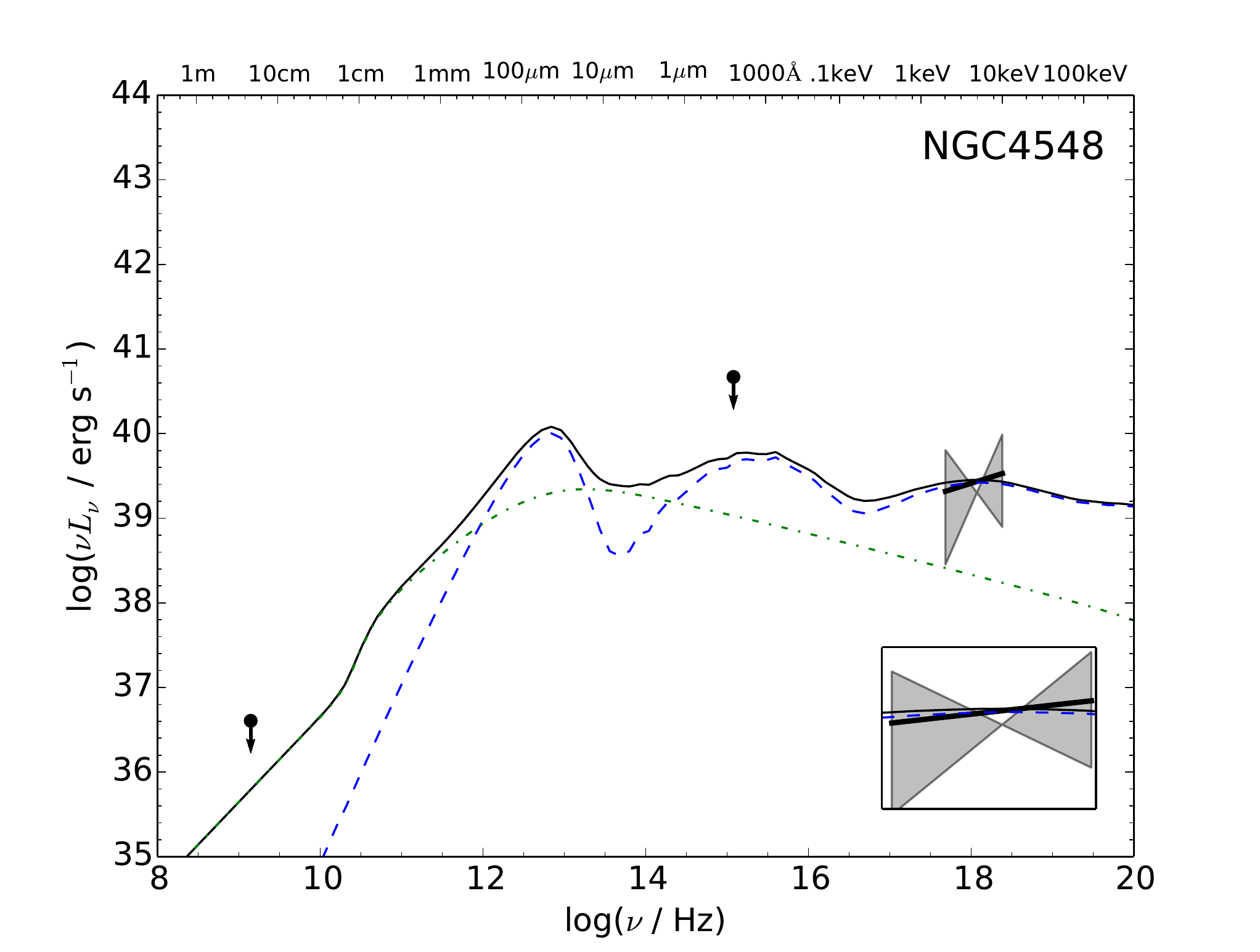}
\hskip -0.3truein
\includegraphics[scale=0.5]{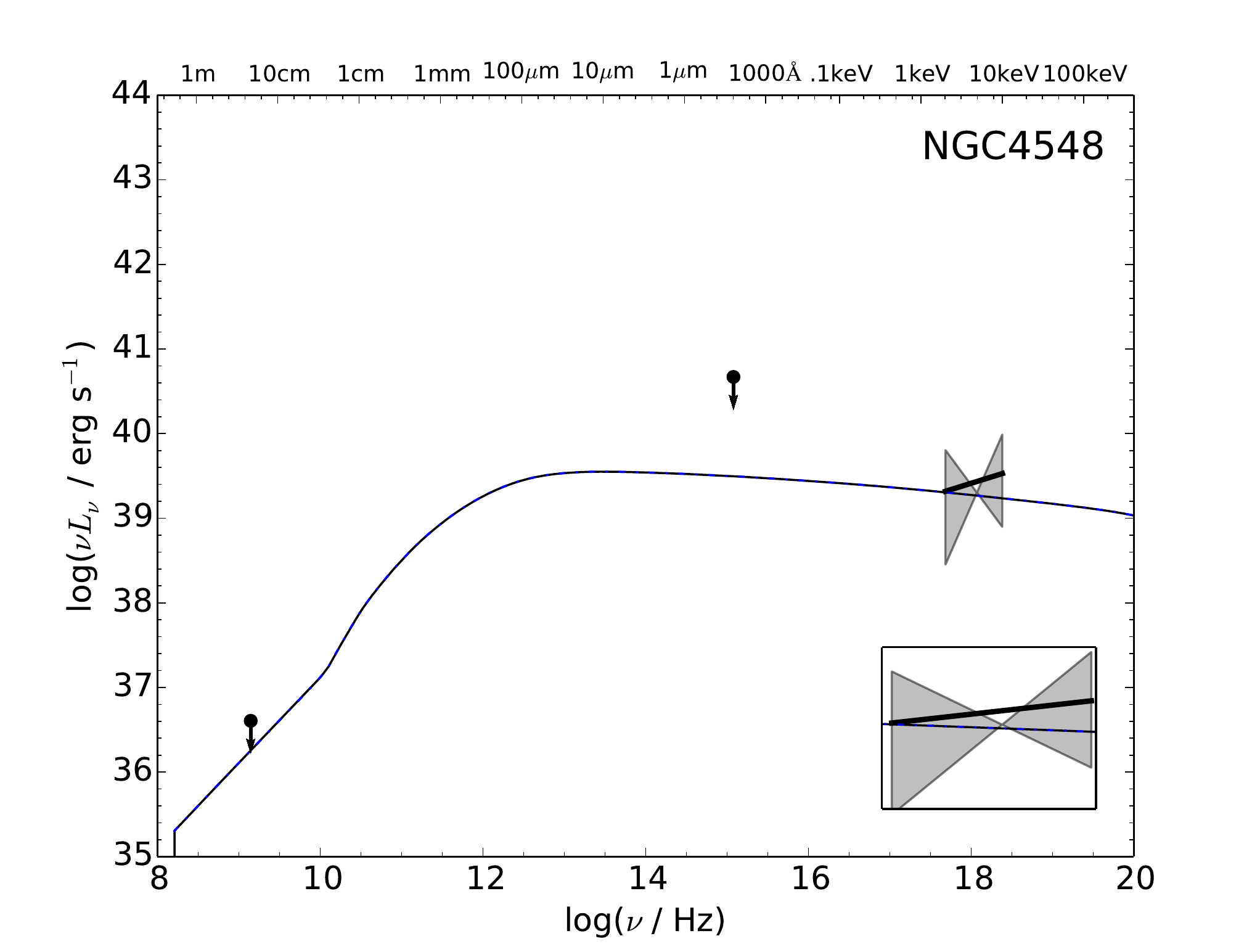}
}
\caption{Same as Figure \ref{fig:seds01} for NGC 4278, NGC 4457 and NGC 4494.}
\label{fig:seds07}
\end{figure*}

For this LLAGN, there are not enough optical observations to fit the truncated thin disk model and estimate the transition radius (Fig. \ref{fig:n2681}).

\subsection{NGC 3169}

As was the case of NGC 2681, for this LLAGN there are not enough optical observations to fit the truncated thin disk model and estimate the transition radius. 
The JD model accounts well for the available data. The inferred extreme values of $\epsilon_e$ and $\epsilon_B$ imply that essentially all the energy in the post-shock region of the jet is carried by the particles.

\subsection{NGC 3226}

The SED of this LLAGN is well fitted by both an AD and JD type of models. As is the case of NGC 2681, there are no good optical band constraints on the emission of the truncated thin disk.

\subsection{NGC 3379}

This source has only two data points outside the X-ray band, one in the radio and the other in the optical, both of which correspond to upper limits. Therefore, there are few constraints for the accretion-jet model. The Bondi rate was estimated by \citet{David05}. 

The AD model is able to reproduce the observed SED given the few constraints available, but there are no radio data to constrain the jet model in this case so we don't include it. The available data are not enough to constrain well the transition radius. 
The accretion rate required by the AD model is more than an order of magnitude higher than $\dot{m}_{\rm Bondi}$.  One possible explanation for this result is that the accretion rate is enhanced by gas released by stars. This is in line with the findings of \citet{Soria06, Soria06b} for a sample of quiescent early-type galaxies.

\subsection{NGC 4457}

Since there are only upper limits in the radio band, the jet power for this object was estimated using the observation at $\nu = 1.5 \times 10^{10}$ Hz and the \citet{Merloni07} correlation.

\subsection{NGC 4494}

Same as NGC 4457.

\subsection{NGC 4548}

Same as NGC 4457.
Given the large uncertainty in the photon index, the X-ray spectrum does not provide good constraints for the models.

\begin{table*}
\centering
\caption{Model parameters resulting from the sparsely sampled SED fits. For all models listed below, $R_{\rm o}=10^4$, $s=0.3$ and $i=30^\circ$.}
\scriptsize
\begin{tabular}{@{}cccccccccccc@{}}
\hline
Galaxy & Model & $\dot{m}_{\rm o}$ &  $\delta$ & $\dot{m}_{\rm jet}$ & $p$ & $\epsilon_e$ & $\epsilon_B$ &  $P_{\rm jet}^{\rm mod}$ & $P_{\rm jet}^{\rm obs}$ & $\dot{m}_{\rm Bondi}$ & Refs. \\
& &  &   &  &  &  &  &  &  &  &   and notes \\
\hline
NGC 0266 &  AD & $5.5 \times 10^{-3}$ & 0.3 &  $5 \times 10^{-6}$ & 2.5 & 0.1 & 0.01 &  $1.1 \times 10^{42}$ &  & - &   \\
\hline
NGC 1553 &  AD & $2.5 \times 10^{-3}$ &  0.3 &  $7 \times 10^{-7}$ & 2.5 & 0.1 & 0.08 &  $3.7 \times 10^{41}$ &  & &  \\
\hline
NGC 2681 &  JD &  - & - &  $5 \times 10^{-7}$ & 2.1 & 0.2 & 0.1 &  $4.3 \times 10^{40}$ & $1.4 \times 10^{41}$ & - &  b \\ 
NGC 2681 &  AD & $7.5 \times 10^{-3}$ &  0.01 &  $10^{-7}$ & 2.5 & 0.1 & 0.1 &  $8.7 \times 10^{39}$ &  & - &  \\
\hline
NGC 3169 & JD &  - & - &  $3 \times 10^{-5}$ & 2.8 & 0.99 & $1.5 \times 10^{-7}$ &  $1.3 \times 10^{43}$ &  $10^{42}$ & - &  b \\ 
NGC 3169  & AD & 0.03 &  0.01 &  $2.5 \times 10^{-6}$ & 2.2 & 0.01 & 0.01 &  $10^{42}$ &   & - &   \\ 
\hline
NGC 3226 & JD &  - & - &  $10^{-6}$ & 2.2 & 0.5 & $2 \times 10^{-3}$ &  $8.7 \times 10^{41}$ &  $1.1 \times 10^{42}$ & - &  b \\ 
NGC 3226  & AD & 0.12 &  0.01 &  $5 \times 10^{-7}$ & 2.3 & 0.1 & 0.01 &  $4.3 \times 10^{41}$ &   & - &   \\ 
\hline
NGC 3379 & JD & - & - &  $2 \times 10^{-9}$ & 2.05 & 0.2 & 0.15 &  $2.1 \times 10^{39}$ & - &  $1.6 \times 10^{-5}$ &  2 \\ 
NGC 3379  & AD & $6.5 \times 10^{-4}$ &  0.01 &  - & - & - & - & - & - &   &   \\ 
\hline
NGC 4457 & JD &  - & - &  $1.1 \times 10^{-6}$ & 2.01 & 0.3 & 0.01 &  $6 \times 10^{40}$ &  $<4 \times 10^{41}$ & - &  b \\ 
NGC 4457 & AD & $2.2 \times 10^{-3}$ &  0.1 &  $4 \times 10^{-7}$ & 2.5 & 0.05 & 0.01 &  $2 \times 10^{40}$ &  & - &  \\ 
\hline
NGC 4494 & JD &  - & - &  $10^{-6}$ & 2.01 & 0.6 & 0.001 &  $2.3 \times 10^{41}$ &  $<1.4 \times 10^{41}$ & - &   \\ 
NGC 4494 & AD & $6.5 \times 10^{-3}$ &  0.01 &  $10^{-7}$ & 2.3 & 0.1 & 0.01 & $2.3 \times 10^{40}$ &   & - &  \\ 
\hline
NGC 4548 &  JD &  - & - &  $1.5 \times 10^{-6}$ & 2.2 & 0.1 & 0.01 &  $3.4 \times 10^{41}$ &  $4.2 \times 10^{41}$ & - &  b  \\
NGC 4548 &  AD & $8 \times 10^{-3}$ &  0.01 &  $5 \times 10^{-7}$ & 2.5 & 0.1 & 0.01 &  $1.1 \times 10^{41}$ &   & - &   \\
\hline
\end{tabular}
\label{tab:modelsp}
\end{table*}

\bsp

\label{lastpage}

\end{document}